\documentclass[11pt,onecolumn,superscriptaddress,prb,aps,showpacs,amsmath,amssymb]{revtex4-2}

\makeatletter
\renewcommand{\p@subsection}{}
\renewcommand{\p@subsubsection}{}
\makeatother

\usepackage{graphicx}
\graphicspath{{manuscript/}{manuscript/figures/}}
\usepackage{dcolumn}
\usepackage{bm}
\usepackage{subfigure}
\usepackage{natbib}
\usepackage{multirow}
\usepackage{soul}
\usepackage{float}
\usepackage[normalem]{ulem}
\usepackage[usenames,dvipsnames]{color}
\usepackage{appendix}
\usepackage{mathrsfs}
\usepackage[colorlinks=true,linkcolor=blue,citecolor=blue,urlcolor=blue]{hyperref}
\usepackage{xspace}
\usepackage[most]{tcolorbox}

\usepackage[abbreviations]{glossaries-extra}
\makeglossaries
\newabbreviation{DFT}{DFT}{density-functional theory}

\usepackage{longtable}
\usepackage{booktabs}
\usepackage{array}
\usepackage{gensymb}

\usepackage{siunitx}
\usepackage{placeins}

\usepackage[percent]{overpic} %-> to control position of (a), (b), ... labels

\begin{document}

\raggedbottom % 

\title{Energetics of Coherent Heterophase Interfaces: Separating Chemical and Strain Contributions}

\author{Adolfo Salgado-Casanova} 
\affiliation{Department of Materials Science and Engineering, Northwestern University, Evanston, IL 60208, USA
}

\author{Jacob Pietryga}
\affiliation{Department of Materials Science and Engineering, Northwestern University, Evanston, IL 60208, USA
}

\author{Yi Xia}
\affiliation{Department of Mechanical \& Materials Engineering, Portland State University, Portland, OR 97201, USA
}

\author{Chris Wolverton}
\email{c-wolverton@northwestern.edu }
\affiliation{Department of Materials Science and Engineering, Northwestern University, Evanston, IL 60208, USA
}

\begin{abstract}

Coherent interfacial energies govern the formation and evolution of multiphase microstructures, but their calculation is complicated by the elastic energy required to accommodate lattice mismatch. 
Here, we construct a high-throughput DFT database of 365 coherent \mbox{FCC/FCC}, \mbox{BCC/BCC}, and \mbox{HCP/HCP} interfaces involving 41 metallic elements and lattice mismatches of up to 25\%. 
We use the supercell approach, consisting of constructing superlattices of different thicknesses to separate the coherency strain energy from the chemical contribution to the interfacial energy. 
We classify the results according to the long-range ordering tendency of each binary system and the sign of the calculated interfacial energy. 
Most phase-separating systems exhibit positive interfacial energies, whereas most compound-ordering systems exhibit negative values, reflecting favorable unlike-atom bonding. 
Exceptions show that the chemical preference at a particular interface need not coincide with bulk phase behavior: coherency strain can oppose ordering, and stable compounds can adopt structures distinct from the calculated superlattices.
No clear correlation of interfacial energy with lattice mismatch is observed, whereas coherency strain energy increases with mismatch. 
For phase-separating systems, the lowest interfacial energies generally occur along the closest-packed orientations, while this trend is absent for ordering systems.
Comparisons with reported interfacial energies and binary phase diagrams show qualitative consistency, although quantitative agreement varies with reference states, structural constraints, and temperature.  
The database provides a systematic reference for metallic coherent heterophase interfaces and demonstrates the importance of separating strain and chemical energy contributions when interpreting alloy behavior.

\end{abstract}

\date{\today}

\maketitle

\newpage
\section{Introduction}
\label{sec:intro}

Most computational materials databases focus on bulk, single-phase crystals, although many engineering materials contain multiphase microstructures that are governed by the balance of bulk, interfacial, and elastic energies.
In particular, interfacial energies contribute to the energetic cost of forming phase boundaries and influence their preferred orientations and morphologies.
For example, interfacial energies between the B1-type carbide precipitates and their Fe FCC matrix have been calculated to characterize the 
energetics of carbide--matrix boundaries~\cite{Jia:2020}.
In Al--Cu alloys, interfacial energy anisotropy and elastic strain energy play critical roles in determining the morphology and evolution of metastable \(\theta'\mathrm{-Al_2Cu}\) precipitates, as demonstrated by first-principles-informed phase-field modeling ~\cite{Vaithyanathan:2002, Vaithyanathan:2004}. 
Therefore, quantitative interfacial energies are essential for modeling microstructural evolution and understanding their effects on material properties.

Reliable interfacial energies are difficult to determine experimentally. Reported values are frequently obtained indirectly by fitting measurements of nucleation, precipitate coarsening, or equilibrium shapes to kinetic or thermodynamic models~\cite{Servi:1966, Li:2002, Ardell:1972, Hattenhauer:1993, Stowell:2002, Bulou:2013}.
This is especially important for coherent interfaces, where the measured evolution of the microstructure reflects contributions from both interfacial and elastic energies~\cite{Sonderegger:2009, Louchez:2017}.
Thermodynamic and broken-bond models provide lower-cost alternatives. 
Kaptay developed a model to estimate coherent interfacial energies from the Gibbs energies and molar volumes of the adjoining phases, and later extended this framework to alloys, strained coherent interfaces, and coherent nanocomposites~\cite{Kaptay:2012, Yang:2019, Kaptay:2023}.
Sonderegger and Kozeschnik developed generalized broken-bond models for multicomponent FCC and BCC alloys, including an extension from sharp to diffuse interfaces with temperature-dependent entropic contributions~\cite{Sonderegger:2009, Sonderegger:2010}.
These approaches can be coupled efficiently with CALPHAD thermodynamic data, providing valuable inputs for mesoscale modeling. However, their predictive accuracy depends on the quality of the thermodynamic descriptions and the assumptions concerning atomic interactions and interfacial structure.

First-principles calculations provide atomic-scale and orientation-dependent information, although different computational approaches and reference states can lead to distinct interpretations of the resulting interfacial energies.
One commonly used strategy is the ``direct bulk-referenced excess-energy approach", in which researchers construct a periodic supercell with two interfaces and subtract the energies of the corresponding amounts of the two unstrained bulk phases.
For lattice-matched interfaces, this can provide a useful interfacial excess energy, as reported by Lu et al.\ for Fe/Cr systems~\cite{Lu:2010}.
For coherent interfaces with finite mismatch, however, the result also includes the energy required to strain the constituent phases, resulting in an apparent interfacial energy that depends on superlattice thickness.
This issue is relevant, for example, to the Au/Rh interfacial energy reported by Chen et al.~\cite{Chen:2024}, who used the same bulk-referenced approach without explicitly separating the coherency-strain contribution.
Another strategy employs vacuum-terminated interface slabs and their corresponding isolated surface slabs, usually (though not necessarily) under the same in-plane constraints. The free-surface contributions are then removed to obtain an interfacial energy~\cite{Lu:2013, Li:2014, Gong:2022, Ding:2023}.
To account for the effects of lattice mismatch, several approaches explicitly separate the strain energy from the interfacial energy~\cite{Ferreira:1987, Wood:1987, Ferreira:1988, Lu:1994}.
These include determining epitaxial strain energies from isolated bulk phases~\cite{Ozolins:1998b, Wolverton:2000}, 
analyzing sequences of superlattices of different thickness~\cite{Wei:1988, Dandrea:1990, Lu:1995, Ozolins:1998, Vaithyanathan:2004, Louchez:2017},
and systematically increasing the thickness of one constituent and relaxing while keeping the substrate lattice fixed~\cite{Lepley:2015, Oh:2024, Oh:2025}.

In this work, we construct a high-throughput database of coherent interfacial energies for metallic binary systems using a strain-separated framework.
Specifically, we use a series of superlattices with different thicknesses to separate the coherency strain energy from the chemical interfacial energy, following the supercell approach~\cite{Lu:1995, Ozolins:1998}.
The database is restricted to coherent heterophase interfaces between metals with the same crystal structure, including FCC/FCC, BCC/BCC, and HCP/HCP systems.
We report 365 coherent interfacial energies involving 41 metallic elements and lattice mismatches of up to 25\%, considering three principal crystallographic orientations for FCC- and BCC-based interfaces and the basal plane for HCP interfaces.
We performed more than 5,800 density functional theory (DFT) relaxations to identify the lowest-energy superlattice configurations and obtain reliable strain-separated interfacial energies. 
Based on the long-range order (LRO) type from OQMD and the sign of the interfacial energies, we classify the results into four quadrants to disentangle the resulting energetic trends.
In particular, 86\% of the calculated interfacial energies fall into two main quadrants: quadrant I contains 128 phase-separating LRO-type systems with only positive interfacial energies. 
In contrast, quadrant III comprises 186 ordering LRO-type systems with negative interfacial energies. 
We also compile reported interfacial energies from the literature for comparison and propose a phase-diagram-based classification scheme to relate the calculated interfacial energies to known alloy behavior. 
Overall, the database shows good qualitative agreement with the available literature data and provides a systematic reference for coherent metallic interfaces.
% Authors

\newpage
\section{Methods}
\label{sec:methods}

\subsection{Interfacial Energy approach}
\label{sec: methods_IE_approach}
We begin by establishing the definition of the formation energy of any binary phase $\gamma$---ordered, random alloy, or phase-separated structure---with composition $A_{1-x}B_{x}$ at its equilibrium lattice constants $a_{\gamma}$ as the (zero-pressure) total energy per atom that is gained or lost relative to the bulk, unstrained, phase-separated constituents at their equilibrium lattice constants $a_A$ and $a_B$, respectively:
\begin{equation}
\label{eq:fe_general}
\Delta H_\gamma(x)=E(x,\bm{a_{\gamma}})- \left[(1-x)\,E_A(a_A) + x\,E_B(a_B)\right]
\end{equation}
This is the basic equation used to construct convex hulls in DFT databases such as OQMD~\cite{Kirklin:2015}, and its sign marks the low-temperature long-range order (LRO) that corresponds to compound formation when $\Delta H_\gamma(x)<0$, or to phase separation if $\Delta H_\gamma(x)>0$~\cite{Lu:1994, Wolverton:2000}. 

Following equation (\ref{eq:fe_general}), the formation energy for an \(N-\)atom superlattice \((SL)\) consisting of equiatomic regions of phases \(A\) and \(B\) $(x=0.5)$ separated by a sharp, coherent interface at equilibrium lattice parameter $a_{SL}$ and stacked along the direction $\hat{k}$ is:
\begin{equation}
\label{eq:fe_sl}
\Delta H_{SL}(N,\hat{k})=E(A_{N/2}B_{N/2},\hat{k},\bm{a_{SL}})- \left[\frac{1}{2}E_A(a_A) + \frac{1}{2}E_B(a_B)\right]
\end{equation}
Assuming that the superlattice grows coherently on a substrate characterized by the in-plane lattice vector $\bm{a_s}$, the epitaxial formation energy can then be expressed \cite{Lu:1995, Ozolins:1998} with respect to the energy of the epitaxial constituent that is elastically deformed in the growth plane to match the substrate unit-cell vectors $\bm{a_1}$ and $\bm{a_2}$, and subsequently fully relaxed along the out-of-plane lattice vector $\bm{c}$, which is oriented parallel to $\hat{k}$:
\begin{equation}
\label{eq:fe_sl_ep}
\delta H_{SL}(N,\hat{k},\bm{a_s})=E(A_{N/2}B_{N/2},\hat{k},\bm{a_s})- \left[\frac{1}{2}\underset{c}{min}\left[E_A(\bm{a_1},\bm{a_2},\bm{c})\right] + \frac{1}{2}\underset{c}{min}\left[E_B(\bm{a_1},\bm{a_2},\bm{c})\right]  \right]
\end{equation}
Then, if the substrate lattice plane matches the interface plane dimensions of the equilibrium superlattice $(\bm{a_s}=\bm{a_{SL}})$, the difference between (\ref{eq:fe_sl}) and (\ref{eq:fe_sl_ep}) corresponds to the constituent strain energy along the interface direction $\hat{k}$ and the composition $x=0.5$: 
\begin{equation}
\label{eq:fe_diff}
\Delta H_{SL}(N,\hat{k})-\delta H_{SL}(N,\hat{k},\bm{a_{SL}}) =\Delta E_{CS}(\hat{k},x=0.5)
\end{equation}
This represents the elastic energy cost due to the deformation of the two solids \(A\) and \(B\) from their equilibrium lattice constant \(a_A\) and \(a_B\) to a common lattice plane $\bm{a_{SL}}$, when generating a coherent interface between the two semi-infinite slabs along $\hat{k}$. The aforementioned energy is also known as the coherency strain energy: 
\begin{equation}
\label{eq:ecs}
\Delta E_{CS}(\hat{k},x=0.5)= \underset{\bm{a_1},\bm{a_2}}{min}\left[\frac{1}{2}
\Delta E_A^{epi}(\hat{k},\bm{a_1},\bm{a_2})+\frac{1}{2}\Delta E_B^{epi}(\hat{k},\bm{a_1},\bm{a_2})\right]
\end{equation}
where 
\begin{equation}
\label{eq:epi}
\Delta E_P^{epi}(\hat{k},\bm{a_1},\bm{a_2}) = \underset{c}{min}\left[E_P(\hat{k},\bm{a_1},\bm{a_2},\bm{c})\right] - E_P(a_P)\text{, for }P=A,B    
\end{equation}
Thus, for equiatomic systems with a small lattice mismatch, $\left|\Delta a\right| / \bar{a} \approx 0 \rightarrow a_A\approx a_B$, we expect $\Delta E_{CS}\approx0$ and $\Delta H_{SL}\approx\delta H_{SL}$. On the other hand, for lattice-mismatched systems, $\Delta E_{CS}>0$ and $\delta H_{SL}<\Delta H_{SL}$. This could potentially lead to epitaxially stable superlattices $(\delta H_{SL}<0)$ that are bulk unstable $(\Delta H_{SL}>0)$ \cite{Lu:1995}.

Now, if we consider the infinite-period supercell limit, $(N\rightarrow\infty)$, the superlattice formation energy will have a negligible energetic contribution from the $A/B$ interface interactions, and the remaining energy derives only from the deformed constituents \cite{Lu:1995, Ozolins:1998, Wei:1988}:
\begin{equation}
\label{eq:fe_sl_inf}
\Delta H_{SL}(N\rightarrow\infty,\hat{k}) = \Delta E_{CS}(\hat{k},x=0.5) 
\end{equation}
In this way, the bulk formation energy for finite-period superlattices will have two components, one determined by the elastic strain energy and the other by the interaction of the unlike atoms at the A/B interfaces. The latter is defined as the interfacial energy $I_{AB}(\hat{k})$ and corresponds to the total energy per layer of a single interface between the materials \(A\) and \(B\) along $\hat{k}$. Taking into account (\ref{eq:fe_diff}) and (\ref{eq:fe_sl_inf}), its equation is:
\begin{equation}
\label{eq:ie_def}
\Delta H_{SL}(N,\hat{k})-\Delta H_{SL}(N\rightarrow\infty,\hat{k})=\delta H_{SL}(N,\hat{k},\bm{a_{SL}})\equiv\frac{I_{AB}(\hat{k})}{N}
\end{equation}
We can finally reorder (\ref{eq:ie_def}) and obtain the formation energy of these $A_{N/2}B_{N/2}$ supercelattices, represented by supercells with an energetic contribution from a double interface area due to the periodic boundary conditions and the elastic penalty cost as: 
\begin{equation}
\label{eq:fe_final}
\Delta H_{SL}(N,\hat{k})=2A(\hat{k})\,\sigma_{AB}(\hat{k})\frac{1}{N}+\Delta E_{CS}(\hat{k},x=0.5)
\end{equation}
It is important to emphasize that a positive value of the concentration-independent interfacial energy, \(\sigma_{AB}(\hat{k})\), indicates that long-period superlattices are more stable than short-period ones, thereby promoting the formation of like-atom pairs and, consequently, a phase-separating tendency. Conversely, a negative \(\sigma_{AB}(\hat{k})\) indicates that the creation of interfaces is energetically favorable, corresponding to a thermodynamic preference for unlike-atom pairs and thus an ordering tendency \cite{Ozolins:1998}. This interpretation is analogous to that of the Warren\(-\)Cowley short-range order parameter, for which \(\alpha_{lmn}>0\) reflects A--A or B--B clustering tendencies, whereas \(\alpha_{lmn}<0\) reflects A--B ordering \cite{Wolverton:1998}. Nevertheless, this interpretation, based solely on the sign of \(\sigma_{AB}(\hat{k})\), is not universally valid, as elaborated in the discussion section.

To separate the energetic contributions of $\Delta H_{SL}(N,\hat{k})$ in equation~(\ref{eq:fe_final}),  \(\sigma_{AB}(\hat{k})\) in $[meV/\mathring{A}^2]$ (or traditionally expressed in $[mJ/m^2]$\cite{Porter:2009}) and $\Delta E_{CS}(\hat{k})$ in $[meV/atom]$, we need to directly calculate the coherency strain energy via isotropic strain calculations (also known as direct method) and subtract it from the supercell formation energy (\textit{cf.} \cite{Laks:1992, Ozolins:1998, Ozolins:1998b}). Alternatively, we can relax a series of different-sized reoriented supercells with the same interface system and, after a linear regression, obtain both $\sigma_{AB}(\hat{k})$ from the slope and $\Delta E_{CS}(\hat{k})$ from the y-intercept; this latter methodology is called the supercell approach (\textit{cf.} \cite{Lu:1995, Ozolins:1998}), and is the method used to build the interfacial energy database shared in this work.
In particular, we created a series of oriented prototypes of varying sizes, each decorated with the elements shown in Figure~\ref{fig:interfaces-count}. 
We use these prototypes as a basis for DFT relaxations unless explicitly stated in the text; see Table S.1 in the Supporting Information (S.I.) for more details of these prototypes.
Finally, all the formation energies, $\Delta H_{SL}(N,\hat{k})$, were estimated using the reference elemental energies listed in Table S.2, which were extracted from OQMD and represent the DFT ground state of each element in their respective crystal structures (FCC, BCC, or HCP) at standard temperature and pressure (STP) \cite{West:2014}. No corrections to these chemical potentials were included, as in the OQMD database \cite{Kirklin:2015}. For more details on the supercell approach and some examples, see Figures S.1-S.3.

\subsection{DFT scheme and error tolerance}
All DFT calculations were performed within the OQMD framework, and the details of the settings can be found in \cite{Saal:2013, Kirklin:2015}. In summary, these calculations were performed using the Vienna Ab initio Simulation Package (VASP)\cite{Kresse:1996, Kresse:1996b} with the projector-augmented-wave (PAW) method\cite{Blochl:1994}, and the Perdew-Burke-Ernzerhof generalized gradient approximation (GGA-PBE) was used as the exchange-correlation functional\cite{Perdew:1996, Kresse:1999}. To capture possible magnetism, all elements with partially filled d and f orbitals are considered for spin-polarization calculations with initial magnetic moments of 5 $\mu$B and 7 $\mu$B, respectively, and ferromagnetic alignment. The OQMD convergence criteria were used for the DFT convergence calculations, which include an additional 5\% relative volume tolerance check in each converged relaxation calculation. 

The linear regression (LR) approach and the Theil-Sen (TS) regression were applied to eq.~\ref{eq:fe_final} using \(\Delta H_{SL}(N,\hat{k})\) vs \(1/N\) as inputs. TS regression is the standard approach because it allows us to detect and discard outliers arising from, for example, the interface-interface interactions in the supercells \cite{Ozolins:1998}. 
The linear regression error was calculated for each orientation within each system after the first 4 supercells had converged. This allowed us to estimate the error of the interfacial and coherency strain energies from the slope and the y-intercept, respectively. Since there is a wide range of formation energy results, we defined the relative interfacial energy error as:
\begin{equation}
\label{eq:ie_error}
\sigma_{AB}^{error}(\hat{k}) = |S_{\sigma_{AB}(\hat{k})}/\sigma_{AB}(\hat{k})|\times100
\end{equation}
Here, \(S_{\sigma_{AB}(\hat{k})}\) denotes: (i) for the TS regression, the mean variability of the slope estimates obtained via a bootstrap resampling procedure; and (ii) for the LR, the square root of the corresponding diagonal element of the variance–covariance matrix of the model. The objective in constructing this database was to minimize \(\sigma_{AB}^{\text{error}}(\hat{k})\) as far as possible, with particular emphasis on systems whose formation energy spans exceed 3 meV/atom. 

Figure S.4 shows the results of this approach, indicating that we reduced the error or uncertainty to an average of \(6.5\%\), with a standard deviation of \(5.4\%\), for the 313 systems with a formation energy range greater than or equal to \(3.0~meV/at\). 
This uncertainty originates predominantly from small variations in the final formation energies associated with the multiple degrees of freedom involved in superlattice relaxation, including atomic positions, cell shape, cell volume, and spin configuration. 
These variations reflect the different ways in which each superlattice balances strain relaxation with favorable and unfavorable chemical interactions at the interface.
An example of the importance of detecting this variability in the HT database and reducing calculation uncertainty is shown in Figure S.5.

Finally, it is worth noting that the calculated coherency strain energies using the supercell approach are obtained by extrapolating the formation energies, as we explained above in Section~\ref{sec: methods_IE_approach} and detailed in Fig.~S.1.
Therefore, these values are not strictly positive, unlike those obtained from the direct approach based on the epitaxial relaxation curves~\cite{Wolverton:2000}. 
In our dataset, only seven coherency strain energies exhibit slightly negative values, with \(-1.0<\Delta E_{CS}(\hat{k})<0.0~meV/at\); these cases are listed in Table S.3. 
Since all of them fall within the regression uncertainty, they were set to \(0.0~meV/at\) for plotting purposes, without altering the interfacial energy results.

\subsection{Scope of the database}

\begin{figure}[t]
\centering
\includegraphics[width=1.0\textwidth]{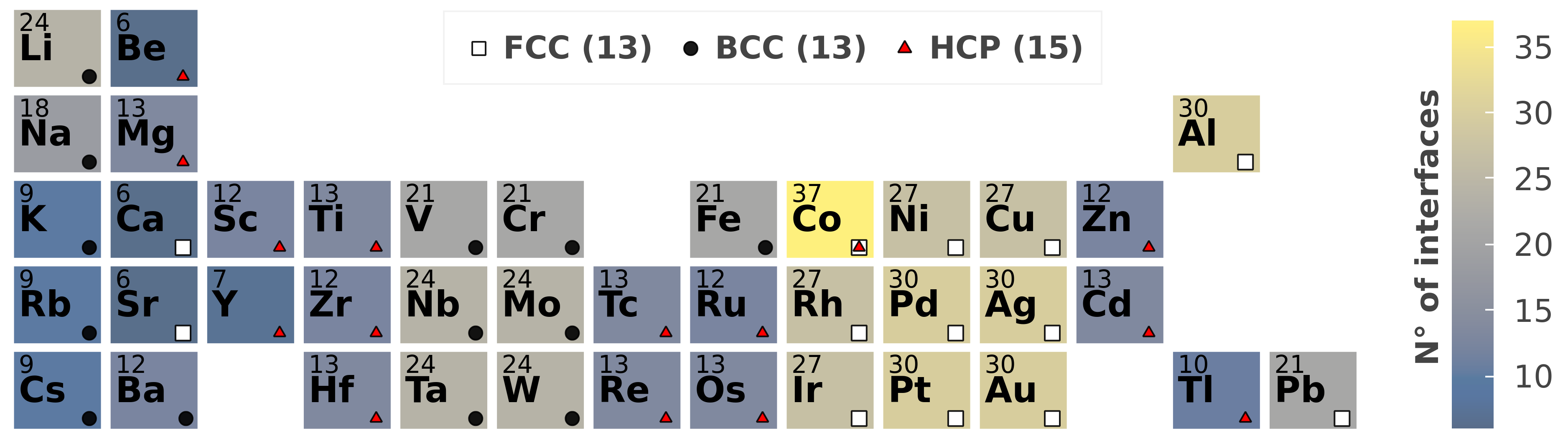}
\vspace{2pt}

\begin{center}
\begin{tcolorbox}[
  enhanced,
  hbox,                
  colback=gray!2,
  colframe=gray!10,
  boxrule=0.6pt,
  arc=2pt,
  left=8pt,right=8pt,top=7pt,bottom=8pt,
]
\begin{minipage}{0.65\textwidth} 
\centering
\textbf{Metal/Metal isostructural combinations with $\dfrac{\left|\Delta a\right|}{\bar{a}} \le 0.25$:}\\[1pt]
\(
\begin{aligned}
\text{FCC/FCC} &= 53 \text{ of } 78 \text{ pairs} \times 3 \text{ orientations } [100],[110],[111]\\
\text{BCC/BCC} &= 40 \text{ of } 78 \text{ pairs} \times 3 \text{ orientations } [100],[110],[111]\\
\text{HCP/HCP} &= 86 \text{ of } 105 \text{ pairs} \times 1 \text{ orientation } [0001]
\end{aligned}
\)\\[12pt]
\textbf{Total coherent interfaces = 365 (of 573)}
\end{minipage}
\end{tcolorbox}
\end{center}
\vspace{-12pt} 
\caption{Selected elements with their crystal structure (bottom right of each square) and occurrence frequencies (top left) in the Interfacial Energy Database. The coherent interfaces are restricted to a 25\% lattice mismatch, resulting in 365 isostructural combinations, as summarized in the bottom panel.} 
\label{fig:interfaces-count}
\end{figure}

For this database, we focus on the isostructural combination of FCC, BCC, and HCP metals with lattice mismatch, $100\cdot\left|\Delta a\right|/\bar{a}$, smaller than or equal to 25\%, covering the principal directions for FCC and BCC ([100], [110], and [111]), and the basal plane for the HCP interfaces. 
In Figure~\ref{fig:interfaces-count}, we highlight the elements selected to build these heterointerfaces, their preferred crystal structure under STP conditions, and their frequency in the database.
The element with the highest frequency is Co because it is dynamically stable in both the FCC and the HCP phases \cite{Lizarraga:2017}, and it was used as a test for the predictions of the supercell approach.
For all orientations, we construct coherent interfaces prior to DFT relaxation. Here, we consider cases with $100\cdot\left|\Delta a\right|/\bar{a} > 25\%$ to potentially yield unreliable interfacial energy estimates because such large lattice mismatches would require supercells containing three to four unit cells of each material along the interface plane to more realistically represent an incoherent interface; this type of interface is cost-prohibitive with the supercell approach.
Therefore, based on the lattice parameters in Table S.2, we reduce the 78 possible combinations of FCC/FCC interfaces to 53, and considering the 3 principal orientations, we estimate 159 interfacial energies for FCC systems. A similar reduction approach was applied for the BCC/BCC and HCP/HCP interfaces, as summarized in the bottom panel of Fig.~\ref{fig:interfaces-count}. 

\newpage
\section{Results and Discussion}
\label{sec:discussion}

\FloatBarrier

\begin{figure}[t]
\centering
\includegraphics[width=1.0\textwidth]{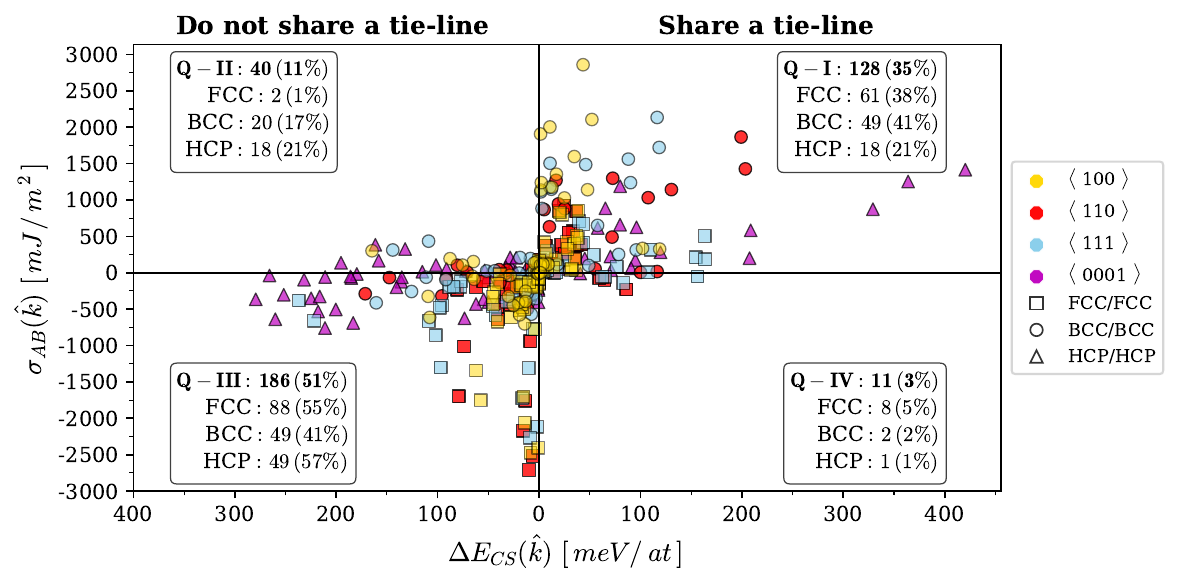}
\caption{Quadrants generated with the couple of the interfacial energies, $\sigma_{AB}(\hat{k})$, and coherency strain energies, $\Delta E_{CS}(\hat{k})$, for all the selected systems and their respective orientations in the database (cf. Fig.~\ref{fig:interfaces-count}). Systems that have one or multiple ordered compounds on the OQMD convex hull are on the left-hand side of the plot (Q-II and Q-III), while systems that show a tie-line are on the right-hand side (Q-I and Q-IV). The different colors represent different supercell orientations, and the different shapes represent the types of structures that formed the interfaces. Note that $\Delta E_{CS}(\hat{k})$ is positive in both directions. Small boxes represent the distribution of the results in each quadrant.} 
\label{fig:quads_all}
\end{figure}

As previously mentioned, the supercell approach allows us to separate the energetic contributions to $\Delta H_{SL}(\hat{k})$ into the interfacial energy, $\sigma_{AB}(\hat{k})$, and the coherency strain energy, $\Delta E_{CS}(\hat{k})$.
The results of applying this methodology to all coherent interfaces highlighted in the bottom panel of Fig.~\ref{fig:interfaces-count} are shown in Fig.~\ref{fig:quads_all}. This figure is divided into four quadrants (I--IV), labeled in counterclockwise order, with the results of $\sigma_{AB}(\hat{k})$ along the vertical axis and the positive-definite values of $\Delta E_{CS}(\hat{k})$ along the horizontal axis.
Figure~\ref{fig:quads_all} is constructed by partitioning the \(A\text{--}B\) systems into those that share a tie-line in the OQMD convex hull (right, Q-I and Q-IV) and those that do not (left, Q-II and Q-III). 
If a system shares a tie-line, i.e., if \(\Delta H_\gamma(x) > 0\) (eq.~(\ref{eq:fe_general})) for every phase \(\gamma\), then its low-temperature LRO corresponds to a phase-separating (PS) system. In contrast, if a system contains one or more ordered compounds on the convex hull, (\(\Delta H_\gamma(x) < 0\)), then its LRO is of the ordering (O) type.

This categorization is analogous to the classification of alloy systems proposed by Lu and Zunger\cite{Lu:1994}, with the inclusion of an additional regime in which the formation enthalpy, $\Delta H_f$, is negative, whereas both the spin-flip chemical energy, $\varepsilon$, and the elastic energy, $G(x)$, are positive. 
This regime corresponds to Q-II in the present work and is caused by one or more compounds on the hull that have a structural ordering different from that of the calculated coherent interfaces.

An initial observation from Fig.~\ref{fig:quads_all} is that the majority of data points are located in quadrants Q-I and Q-III. 
Q-I represents the PS systems characterized by $\sigma_{AB}(\hat{k})>0$ and accounts for 35\% of all entries in the dataset, including more than one-third of the data for the FCC (38\%) and BCC (41\%) systems. 
Q-III, in turn, contains O systems with $\sigma_{AB}(\hat{k})<0$, representing 51\% of the database and including more than half of the results of the FCC (55\%) and HCP (57\%) interface structures. 
In this figure, the BCC-based structures dominate the largest positive values of $\sigma_{AB}(\hat{k})$ in Q-I (represented by filled circles), whereas the FCC-based structures predominate at the negative values in Q-III (represented by filled squares). 
In contrast, the results for the HCP-based interface structures span a narrower range, $\sigma_{AB}(\hat{k})\lesssim|1000|$, and exhibit the highest coherency strain energies.

The remaining quadrants, Q-II and Q-IV, contain comparatively fewer data points, accounting for 11\% and 3\% of the results, respectively. 
Quadrant Q-II denotes systems that exhibit at least one stable ordered compound on the convex hull and satisfy $\sigma_{AB}(\hat{k})>0$, whereas Q-IV comprises PS systems (i.e., no stable compounds on the hull) with $\sigma_{AB}(\hat{k})<0$. 
In the following subsections, we provide a more detailed analysis of the systems associated with each quadrant.
\subsection{Quadrant I \& IV: Systems that share a tie-line }
The binary systems classified as phase-separating LRO-type are compiled on the right-hand side of Fig.~\ref{fig:quads_all}. 
Among these, 92\% of the interface structures, corresponding to 128 Q-I cases, have positive $\sigma_{AB}(\hat{k})$ values.
Almost half of these cases are FCC/FCC interfaces, representing 38\% of all calculated FCC systems. 
In contrast, only 11 PS cases fall in Q-IV, with negative $\sigma_{AB}(\hat{k})$, and most of them are FCC systems (8 cases).
The other dominant contribution in these quadrants comes from BCC/BCC interfaces, which account for 51 results.
Only two of these BCC cases fall in Q-IV, while the remaining 49 have positive \(\sigma_{AB}(\hat{k})\) values in Q-I, representing 41\% of all calculated BCC interfaces in the dataset. 
The remaining results correspond to HCP/HCP interfaces, with 19 cases in total; only one of them has a negative \(\sigma_{AB}(\hat{k})\) value and therefore belongs to Q-IV. A more detailed discussion of each group is provided below.

\subsubsection{FCC/FCC interfaces}
\label{sec:Q_I_fcc_interfaces_subsub}
\begin{figure}[t]
\centering
\includegraphics[width=1.0\textwidth]{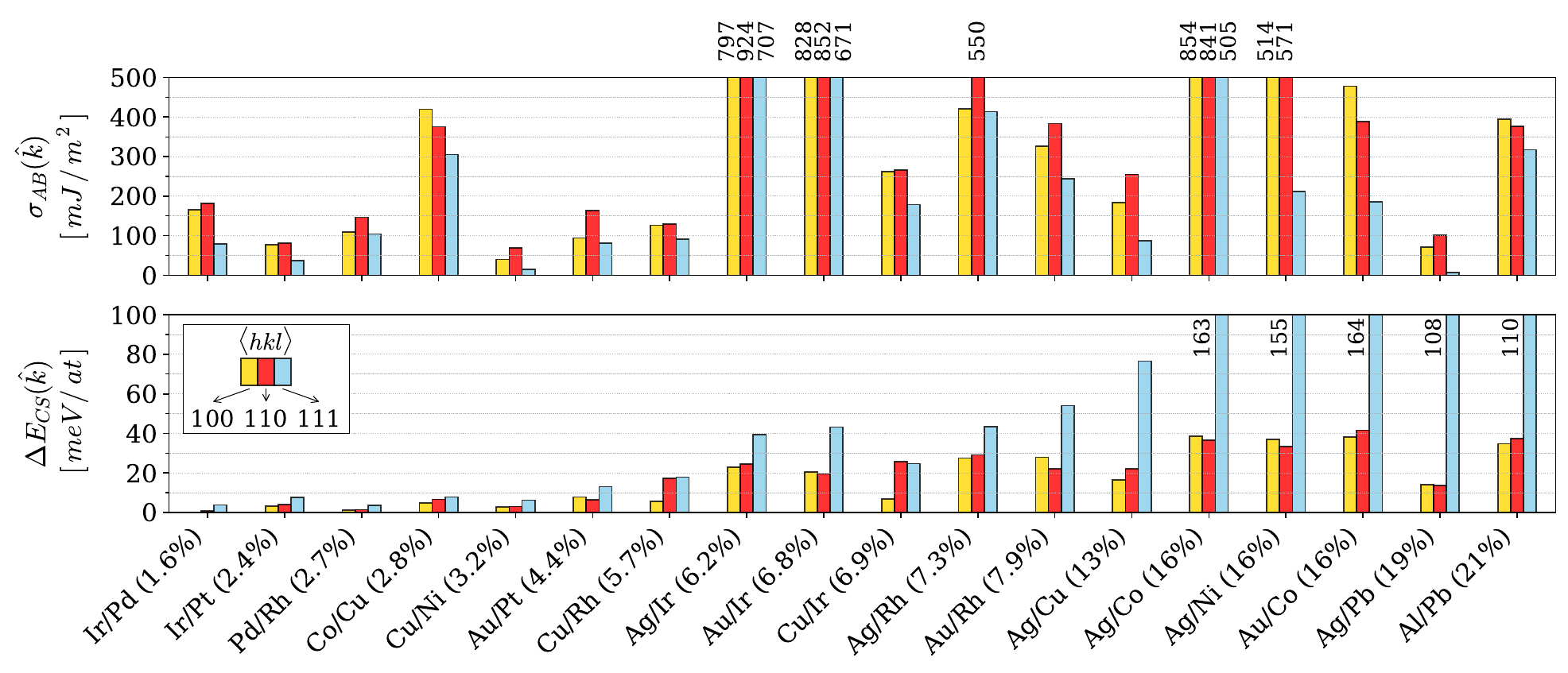}
\caption{Interfacial energies (top) and coherency strain energy (bottom) for all the phase-separating FCC systems with positive $\sigma_{AB}(\hat{k})$ for the three calculated orientations \(\hat{k}\) (Q-I). The interfaces are sorted by lattice mismatch (value in parentheses), and the colored bar represents the results for each principal orientation, $(\langle100\rangle,\langle110\rangle,\langle111\rangle)$, from left to right. The elements in the \(A/B\) labels are sorted alphabetically.}
\label{fig:PS_FCC_IEpos}
\end{figure}

All the \(\sigma_{AB}(\hat{k})\) and \(\Delta E_{CS}(\hat{k})\) results for the PS FCC systems that contain positive \(\sigma_{AB}(\hat{k})\) for all \(\hat{k}\) are compiled in Fig.~\ref{fig:PS_FCC_IEpos}. 
In this figure, the systems are arranged in order of increasing lattice mismatch from left to right, and the differently colored bars denote the principal orientations (\(\langle100\rangle\), \(\langle110\rangle\), and \(\langle111\rangle\)) along which the supercell approach was applied. 
In the top panel of Fig.~\ref{fig:PS_FCC_IEpos}, the plotted values of \(\sigma_{AB}(\hat{k})\) were capped at 500 \(mJ/m^2\) for visualization purposes. 
For all systems in this quadrant, no systematic dependence of \(\sigma_{AB}(\hat{k})\) on increasing lattice mismatch is observed. 
A consistent feature, however, is that the interfacial energy in the close-packed plane \(\{111\}\) is the lowest among all systems. 

This is in line with the coherency strain energy shown in the bottom panel of the same figure, where the highest value in almost all interfacial systems is along the \(\langle111\rangle\) orientation. 
As expected, the strain energy increases with the size mismatch. 
(See the full trends of $\Delta E_{CS}(\hat{k})$ versus the size mismatch in Figures~S.6-S.9.)
Although we did not calculate \(\Delta E_{CS}(\hat{k})\) for other directions, the current results are in good agreement with the harmonic theory, which always predicts \(\langle111\rangle\) as the hardest directions and \(\langle100\rangle\) as the softest direction, or vice versa \cite{Ozolins:1998, Bottomley:1996}.

There is little literature reporting interfacial energy estimates from experimental or theoretical studies that can be directly compared with our findings at 0K. The existing literature originates from diverse sources and employs a range of methodological approaches. Nevertheless, several cases are worth mentioning here: 
\begin{itemize}
    \item For Co/Cu we obtain values within 305--420 $mJ/m^2$, with \(\langle111\rangle\) being the minimum and \(\langle100\rangle\) the maximum. 
    For comparable \(\langle111\rangle\) interfaces, Li et al. \cite{Li:2022} reported 195--294 $mJ/m^2$, depending on the underlying lattice and the relaxation and magnetic constraints. 
    Experimentally inferred values, which strongly depend on the chosen kinetic models and thermodynamic parameters, include 190 $mJ/m^2$ from nucleation kinetics \cite{Servi:1966},  154--318 $mJ/m^2$ from coarsening data \cite{Ardell:1972}, 240 $mJ/m^2$ from a modified Langer--Schwartz precipitation model\cite{Hattenhauer:1993}, and 181--227 $mJ/m^2$ from the classical capillarity theory applied to nucleation-rate data \cite{Stowell:2002}.
    As we expected, our 0K results overestimate the interfacial free energy reported in the experiments because we do not account for entropic or diffusive effects \cite{Mao:2012}.
    \item For Au/Pt we obtain values within 81--164 $mJ/m^2$, with the minimum along \(\langle111\rangle\). 
    This is about five times smaller than the mixing-enthalpy-based estimate of $\gamma_{mix}$ = 430 $mJ/m^2$ reported for Au/Pt(111) by Bulou et al.~\cite{Bulou:2013}. 
    Similarly, for Au/Co(111), our value of 185 $mJ/m^2$ is significantly smaller than their estimate of 660 $mJ/m^2$.
    \item For Cu/Ir we obtain values within 178--266 $mJ/m^2$, with the minimum along $\sigma_{AB}({111})$. 
    Using a different approach, Kwon et al.~\cite{Kwon:2020} considered a 7-layer superlattice with vacuum on the extrema and reported a considerably larger interfacial energy of $\sim0.066~eV/$\AA$^2$ $\left(\sim1,056~mJ/m^2\right)$ for the same system along \(\langle111\rangle\).  
    \item For Au/Rh we obtain values within 243--383~$mJ/m^2$, with $\sigma_{AB}({111})$ being the lowest value. Recently, Chen et al.~\cite{Chen:2024} explained the miscibility of these metals at nanoscale sizes using a model in which $\sigma_{AB}({111}) = 7.2~eV/nm^2 \left(\sim 1,153~mJ/m^2\right)$. 
    This significantly larger value is likely attributable to strain-energy contributions that were not explicitly separated from the interfacial energy calculations. 
    Since we obtain $\Delta E_{CS}(111) = 54.0~meV/at$ for this system, we estimate that 7.78 of 10.31 eV (from their Eq. S11) corresponds to the elastic energy contribution for their equiatomic 144-atom supercell. Reducing the interfacial estimation to about \(1.76~eV/nm^2\), which is closer to our result of \(1.52~eV/nm^2~(243.3 ~mJ/m^2)\). 
    \item Al/Pb is distinctive among the Al-containing FCC interface systems considered here.
    Whereas Al forms stable intermetallic compounds with other FCC elements in the database and exhibits negative interfacial energies for all calculated orientations---see Fig.~\ref{fig:phase_diag_comp}(a) below and references in Table~S.4---Al/Pb is a phase-separating system and has positive interfacial energies of 317--395  $mJ/m^2$. 
    Here, the strain-free anisotropy ratio is $\sigma_{AB}({100})/\sigma_{AB}({111})=1.25\pm0.03$. 
    These results are comparable with the experimental estimates by Shi and Wynblatt~\cite{Shi:2002, Shi:2002b}.
    From contact-angle measurements of equilibrated Pb crystallites on Al(111) and Al(100), they obtained $217\pm35$ and $330\pm40$ $mJ/m^2$, respectively, corresponding to $\sigma_{AB}({100})/\sigma_{AB}({111})=1.52\pm0.46$.
    Earlier, Xiao et al.~\cite{Xiao:1995} used in-situ HREM to measure the relative separations of (100) and (111) facets in equilibrium cuboctahedral Pb inclusions embedded in Al. 
    Application of the Gibbs–Wulff relation yielded a ratio of $1.24\pm0.07$, in excellent agreement with our calculated value.
\end{itemize}

In summary, Fig.~\ref{fig:PS_FCC_IEpos} shows the Q-I results, where $\sigma_{AB}({111})$ is consistently the lowest interfacial energy across the systems examined, although the calculated magnitudes differ substantially from previous theoretical and experimental estimates. 
These discrepancies likely reflect differences in interface models, reference states, and relaxation constraints~\cite{Li:2022, Bulou:2013, Kwon:2020}, as well as the inclusion of strain-energy contributions~\cite{Chen:2024} and finite-temperature, kinetic, and thermodynamic effects in experimentally derived values~\cite{Servi:1966, Ardell:1972, Hattenhauer:1993, Stowell:2002, Mao:2012}.
Notably, the calculated Al/Pb anisotropy ratio agrees closely with experiment~\cite{Shi:2002, Shi:2002b, Xiao:1995}, supporting the predicted orientation dependence of interfacial energies calculated by the supercell approach.
\begin{figure}[t]
\centering
\includegraphics[width=1.0\textwidth]{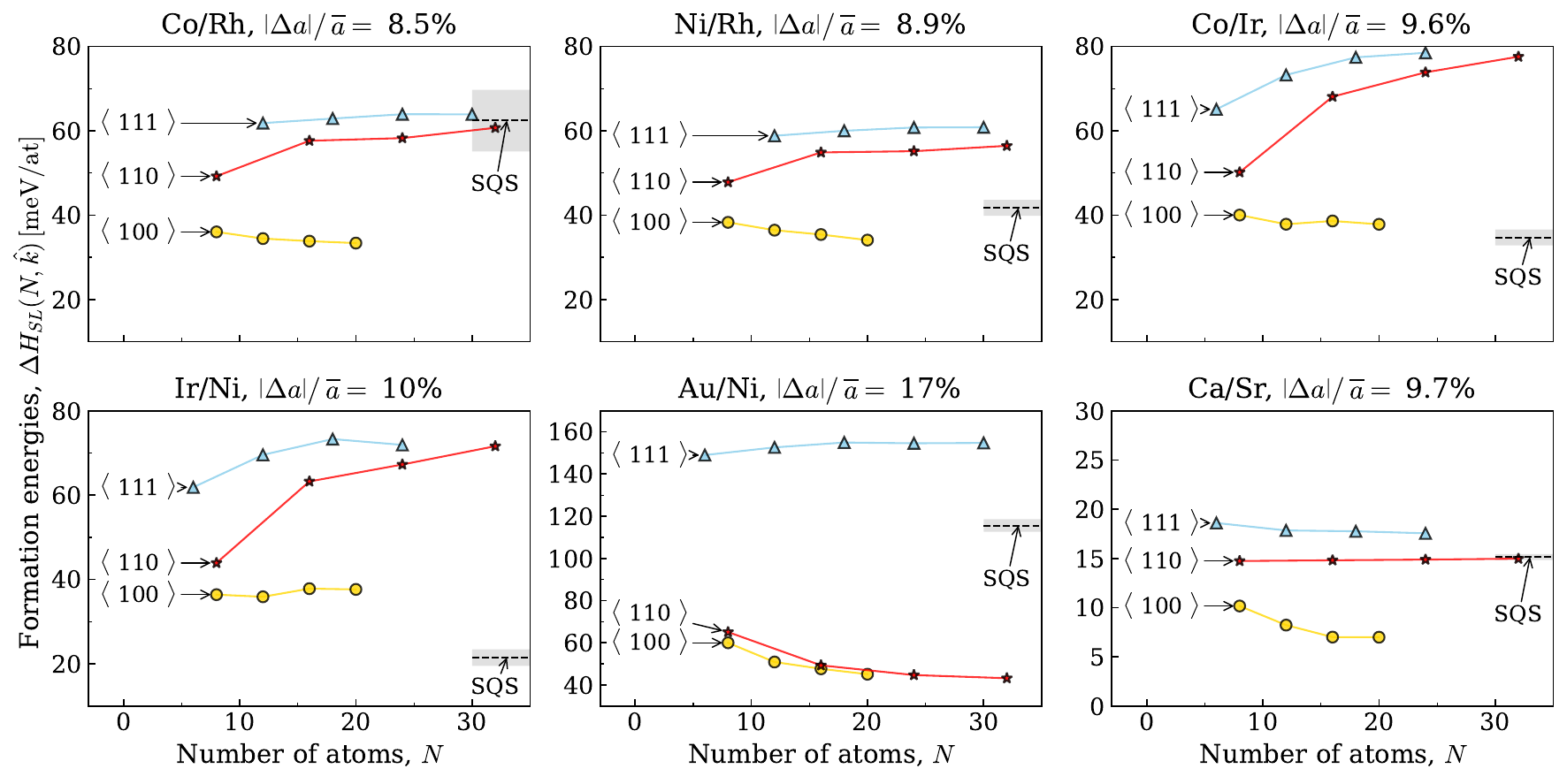}
\caption{Superlattice formation energies of each of the interface orientations for the five phase-separating FCC systems in quadrant Q-IV sorted by size mismatch. SQS approximates the equiatomic random alloy, and the gray shadow shows the standard deviation across five different seeds of 32-atom SQS calculations. The Ir/Ni system is part of Q-III (ordering systems) and was added for comparison.}
\label{fig:PS_FCC_IEmix}
\end{figure}

The 8 remaining phase-separating FCC cases in Q-IV (cf.~Fig.~\ref{fig:quads_all}) occur in five interfacial systems. They deviate from the general behavior of PS systems because they show negative \(\sigma_{AB}(\hat{k})\) for some orientations.
The formation energies for each oriented supercell of these systems are plotted in Fig.~\ref{fig:PS_FCC_IEmix}, where we also include the energy of their random alloys per plot---calculated through equiatomic SQS structures; see S.I. for details---and the ordering system Ir/Ni for comparison purposes. 
Here, the upward trend of the $\Delta H_{SL}(N,\hat{k})$ curves is driven by negative interfacial energies.
All of them have an average size mismatch greater than 8\%, contain a magnetic element, either Co or Ni, and have coherency strain energy results higher than \(59~meV/at\). The only exception is the Ca/Sr system \((15~meV/at)\). 
Due to the high level of strain energy for these cases, we refer to this quadrant as ``strain inhibits ordering".

The first row of plots shows the PS Co/Rh, Ni/Rh, and Co/Ir systems, which exhibit negative interfacial energies along the \(\langle110\rangle\) and \(\langle111\rangle\) directions and a positive value along \(\langle100\rangle\). 
In these three systems, the long-period superlattices along \(\langle100\rangle\) are stable with respect to the random alloy. 
Even for the Co/Ir system, where the long-period limit energy, $\Delta E_{CS}(100) = 36.5\pm0.6~meV/at$, competes with $\Delta H_{SQS}=34.7\pm1.9$.
However, these results consider only equiatomic FCC structures---superlattices and SQSs---and are limited to the FCC incoherent phase-separation space.
Co-rich HCP structures may therefore remain thermodynamically competitive or lie on the global convex hull and need to be further explored. 
This is consistent with experimental observations, which indicate solid-solution formation and allotropic transformations, with no evidence of LRO development (see comments and references for Co–Ir and Co–Rh in Table S.4.)

The next row of plots includes the O system Ir/Ni, for which all calculated interfacial energies are negative, and the PS Au/Ni system, which displays an upward trend along \(\langle111\rangle\), indicating a negative interfacial energy for this coherent interface.
Finally, the PS Ca/Sr system exhibits formation-energy variations smaller than 3.2 \(meV/at\) for all its superlattices, implying relatively small \(\sigma_{AB}(\hat{k})\) results.
In fact, the system has a slightly negative value (\(-1.2\pm0.4~mJ/m^2\)) along \(\langle110\rangle\) and relatively small positive values for its other interfaces: \(\sigma_{AB}(100)=21.7\pm2.2~mJ/m^2\) and \(\sigma_{AB}(111)=4.6\pm0.3~mJ/m^2\).
The long-period superlattices along \(\langle100\rangle\) are stable with respect to the random alloy by \(11~meV/at\). 
These small energy differences are consistent with the solid solution reported in the full composition range~\cite{Aljarrah:2008}.
Interestingly, the CaTCalc database predicts a very small miscibility gap with a critical temperature of around 25~\degree C for Ca-Sr
~\cite{RICT:2026, Shobu:2009}. 

The Ir/Ni system was originally classified as PS in the OQMD database. 
However, the presence of negative interfacial energies across the three calculated orientations, the instability of its long-period superlattices to the formation of a random alloy, and the experimental observation that Ir-Ni formed a continuous series of disordered substitution solid solutions \cite{Shubin:2022} prompted us to re-examine its phase behavior and to search for ordered intermetallic compounds. 
Through this analysis, we identified previously omitted phases that have negative formation energy; in particular, NiIr, which adopts the NbP-type (``40”) structure (SG=I4\(_1\)/amd), is on the Hull.

Au-Ni alloys have been widely studied both experimentally~\cite{Nielsen:1993, Jacobsen:1995, Bienzle:1995} and computationally~\cite{Wolverton:1997, Ozolins:1998, Ozolins:1998b, Colinet:2000, Wang:2005}. 
This system has an asymmetric miscibility gap, with a critical temperature of around 816~\degree C and 73 at.\% Ni~\cite{Wang:2005}. 
Since we use the same approach as Ozoliņš et al.~\cite{Ozolins:1998}, our results in terms of $\sigma_{AB}(\hat{k})$ and $\Delta E_{CS}(\hat{k})$ are similar. 
However, their energetics were calculated using an LDA-based cluster expansion, while we use the energy of discretized prototypes relaxed with GGA potentials. 
The two main discrepancies between the results are: (i) we obtain $\sigma_{AB}(111)=-52.4\pm16.8~mJ/m^2$, indicating mixing tendencies, whereas their results yield negative values only for very short-period superlattices; and (ii) in our case, the formation energies associated with the \(\langle100\rangle\) and \(\langle110\rangle\) orientations converge to similar values, while their results show different trends.
Interestingly, it has been experimentally observed that this immiscible system forms surface alloys along the (110) and (111) planes~\cite{Nielsen:1993, Jacobsen:1995}. 
On Ni(111), driven by local Au–Ni bonding rather than strain relief, Au forms a nearly unstrained monolayer that ends in partial-dislocation loops, with expelled Ni atoms alloying in the Au layer~\cite{Jacobsen:1995}. 
This observation is consistent with the chemical trend described in (i). 
With respect to (ii), the convergence of \(\Delta H_{SL}(100)\) and \(\Delta H_{SL}(110)\) arises from the combined effects of the substantial atomic size mismatch of 17\% in this system and the intrinsic relaxation behavior of the superlattice.  
This phenomenon is also observed in both phase-separating and ordering systems exhibiting a large size mismatch (\(\gtrapprox13\%\)) and can be attributed to the fact that the reoriented superlattices are structurally connected via a Bain-path–type transformation. A more comprehensive investigation of this effect is needed.

The remaining interface structures in Q-IV of Fig.~\ref{fig:quads_all} are the HCP Sc/Y, for which we predict $\sigma_{AB}(0001)=-11.2\pm2.8~mJ/m^2$, and the BCC K/Rb. 
The K/Rb system has a behavior similar to that of Ca/Sr, i.e., narrow formation energies with variability below \(1~meV/at\) and very small interfacial energies: \(0.3\pm0.6\), \(-1.1\pm2.3\), and \(-1.9\pm0.3~mJ/m^2\) for $\sigma_{AB}(100)$, $\sigma_{AB}(110)$, and $\sigma_{AB}(111)$, respectively.
It is curious that a very low-temperature miscibility gap is also predicted for this PS system around \(-220\)~\degree C~\cite{Ren:2011}.

\subsubsection{BCC/BCC Interfaces}

\begin{figure}[t]
\centering
\includegraphics[width=1.0\textwidth]{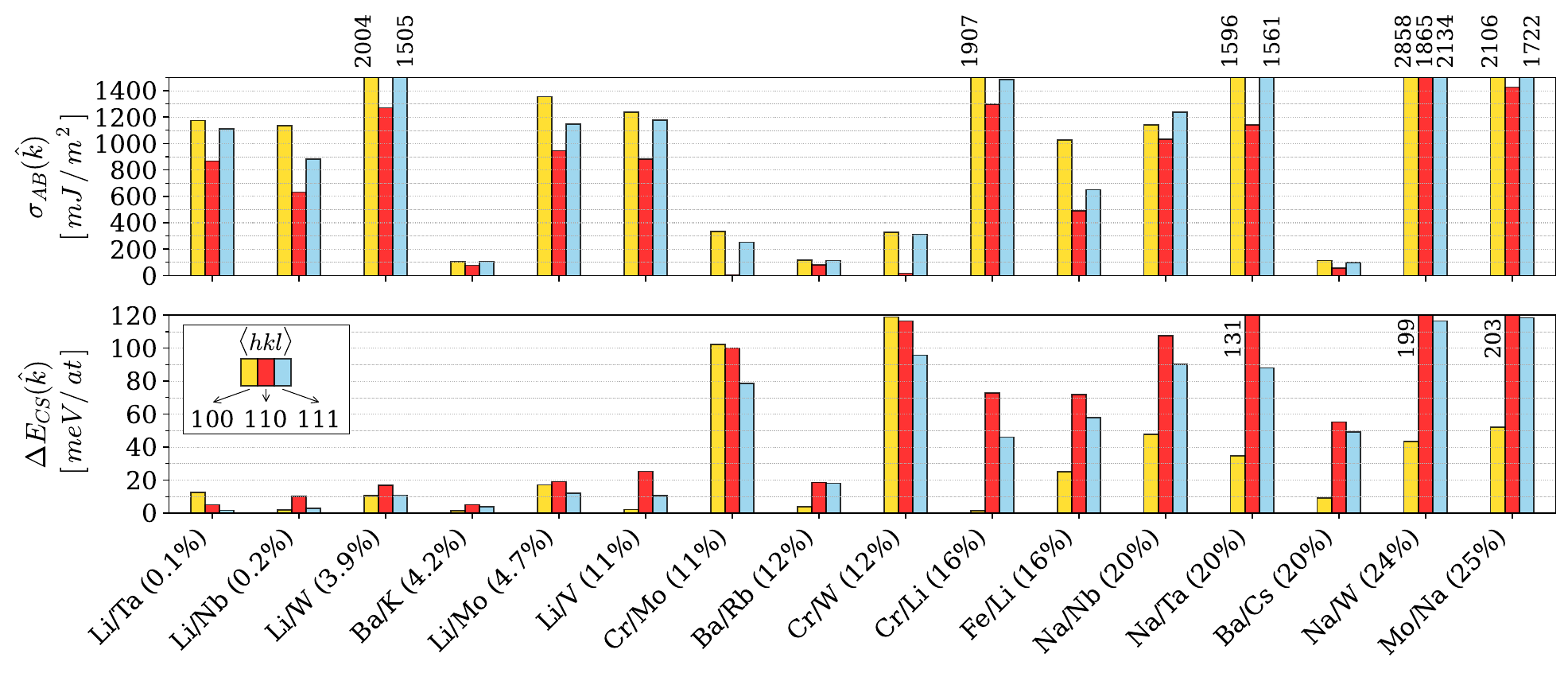}
\caption{Interfacial energies (top) and coherency strain energy (bottom) for all the phase-separating BCC systems with positive $\sigma_{AB}(\hat{k})$ for the three calculated orientations \(\hat{k}\) (Q-I). This basically excluded only the K/Rb system. The interfaces are sorted by lattice mismatch (value in parentheses), and the colored bar represents the results for each principal orientation, $(\langle100\rangle,\langle110\rangle,\langle111\rangle)$, from left to right. The elements in the \(A/B\) labels are alphabetically sorted.}
\label{fig:PS_BCC_all}
\end{figure}

As mentioned above, the other major group in Q-I of Fig.~\ref{fig:quads_all} consists of phase-separating BCC/BCC interfaces, which represent 41\% of all calculated BCC systems in the database. Their results are compiled in Fig.~\ref{fig:PS_BCC_all}, with the exception of the K/Rb system discussed above.
They span values larger than the interfacial energies of PS FCC/FCC and PS HCP/HCP in Q-I, which have average interfacial energies of \(\bar\sigma_{AB}=288~mJ/m^2\) and
\(\bar\sigma_{AB}=529~mJ/m^2\), respectively.
Here, the average is \(961~mJ/m^2\), with 26/48 results containing Li or Na exceeding \(1000~mJ/m^2\) and reaching the order of magnitude of surface energies \cite{Tran:2016}.
By comparison, interfaces formed only with alkali and alkaline earth metals have very small interfacial energies, spanning \(|150|~mJ/m^2\) see Fig.~S.9).
From surface science studies of alkali adatoms on transition-metal substrates, the higher energy penalty associated with forming the alkali-metal/transition-metal interfaces may arise from the need to accommodate partial charge transfer, interface dipole formation, and screening.
In addition, hybridization between diffuse s orbitals and localized d orbitals is expected to be sensitive to both the specific chemical species at the interface and the crystallographic orientation of the system \cite{Albano:1984, Trioni:2013}.
Consistent with this, V, Cr, and Fe atoms in contact with Li show distinct magnetic-moment responses that depend on their orientation and distance from the interface.
For example, Cr develops an antiferromagnetic alignment along \(\langle100\rangle\) and \(\langle111\rangle\), while maintaining a zero magnetic moment along \(\langle110\rangle\), despite being initialized ferromagnetically, as we described in section~\ref{sec:methods}.
See Figure S.10 for more detail on these examples.

It is important to emphasize that, similar to the close-packed \((111)\) planes in PS FCC systems, the relatively close-packed \((110)\) planes in PS BCC systems exhibit the lowest \(\sigma_{AB}(\hat{k})\) values.
This behavior is consistent with the generally lower surface energies of densely packed planes, which minimize coordination loss upon surface formation.
At a coherent interface, bonding across the interface partially restores this coordination through unlike-atom interactions, suggesting a qualitative connection between crystallographic packing, surface energy, and interfacial energy.
Comparison with the theoretically predicted surface energies from Crystalium~\cite{Tran:2016}, shown in Figures~S.11-S.13, supports this correspondence primarily for phase-separating systems.

Conversely, the coherency strain energy of these systems---bottom panel of Fig.~\ref{fig:PS_BCC_all}---does not follow a clear dependence on size mismatch. For example, some systems with size mismatches of 11, 12, 16, and 20\% have \(\Delta E_{CS}(100) \leq10~meV/at\), whereas Cr/Mo and Cr/W, with mismatches of 11\% and 12\%, respectively, exhibit \(\Delta E_{CS}(\hat{k}) = 78 -116 ~meV/at\) for all orientations considered.
This behavior can also be seen by comparing Fig.~S.6 with Fig.~S.7, which show the coherency strain energy for all FCC and BCC systems. 
In the BCC case, the similar lattice parameters of Cr and Fe, Mo and W, and Nb and Ta lead to similar coherency-strain results at specific lattice-mismatch values in Fig.~S.7; this clustering is not observed in the FCC or HCP systems, shown in Fig.~S.6 and Fig.~S.8, respectively. 

\vspace{1em}
\begin{table}[htbp]
\centering
\caption{Comparison between the interfacial energy, \(\sigma_{AB}\), from Oh et al.~\cite{Oh:2024, Oh:2025} and this work.}
\label{tab:IE_oh-comparison}
\begin{tabular}{l @{\hspace{8pt}} c @{\hspace{16pt}} c @{\hspace{16pt}} c @{\hspace{8pt}}c}
\toprule
System & $\langle hkl \rangle$ & \(\sigma_{AB}\,[mJ/m^2]\) (from their S.I.~\([meV/at]\)) & This work, \(\sigma_{AB}\,[mJ/m^2]\) & $\Delta$\\
\midrule
Li/Cr & $\langle 100 \rangle$ & \(2034~(126.96)\) & \(1907\pm39\) & 127\\
Li/Cr & $\langle 110 \rangle$ & \(901~(56.244)\) & \(1299\pm6.5\) & -398\\
Li/Cr & $\langle 111 \rangle$ & \(1601~(99.907)\) & \(1486\pm23\) & 115\\
Li/Mo & $\langle 100 \rangle$ & \(1310~(81.742)\) & \(1355\pm51\) & -45\\
Li/Mo & $\langle 110 \rangle$ &  \(942~(58.786)\) & \(946\pm24\) & -4\\
Li/Mo & $\langle 111 \rangle$ & \(1151~(71.87)\) & \(1148\pm31\) & 3\\
Li/W & $\langle 100 \rangle$ &  \(2100~(131.11)\) & \(2004\pm4.4\) & 96\\
Li/W & $\langle 110 \rangle$ & \(1362~(85.027)\)  & \(1270\pm45\) & 92 \\
Li/W & $\langle 111 \rangle$ & \(1600~(99.872)\) & \(1505\pm11\) & 95\\
\bottomrule
\end{tabular}
\end{table}

As mentioned above, no systematic study has estimated interfacial energies for FCC systems, and we found no corresponding systematic work for BCC-based interfaces, either experimental or computational. 
As a result, the available methods and values vary considerably. 

We identify two cases worth discussing.
First, for Cr/Mo, we obtain \(336\pm45\) and \(5.3\pm14\) for \(\langle100\rangle\) and \(\langle110\rangle\), respectively. 
The large uncertainty for the \(\langle110\rangle\) result arises from a very flat formation-energy landscape, where the \(\Delta H_{SL}\) range is only \(1.1~meV/at\).
Gong et al.~\cite{Gong:2022}, who studied the stability and adhesion of Cr coatings on Mo for accident-tolerant nuclear cladding, estimated the interfacial energies of coherent Mo(100)/Cr(100), Mo(110)/Cr(110), and Mo(110)/Cr(100) interfaces.
Using a method involving interface slabs terminated by vacuum, they obtain values much higher than ours, about \(2280~mJ/m^2\) and \(1270~mJ/m^2\) for \(\langle100\rangle\) and \(\langle110\rangle\), respectively.
They also suggested that Mo(110)/Cr(110) is the most stable Mo/Cr interface, which matches our predictions.
Second, Oh et al.~\cite{Oh:2024, Oh:2025} investigated lithiophobic interfacial engineering as a strategy to suppress Li dendrite growth and promote dense, uniform Li deposition in anode-less and Li-metal batteries. 
To quantify the repulsive interactions underlying this behavior, they calculated coherent metal–Li interfacial energies, with an emphasis on Li/X interfaces (X = Cr, Mo, W, Ti, Zr). 
They used the approach developed by Lepley and Holzwarth~\cite{Lepley:2015}, namely, a constrained-relaxation version of the supercell approach.
Their isostructural results, compared with ours, are compiled in Table~\ref{tab:IE_oh-comparison}. 
In agreement with our results, the values along \(\langle110\rangle\) are the smallest. 
Some values are very close; in particular, the difference for Li/Mo is within our estimated interfacial-energy uncertainty. 
For Li/W, they obtain values \(\sim95~mJ/m^2\) higher than ours.
The largest discrepancies occur for the Li/Cr system: for example, their \(\sigma_{AB}(110)\) value is about \(\sim400~mJ/m^2\) lower than ours. 
We suspect that this difference may be attributed to the oscillatory magnetic behavior of Cr, which is affected by strain, interface geometry, and the presence of unlike atoms at the interface.

Cr systems are of particular interest because they are the only systems in the entire database that exhibit antiferromagnetic (AFM) behavior in the ground state (see some examples in Fig.~S.10.) 
As we mentioned above, all calculations are initialized with ferromagnetic alignment; however, under certain strain levels and specific interfacial interactions that depend on chemical species and orientation, Cr relaxes toward an AFM state.
In the S.I., we include Fig.~S.14 as another example showing this behavior, where all DFT-converged calculations for the Cr/W system along the \([100]\) orientation are compiled.
On the one hand, if we consider the largest superlattices, which show an AFM behavior (see Fig.~S.14(b)\&(c)), we obtain an interfacial energy of \(326\pm36~mJ/m^2\), indicating phase separation between unlike atoms, and a coherency strain energy of about \(119\pm2~meV/at\). 
On the other hand, when the smallest superlattices with weak or negligible AFM behavior are considered (see Fig.~S.14(d)\&(e)), the interfacial energy shifts to \(-228\pm16~mJ/m^2\), indicating a mixing tendency, while the coherency strain energy increases to \(153\pm2~meV/at\). 
In recent work, Smith et al.~\cite{Smith:2024} studied short-range order (SRO) of binary and ternary alloys in the Cr–Mo–W system using the direct method, i.e., isotropic strain calculations. 
Because they considered magnetism unlikely to be significant near the experimental phase boundaries, they did not use spin-polarized calculations.
From their results, the coherency strain energy for Cr-W along \(\langle100\rangle\) is about \(155~meV/at\), and the Cr-W SRO remains ordering-type, in agreement with our weak or negligible AFM results. 
Since Cr undergoes an antiferromagnetic-to-paramagnetic transition at a Néel temperature of 311 K, which can shift with alloying with metals such as V, W, and Mn~\cite{Schwebke:2025}, these effects may still influence the final material under ambient conditions. 
The connection among magnetism, strain, and interfacial-energy trends therefore warrants further study.

\subsubsection{HCP/HCP Interfaces}
Finally, the results for the 18 phase-separating HCP systems are shown in the complementary Figure~S.15 in the S.I. 
Since no additional crystal orientations were calculated due to computational constraints, a comparative orientation analysis cannot be made. However, a few observations are worth noting.
First, Cd/Re is the HCP phase-separating system with the lowest size mismatch; nevertheless, its mismatch \((\approx8\%)\) is relatively high compared with the first phase-separating FCC and BCC systems (cf. Fig.~\ref{fig:PS_FCC_IEpos} and Fig.~\ref{fig:PS_BCC_all}). 
Second, as in the previous cases, interfacial energies do not clearly correlate with the size mismatch of interface-conforming elements.
Third, there is also no clear trend between coherency strain energy and size mismatch when considering only the systems in Figure~S.15. However, a clearer trend appears when all PS and O systems are compiled into a single plot, as shown in Fig.~S.8: in general, a larger size mismatch leads to higher coherency strain energy in the HCP basal plane. 
This comparison also shows that HCP/HCP interfaces reach the highest \(\Delta E_{CS}(\hat{k})\) values among the crystal-structure types and orientations considered.
Lastly, in contrast to the FCC and BCC systems, almost no previous information regarding interfacial energies was found for these HCP systems. 
The only case identified was Cd/Zn, for which Martin, Doherty, and Cantor~\cite{Martin:1997} reported a value of \(110~mJ/m^2\) in their Table 5.2, whereas we obtain a \(\sigma_{Cd/Zn}(0001)=184\pm9~mJ/m^2\).
Moreover, little or no phase-diagram information in ASM~\cite{ASM:2026}, or ordered- or disorder-compound information in the experimentally reported structures in ICSD~\cite{ICSD:2026}, was found for 10 of the 18 systems in this figure, namely Cd-Re, Cd-Tc, Cd-Os, Hf-Tl, Cd-Ru, Mg-Re, Cd-Co, Re-Tl, Tc-Tl, and Os-Tl (sorted by increasing size mismatch).

\subsection{Quadrant II \& III: Systems that do not share a tie-line }
\label{sec:QII-III}

\begin{figure}[t]
    \centering
    \includegraphics[width=1.0\linewidth]{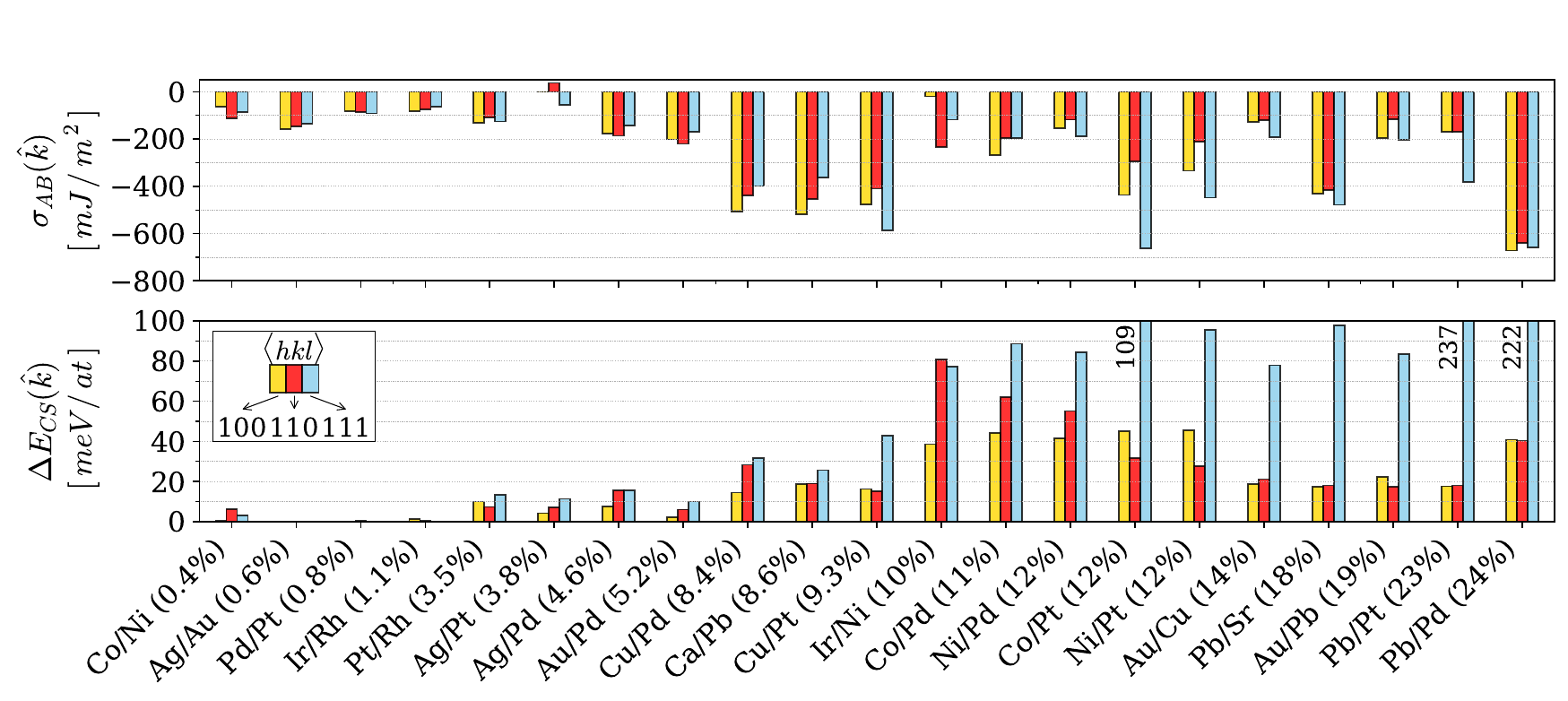}
    \caption{Interfacial energies (top) and coherency strain energy (bottom) for all the ordering FCC systems for the three calculated orientations \(\hat{k}\) in quadrants Q-II and Q-III of Fig.~\ref{fig:quads_all}. For visualization purposes, we exclude systems containing Al. Their results are shown in Fig.~S.16.
    The interfaces are sorted by lattice mismatch (value in parentheses), and the colored bar represents the results for each principal orientation, $(\langle100\rangle,\langle110\rangle,\langle111\rangle)$, from left to right. The elements in the \(A/B\) labels are sorted alphabetically.}
    \label{fig:FCC_Ordering_noAl}
\end{figure}

The systems classified as ordering (O) LRO-type are depicted on the left-hand side of Fig.~\ref{fig:quads_all}. 
As established previously, these systems exhibit one or more ordered compounds on the convex hull, i.e., \(\Delta H_\gamma (x)<0\) (Eq.~(\ref{eq:fe_general})). 
The interfacial energies of these O-type systems constitute quadrants Q-II and Q-III.  
Quadrant Q-II encompasses cases with positive interfacial energies and is dominated by BCC- and HCP-type interfaces (38 out of 40 cases).  
Conversely, quadrant Q-III comprises structures exhibiting negative interfacial energies and is predominantly composed of FCC-type interfaces (88 out of 186 cases), corresponding to 55\% of all \(\sigma_{AB}(\hat{k})\) values associated with FCC systems in this database.
Quadrant Q-III is the most populated region in Fig.~\ref{fig:quads_all}, comprising 186 out of 365 cases in total. 

\subsubsection{FCC/FCC interfaces}

Because quadrant Q-III contains mostly FCC-type interfaces, we first focus on these results. Figure~\ref{fig:FCC_Ordering_noAl} shows the ordering systems in quadrants Q-II and Q-III, excluding Al-containing interfaces for visualization purposes.
Since Q-II contains only two FCC results, we refer to this figure as the interfacial-energy results for Q-III hereafter.
As in the phase-separating systems in quadrants Q-I and Q-IV, the interfacial energies in Q-III show no clear dependence on lattice mismatch. 
This is especially evident for the Al-containing interfaces, where some of their values are below \(-2000~mJ/m^2\) for systems with very small size mismatches.
These Al-containing systems have the most negative values in the database, with an average of approximately \(-1400~mJ/m^2\) (cf. the complementary Fig.~S.16.)
In contrast, the coherency strain energy shows some dependence on size mismatch, with a positive correlation that is most evident in the \(0-10\%\) range (see also Fig.~S.6).

\begin{figure}[t]
\centering
\includegraphics[width=1.0\textwidth]{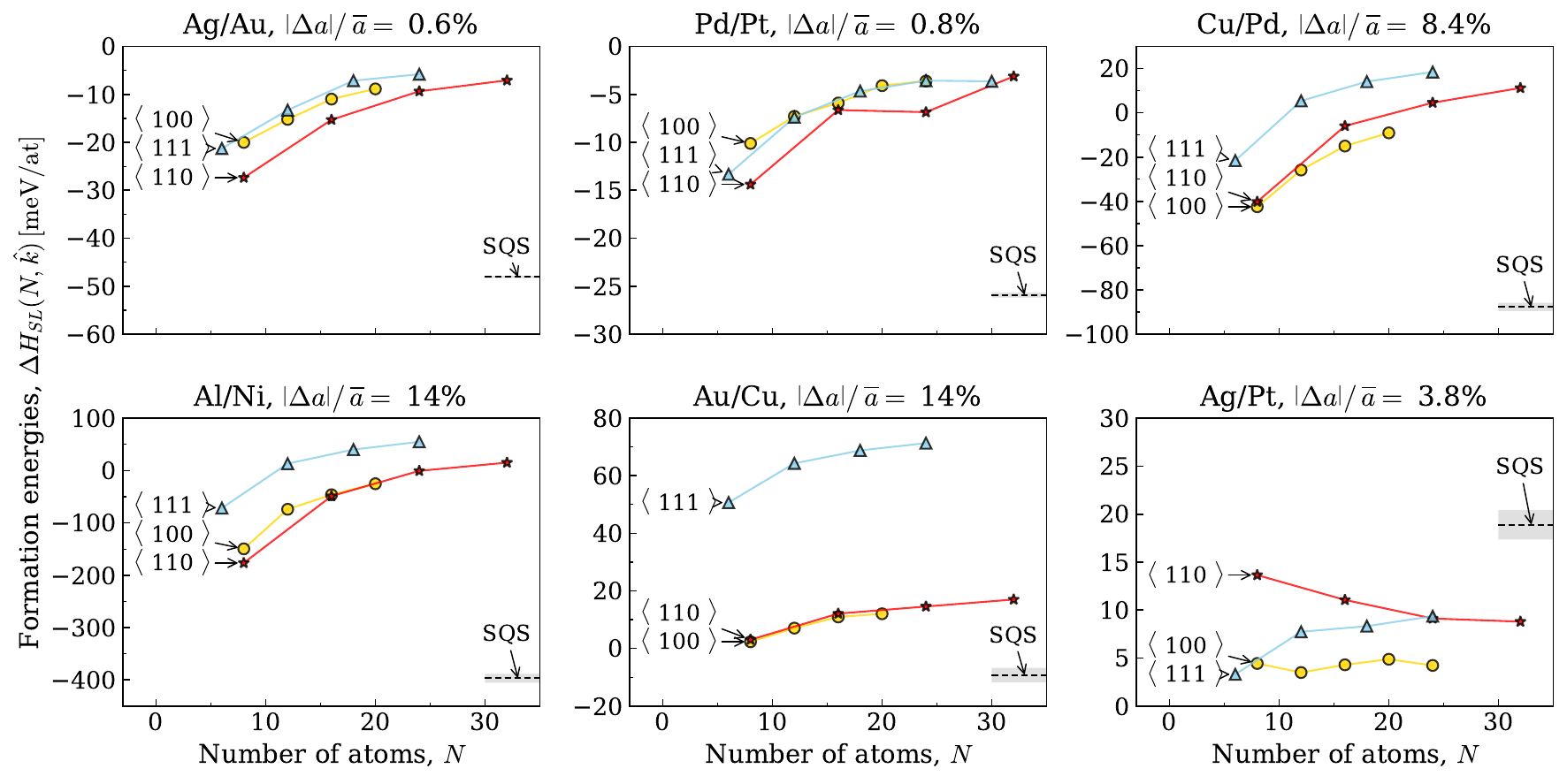}
\caption{Superlattice formation energies for each interface orientation of six selected FCC/FCC cases of quadrant \hyperref[fig:quads_all]{Q-III} (sorted by size mismatch). SQS approximates an equiatomic random alloy, and the gray shadow represents the standard deviation across five different seeds for 32-atom SQS calculations. Note that the last Ag/Pt system is the only FCC case that falls under Q-II and Q-III.}
\label{fig:O_FCC_FEs_AgPt}
\end{figure}

We found no experimental studies reporting interfacial energies for these FCC systems. 
This may be related to their ordering tendency, which makes the formation of thermodynamically stable, sharp interfaces separating two phases less likely.
Furthermore, only a few theoretical studies were available for comparison.
Sakong et al.~\cite{Sakong:2024} separated the elastic compression of pseudomorphic Ag layers from the chemical interface contribution and obtained a weakly attractive Ag/Pt(111) interface of about \(-95~mJ/m^2\).
Their Ag/Pd(111) value was approximately twice as negative, indicating stronger Ag–Pd bonding and a greater driving force for intermixing.
We obtain a similar trend, with \(\sigma_{AB}(111)\) values of \(-55\pm4\) and \(-142\pm2\) for Ag/Pt and Ag/Pd, respectively. For Ag/Pt, we also obtain values of \(37\pm5\) and \(0.3\pm12~mJ/m^2\) along \([110]\) and \([100]\), respectively. 
Ag/Pt is the only ordering-type system that exhibits both positive and negative interfacial energies. 
This behavior is consistent with its complex phase diagram, which shows ordering tendencies at low temperatures and solid-solution and phase-separation regions at higher temperatures~\cite{Durussel:1996, Hart:2017}. 
Hart et al.~\cite{Hart:2017} recently revisited this system and showed, using a combination of DFT, cluster expansion, finite-temperature simulations, and microscopy, that the previously reported 53 at.\% Pt phase in Ag--Pt correspond to the equiatomic ordered \((L1_1)\) AgPt compound.

For further details on the formation-energy trends of Ag/Pt and other systems, we compiled Figure~\ref{fig:O_FCC_FEs_AgPt}. 
This figure also includes Ag/Au and Au/Cu, which were previously studied by Ozoliņš et al.~\cite{Ozolins:1998}. 
We observe similar trends for these systems, particularly the upward behavior of the \(\Delta H_{SL}(N,\hat{k})\) curves for all the calculated orientations.
This behavior, caused by the negative interfacial energies, indicates a chemical preference for unlike atoms at the interface and a tendency to form ordered structures at 0 K.
We also include the random-alloy energies obtained from SQSs, which are lower than those of the long-period superlattices for all selected systems. 
In particular, the Al/Ni SQS is about \(~400~meV/at\) more stable than all corresponding long-period superlattices.

We emphasize again that a distinct pattern is observed for systems with size mismatches \(\gtrsim14\%\) in Fig.~\ref{fig:FCC_Ordering_noAl}: their \(\Delta E_{CS}(\hat{k})\) values for \(\hat{k}=[100]\) and \([110]\) tend to converge. 
Au/Cu is one example of this behavior; as shown in Fig.~\ref{fig:O_FCC_FEs_AgPt}, its formation-energy curves along these orientations tend to overlap.
We attribute this behavior to the Bain-path connection between the superlattice structures and the accumulation of strain energy at large size mismatch, which may help overcome the FCC--BCC energy barrier that connects these two structure types.

\subsubsection{BCC/BCC and HCP/HCP interfaces}

The ordering-type BCC systems in quadrants Q-II and Q-III are shown in the complementary Fig.~S.17. 
Once again, no general trend is observed for the interfacial energies. 
However, most systems with positive interfacial energies in Q-II are aligned along one or two crystallographic directions and correspond to spin-polarized calculations that include either Cr or Fe. The only exception is Cr/Fe.

The Cr/Fe system exhibits positive interfacial energies of \(104-151~mJ/m^2\) across three calculated orientations. 
Lu et al.~\cite{Lu:2010} reported values ranging from \(138-148~mJ/m^2\) for \(\sigma_{AB}(100)\), in agreement with our value of \(149\pm27~mJ/m^2\) for the same orientation.
Although they did not extract the strain energy from their slabs, we estimate coherency strain energies no larger than \(3~meV/at\). 
This small strain contribution may explain the agreement between the results despite the use of different approaches.
It should be emphasized that Cr/Fe has an experimentally confirmed miscibility gap, as reported at the end of Table~S.4.
The placement of this system in Q-II results from a very dilute calculation in OQMD, CrFe\(_{15}\), which is reported to be stable on the hull by \(6~meV/at\). 

Another system that may be misclassified in Q-II is Li/Na, for which we obtain positive and negative values ranging from \(-140\) to \(135~mJ/m^2\). 
On the current hull, another very diluted compound, Na\(_{13}\)Li, shows an ordering tendency of more than \(70 ~meV/at\).
This occurs because the OQMD convex hull was constructed using corrected chemical potentials to reproduce experimental values, which may overcorrect some cases, such as \(\mu_{Na}\)~\cite{Kirklin:2015}. 
When the uncorrected BCC Na reference is used, the same compound lies \(19~meV/at\) above the hull, leading to a phase-separating classification for the Li/Na system.
This interpretation is more consistent with the phase diagram, which shows that Li and Na are immiscible at low temperatures~\cite{Abdessameud:2016}. 

The ordering-type HCP systems in quadrant Q-II are shown in the complementary Fig.~S.18.
The HCP systems in Q-III are divided into two figures for visualization purposes: Fig.~S.19 includes the ordering systems with lattice mismatch \(\leq 10\%\), while Fig.~S.20 includes the remaining cases.

No specific trends are observed for these systems, other than that the BCC- and HCP-based structures produce similar numbers of cases in quadrants Q-II and Q-III and have similar average interfacial energies.
In quadrant Q-II, the average values are \(136\) and \(115~mJ/m^2\), while in Q-III they are \(-240\) and \(-234~mJ/m^2\), for the BCC and HCP systems, respectively.

\subsection{Comparison of the interfacial energy database with phase diagrams}

The general metallurgical view is that interfacial energies are positive. In particular, Porter, Easterling, and Sherif~\cite{Porter:2009} provide a rule of thumb in which coherent interfaces have energies up to \(200~mJ/m^2\), semicoherent interfaces range from \(200-500~mJ/m^2\), and incoherent interfaces have higher energies of approximately \(500-1000~mJ/m^2\).
This view assumes that coherent interfacial energy mainly arises from chemical interactions, whereas semicoherent and incoherent interfaces include additional structural contributions related to lattice mismatch and misfit dislocations.
Here, we present a different perspective. 
We show that strain-separated interfacial energies correlate reasonably well with both phase-separating and ordering systems, as described above through the classification into quadrants Q-I and Q-III. 
In this final section, we extend the analysis by comparing the interfacial-energy and coherency-strain-energy results directly with the dominant features of the binary phase diagrams.

We propose a simple classification of the phase diagrams based on (A) the presence or absence of a solid solution across the full composition range and (B) the presence of intermetallic compounds or complete immiscibility at lower temperatures.
Condition (A) requires a distinct area between the solidus and solvus in a phase diagram. We subdivide this category into three: 
\begin{itemize}
    \item If the phase diagram shows an order-disorder phase transition, we classify this as ``solid solution plus intermetallic" \((SS,INT)\).
    \item If the phase diagram shows the presence of a phase-separating tendency via a miscibility gap, we classify it as ``solid solution plus miscibility gap" \((SS,MG)\).
    \item If the phase diagram does not show any characteristic of the previous subclassification, we classify it as ``solid solution" \((SS)\). 
\end{itemize}

Note that any of these subcategories could include phase diagrams that show allotropic transformations.

The phase diagrams that do not fulfill the condition (A) are classified according to the condition (B). 
If the phase diagram shows any intermetallic compound with a congruent melting point (or not) at low temperatures, it is classified as ``intermetallic" \((INT)\). 
The rest of the phase diagrams, which show eutectic, monotectic, or peritectic reactions, but no presence of intermetallics at low temperatures, were classified as ``immiscible solids" \((IS)\). 
See the complementary Figure S.21 for a schematic of the classification, including some phase diagram examples. 

Based on this proposal, we analyze, classify, and compile phase-diagram references for all FCC systems in this database and selected Cr systems. 
They are listed in Table~S.4 together with some observations. 
All the systems were initially checked in the ASM database~\cite{ASM:2026}, and their references were then examined in depth.
Some systems have been widely studied, and for others, only a few references exist. 
Therefore, in case of any discrepancies or big uncertainties, or if the results were based only on theoretical calculations, we added an extra label to the classification called ``theoretically predicted, or uncertain", symbolized by a superimposed red dot \(\textcolor{red}{\bullet}\). 
Although this is not a complete literature review, it is a point of comparison with the ``most accepted phase diagrams".

\begin{figure}%[htbp]
    \centering

    \subfigure{%
        \begin{overpic}[width=0.98\textwidth]{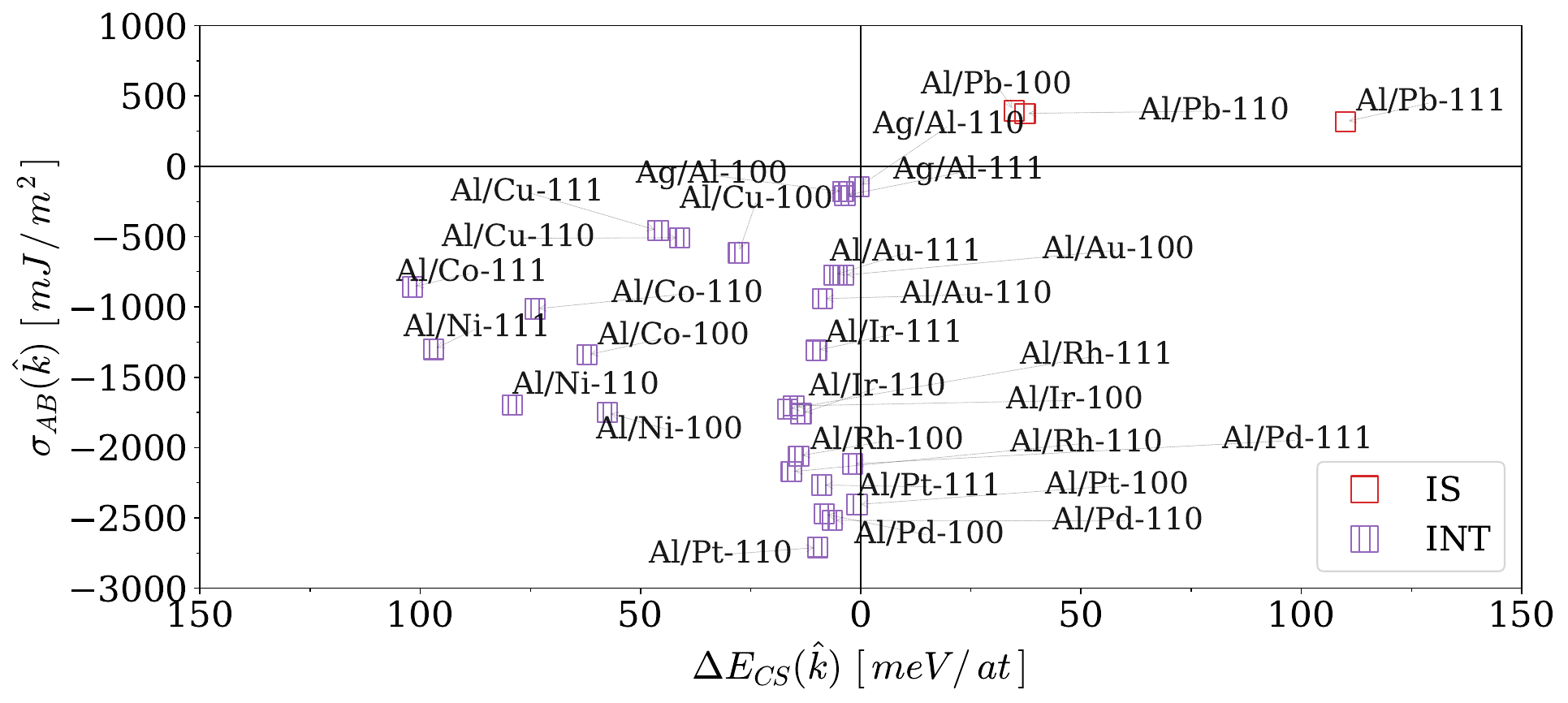}
            \put(0,43){(a)}%{\textbf{(a)}}
        \end{overpic}
    }
    \vspace{0.5em}
    \subfigure{%
        \begin{overpic}[width=0.96\textwidth]{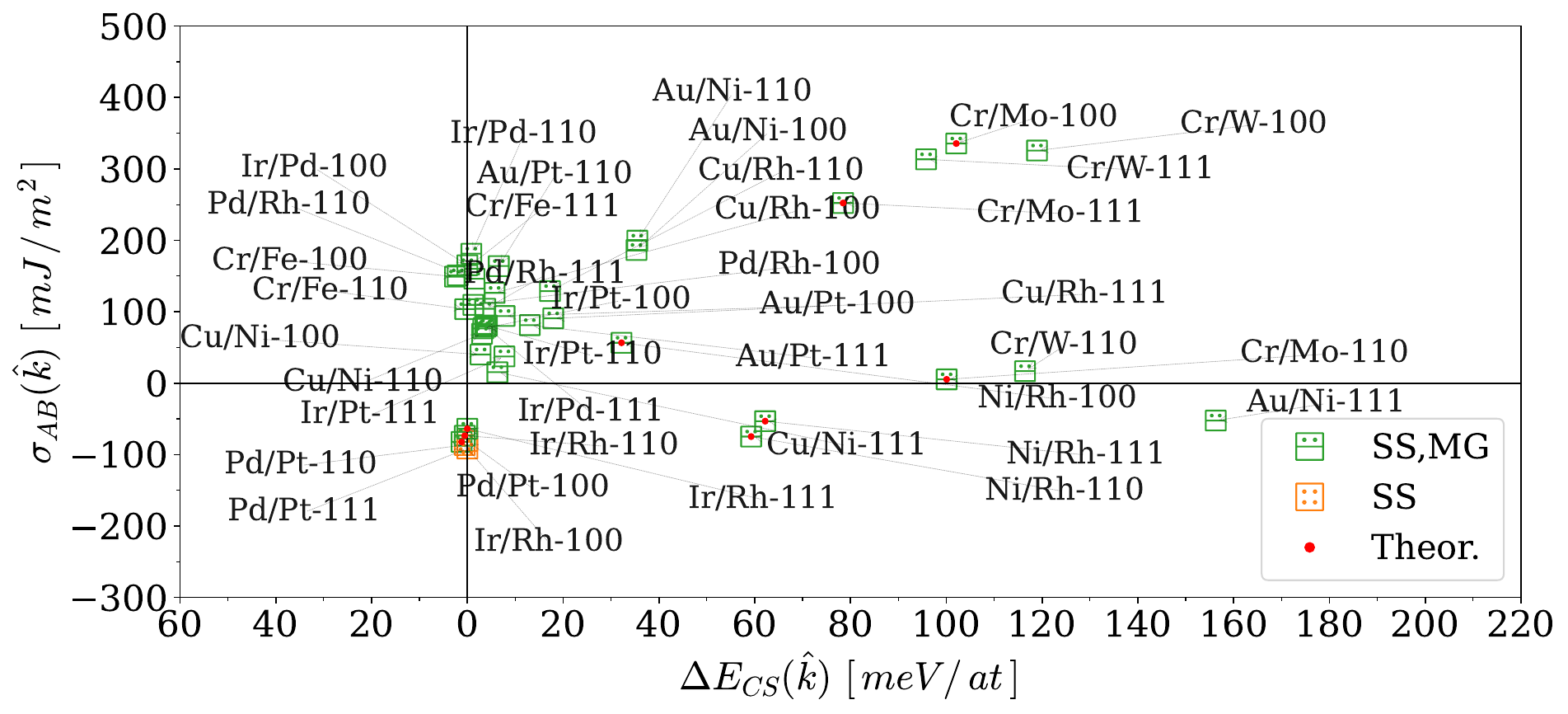}
            \put(-1,43){(b)}
        \end{overpic}
    }

    \caption{Figures showing interfacial and coherent strain energies, similar to Fig.~\ref{fig:quads_all}. (a) All Al-based systems in this database. They can form either intermetallic compounds (INT) or be immiscible (IS). (b) All systems with an experimentally confirmed miscibility gap (SS,MG) or theoretically predicted (SS,MG \(+~\textcolor{red}{\bullet}\)). This comparison was made by cross-referencing our database with the sources compiled in Table~S.4}   
    \label{fig:phase_diag_comp}
\end{figure}

Based on this classification, we present two representative examples in Fig.~\ref{fig:phase_diag_comp}.
In panel (a), we show the interfacial energies and coherency strain energies for all Al-containing interfaces together with their corresponding classifications. 
As discussed previously in Section~\ref{sec:Q_I_fcc_interfaces_subsub}, within FCC systems, the Al phase separates only with Pb. 
Notably, the interfacial energies for the three crystallographic orientations considered in this database are negative for all FCC Al-based intermetallic systems, with the exception of Al/Pb results.

In panel (b), we show all the FCC and BCC systems with experimentally confirmed or theoretically predicted miscibility gaps \((SS,MG)\).
Upon careful review, we note that all experimentally confirmed cases fall within the ranges \(0-20~meV/at\) and \(0-200~mJ/m^2\) for coherency strain energy and interfacial energy, respectively. 
The upper interfacial-energy limit coincides with Porter's rule-of-thumb range for coherent interfaces.
Even if Cr/Fe were reclassified, as discussed above, it would remain within the same region.
The only exceptions are Au/Ni and Cr/W, which show higher interfacial energies and, more notably, higher \(\Delta E_{CS}(\hat{k})\) values.
The theoretically predicted systems with a MG, Ni/Rh and Cr/Mo, exhibit behavior similar to that of the Au/Ni and Cr/W systems.
The only system that falls outside this trend is Ir/Rh, which has a theoretically predicted miscibility gap \(SS,MG^*\).
We also include the Pd/Pt system, for which a \(SS,MG^*\) was previously proposed~\cite{Bharadwaj:1991}, but was later ruled out by theoretical~\cite{Klein:1996} and experimental studies~\cite{Lee:2022}.
Since Ir/Rh and Pd/Pt systems follow similar trends, we suggest that the predicted miscibility gap in Ir/Rh, which exhibits ordering-type LRO and only negative interfacial energies, may also require further reassessment.

Finally, it is worth mentioning that all FCC/FCC systems
forming intermetallic compounds (INT) fall in quadrant Q-III with negative interfacial energy, while all
immiscible solids (IS) are in Q-I, with positive interfacial energies. 
Systems with all labeling types of
solid solution (SS) are present in all quadrants, but with a narrower interfacial energy range than the INT or IS cases. 
See these trends in the complementary Fig.~S.22.

\newpage
\section{Conclusions}
\label{sec:conclusions}

We use the supercell approach to separate the coherency strain energy from the chemical contribution to the interface energy. Using this method, we generate a systematic database of theoretically predicted interfacial energies for coherent isostructural interfaces. 
The database contains 365 FCC/FCC, BCC/BCC, and HCP/HCP interface configurations. FCC/FCC and BCC/BCC interfaces are considered along the principal orientations \(\langle100\rangle\), \(\langle110\rangle\), and \(\langle111\rangle\), while HCP/HCP interfaces are considered along \(\langle0001\rangle\). 
Overall, the database spans 41 metallic elements with size mismatches of up to 25\%. The fitting procedure was refined to reduce the average interfacial-energy uncertainty to \(6.5\%\). To analyze the trends, we divided the results into four quadrants: 
\begin{itemize}
    \item Q-I comprises phase-separating LRO-type systems with positive interfacial energies and contains 128 results. Their average interfacial energies, \(\bar\sigma_{AB}\), are \(288\), \(961\), and \(529~mJ/m^2\) for the FCC/FCC, BCC/BCC, and HCP/HCP interfaces, respectively.
    \item Q-II comprises ordering LRO-type systems with positive interfacial energies. It contains 40 cases that show different orderings relative to the interfacial energy orientations.
    \item Q-III comprises ordering LRO-type systems with negative interfacial energies and is the largest group, containing 186 results. Their \(\bar\sigma_{AB}\) values are \(-608\), \(-240\), and\(-234~mJ/m^2\) for the FCC/FCC, BCC/BCC, and HCP/HCP interfaces, respectively. 
    \item Q-IV comprises 11 phase-separating systems with negative interfacial energies, for which high coherency strain energies \((\gtrsim59~meV/at)\) inhibit the ordering tendency associated with the chemical interface contribution.
\end{itemize}

We analyze these results in detail and compare them with the available literature. 
In general, the calculated magnitudes underestimate previous theoretical and experimental estimates. 
Large discrepancies are found for the FCC systems in Q-I, whereas better agreement is obtained for the BCC systems in the same quadrant. 
These discrepancies likely reflect differences in interface models, reference states, and relaxation constraints, as well as the inclusion of strain-energy contributions and finite-temperature, kinetic, and thermodynamic effects in experimentally derived values.

Despite these quantitative differences with previous literature results, the calculated interfacial energies show consistent qualitative agreement with the phase-diagram classifications. 
For example, all FCC systems classified as immiscible (IS) are in quadrant Q-I, with positive interfacial energies and phase-separation LRO-type, while all FCC systems forming intermetallic compounds (INT) are in quadrant Q-III, with negative interfacial energies and ordering LRO-type.
Also, most systems with a miscibility gap lie within the ranges \(0-20~meV/at\) and \(0-200~mJ/m^2\) for coherency strain energy and interfacial energy, respectively.

We find no clear correlation between interfacial energy and the size mismatch of the elements forming the interface. 
In contrast, coherency strain energy shows a positive correlation with size mismatch. 
We also confirm that, for phase-separating systems, the orientations associated with the lowest close-packed surface energies generally correspond to the lowest interfacial energies. 
This trend is not observed for systems with ordering tendencies.

Overall, the supercell approach provides a practical way to separate coherency strain energy from interfacial energy. 
This separation is necessary to isolate the chemical contribution to interfacial energy and to distinguish whether unlike-atom bonding favors phase separation or ordering.

\newpage
\section*{Acknowledgments}

A.S.-C. acknowledges funding from the Toyota Research Institute, as well as financial support from the National Agency for Research and Development (ANID)/DOCTORADO BECAS CHILE/2018 - 56180024. 
J.P. acknowledges support from the National Science Foundation Graduate Research Fellowship award NSF DGE-2234667.
The authors acknowledge computational resources from the Quest high-performance computing facility at Northwestern University, which is jointly supported by the Office of the Provost, the Office for Research, and Northwestern University Information Technology. 

\section*{Supporting Information}
\label{sec:SI}

Additional calculation details, complementary references, and observations are compiled into long tables, along with complementary plots. The full database of interfacial energies and coherency strain energies, together with OQMD-ID calculations and other relevant data, is attached as a .csv file.

\bibliographystyle{apsrev4-2}
\bibliography{manuscript/references}

\end{document}

\title{Supporting Information \\Energetics of Coherent Heterophase Interfaces: Separating Chemical and Strain Contributions}

\author{Adolfo Salgado-Casanova} 
\affiliation{Department of Materials Science and Engineering, Northwestern University, Evanston, IL 60208, USA
}

\author{Jacob Pietryga}
\affiliation{Department of Materials Science and Engineering, Northwestern University, Evanston, IL 60208, USA
}

\author{Yi Xia}
\affiliation{ Department of Mechanical \& Materials Engineering, Portland State University, Portland, OR 97201, USA
}

\author{Chris Wolverton}
\email{c-wolverton@northwestern.edu }
\affiliation{Department of Materials Science and Engineering, Northwestern University, Evanston, IL 60208, USA
}

\date{\today}

\maketitle

\newpage
\clearpage
\clearpage
% \section{Supporting Information}
\label{sec:SI}

% Wrapping text in longtable columns
\newcommand{\headcell}[2]{\parbox[c]{#1}{\centering #2}}
\newcommand{\notecell}[1]{%
            \parbox[t]{3.8cm}{%
            \raggedright 
            \footnotesize
           #1
           }%
        }
\newlength{\phasediagtablewidth}
\setlength{\phasediagtablewidth}{1.0\textwidth}

\newcommand{\twolinecell}[2]{%
  \begin{tabular}[t]{@{}c@{}}
  #1\\#2
  \end{tabular}%
}

% SI numbering
\renewcommand{\thetable}{S.\arabic{table}}
\renewcommand{\thefigure}{S.\arabic{figure}}
\renewcommand{\theequation}{S.\arabic{equation}}

% Unique hyperref anchors for SI floats
\renewcommand{\theHtable}{SItable.\arabic{table}}
\renewcommand{\theHfigure}{SIfigure.\arabic{figure}}
\renewcommand{\theHequation}{SIequation.\arabic{equation}}

\setcounter{table}{0}
\setcounter{figure}{0}
\setcounter{equation}{0}

\FloatBarrier

\section{Additional details}

All figures in this document have a detailed caption, explaining additional details of the methodology and results from this work.

\subsection{Special Quasirandom Structures}

Direct density functional theory calculations on disordered materials are computationally impractical and statistically inadequate when performed on small random supercells. 
The Special Quasirandom Structure (SQS) method addresses this limitation by identifying a compact periodic configuration that accurately reproduces the correlations of the target random alloy~\cite{Zunger:1990}.

The SQS structures for this work were generated using the \textsc{icet} package, which defines the species, sublattices, cluster cutoffs, composition, and supercell size within a common cluster-space framework \cite{Angqvist:2019}. 
Candidate occupations were optimized to reproduce the most relevant short-range correlations, then relaxed with DFT following the OQMD framework, and used as representative disordered configurations.
The resulting DFT energies approximate the disordered-state energy at a selected composition. 
This approach was used to compare random disorder and superlattice energetics as discussed in the main text.

\subsection{Methodology for phase diagram selection}

The phase diagram references listed in Table~\ref{tab:phase_diagrams} were selected using the following procedure. 
First, we searched the ASM phase diagram database for each binary system and sorted the available references chronologically. 
The most recent experimental phase diagrams, when available, were examined and compared with calculated phase diagrams to assess their consistency, particularly regarding the presence or absence of miscibility gaps. 
For systems with only calculated phase diagrams available in the ASM database, an additional literature search was conducted to identify more recent experimental or computational studies not included in the ASM database. 
When multiple phase diagrams were available and showed no significant discrepancies, preference was given to experimental results. 
Otherwise, the available results were compared to select the most appropriate reference. 
If no additional information was found, the calculated phase diagram from the ASM database was used. 
For some binary systems, neither experimental nor calculated phase diagrams were available.

%%%%%%%%%%%%%%%%%%%%%%%%%%%%%%%%%%%%
%%%%%%%%%%%%% FIGURES %%%%%%%%%%%%%%
%%%%%%%%%%%%%%%%%%%%%%%%%%%%%%%%%%%%

\clearpage
\subsection{Supplementary Figures}

\begin{figure}[H]
\centering
\includegraphics[width=1.0
\textwidth]{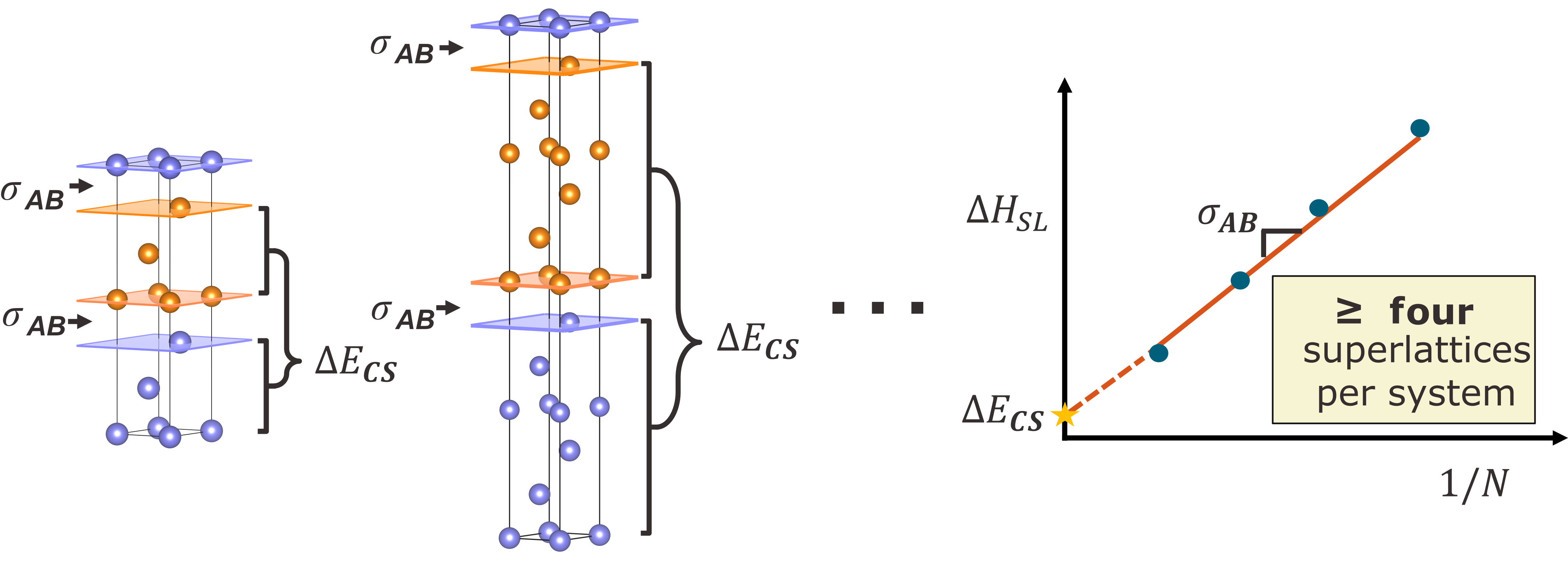}
\caption{Schematic of the supercell approach. The left side shows two FCC superlattices along \(\langle111\rangle\) with the same interface area but different volumes. The right side shows an ideal regression case for the formation energy \((\Delta H_{SL})\) against \(1/N\), being \(N=\) number of atoms per superlattice. 
Here, the coherency strain energy \((\Delta E_{CS})\) is obtained from the y-intercept and the interfacial energy \((\sigma_{AB})\) from the slope of the linear regression.} 
\label{fig:SFig_IE_sup_appr}
\end{figure}

\clearpage

\begin{figure}[htbp]
    \centering
    \subfigure[]{%      
        \includegraphics[width=0.35\textwidth]{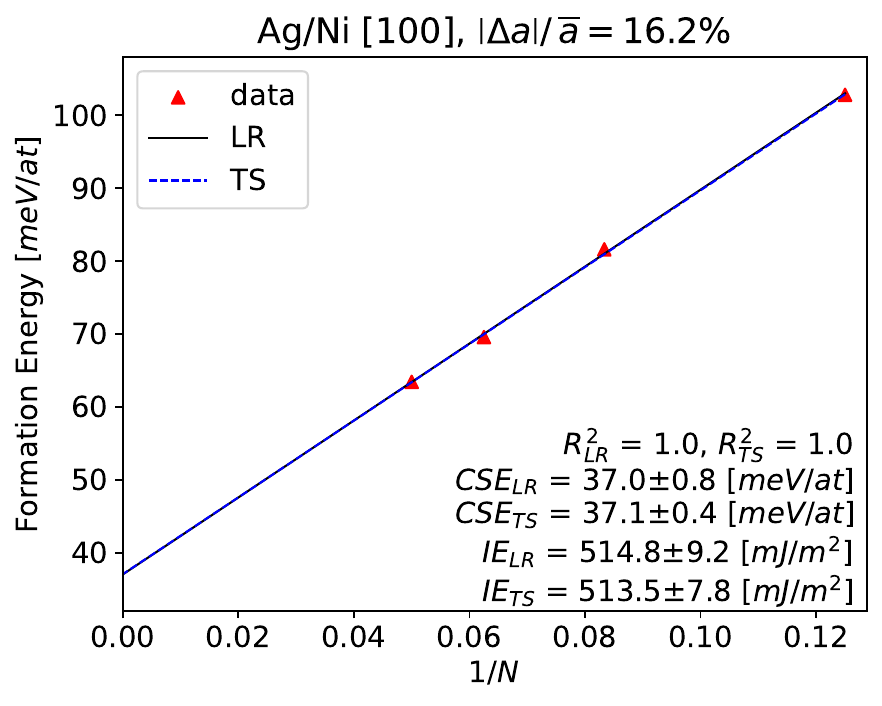}
    }
    \hfill
    \subfigure[]{%      
        \includegraphics[width=0.62\textwidth]{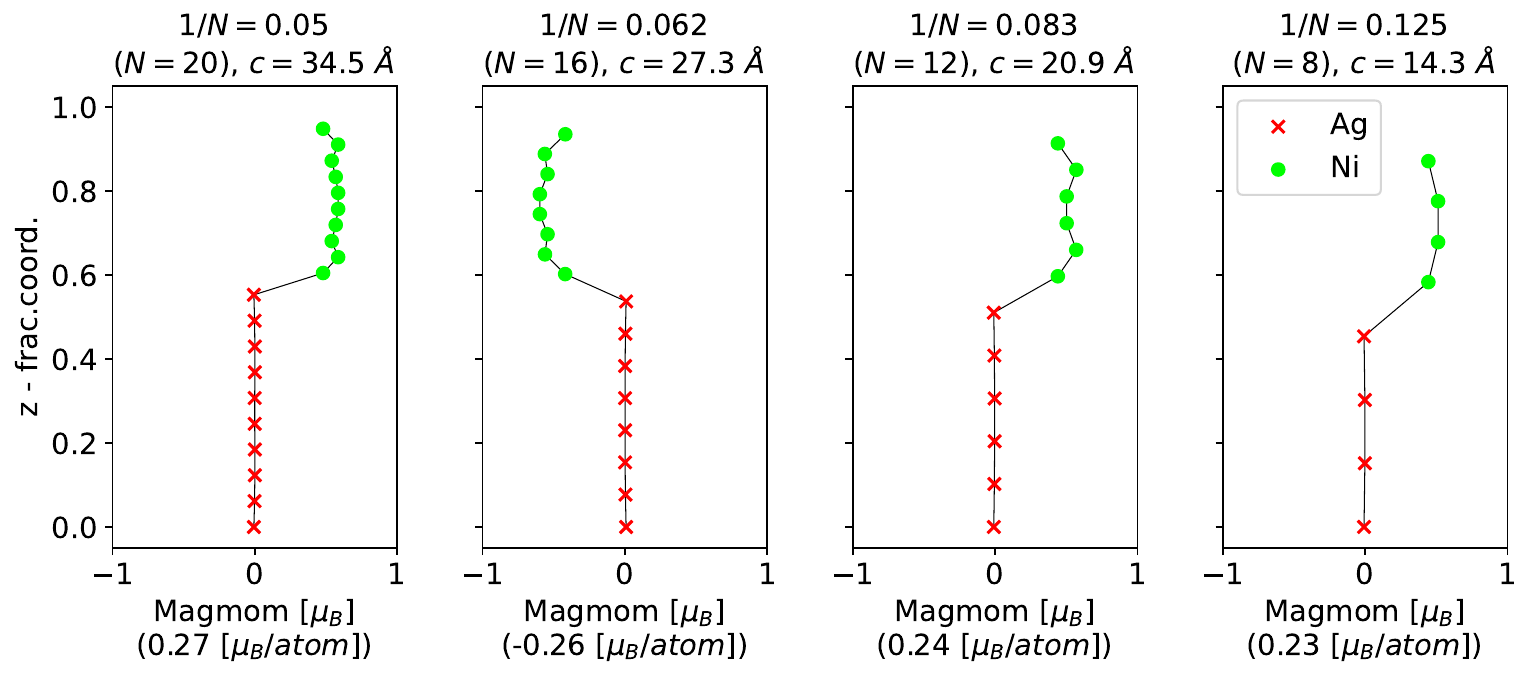}
    }
    \vspace{0.5em}
    \subfigure[]{%      
        \includegraphics[width=0.35\textwidth]{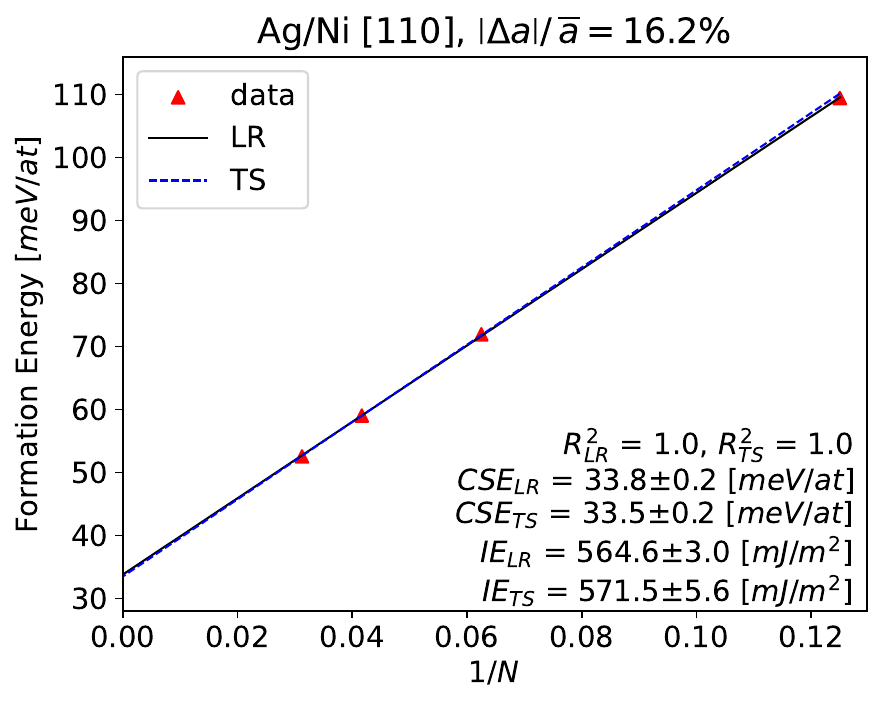}
    }
    \hfill
    \subfigure[]{%      
        \includegraphics[width=0.62\textwidth]{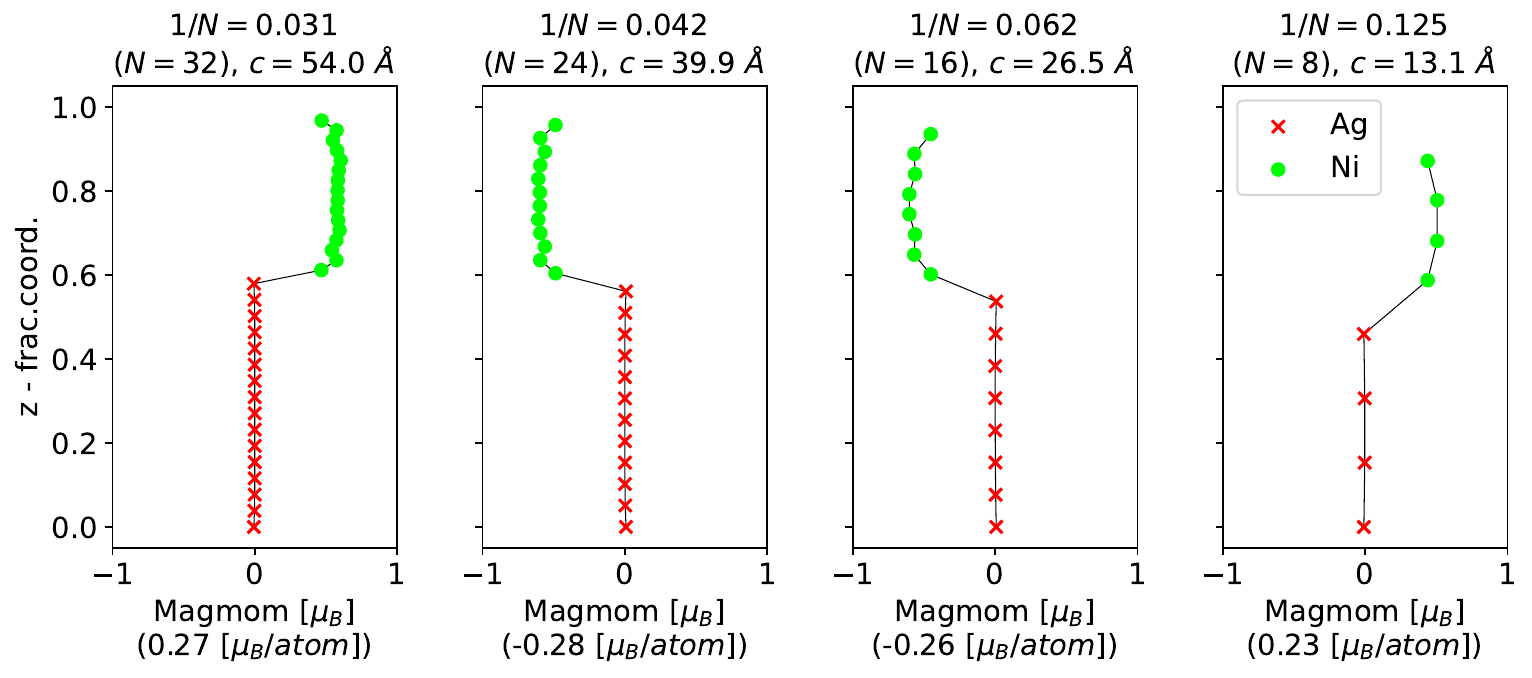}
    }
    \vspace{0.5em}
    \subfigure[]{%      
        \includegraphics[width=0.35\textwidth]{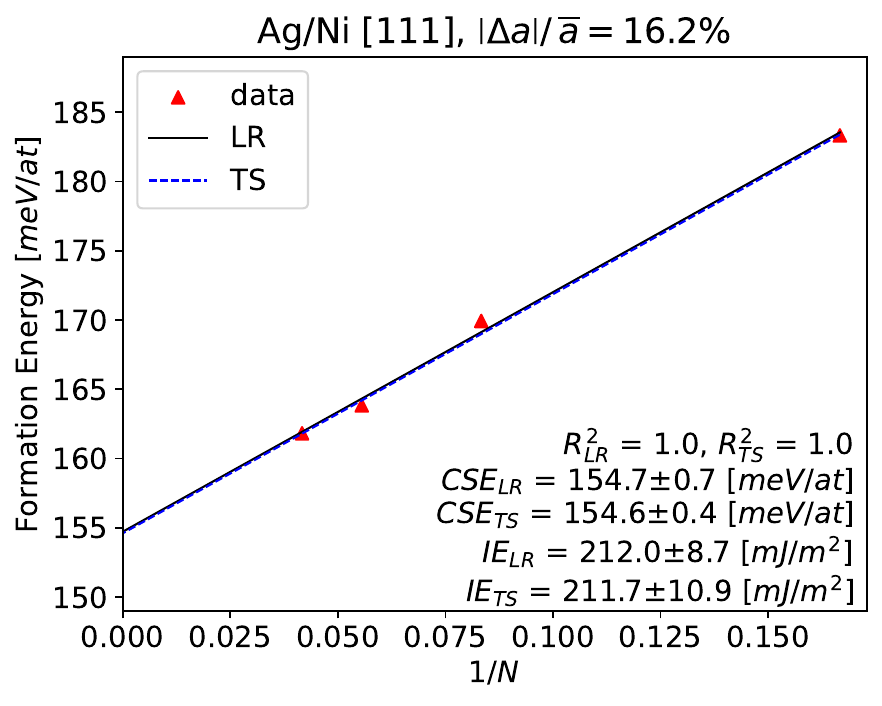}
    }
    \hfill
    \subfigure[]{%      
        \includegraphics[width=0.62\textwidth]{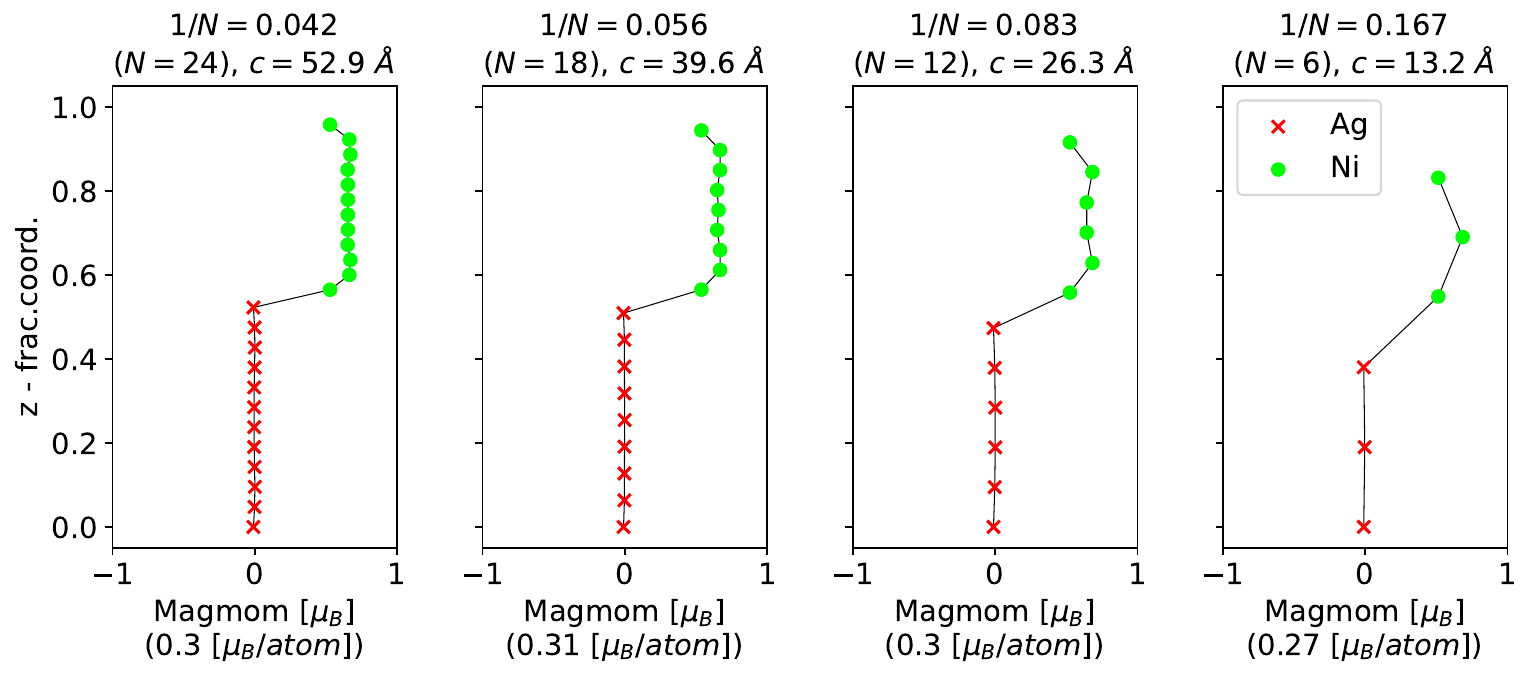}
    }

    \caption{
    Example of converged results for the \textbf{phase-separating} Ag/Ni system. (a),(c),(e): formation-energy calculations used to estimate the interfacial energy (IE) and coherency strain energy (CSE) via Theil-Sen (TS) and linear (LR) regressions along the 3 calculated orientations: [100], [110], and [111]. (b),(d),(f): Magnetic-moment-per-atom profiles along the z-direction---perpendicular to the interface plane---for each superlattice in the formation energy plots. 
    }
    \label{fig:Example_FCC_PS}
\end{figure}

\clearpage
\begin{figure}[htbp]
    \centering
    \subfigure[]{%      
        \includegraphics[width=0.35\textwidth]{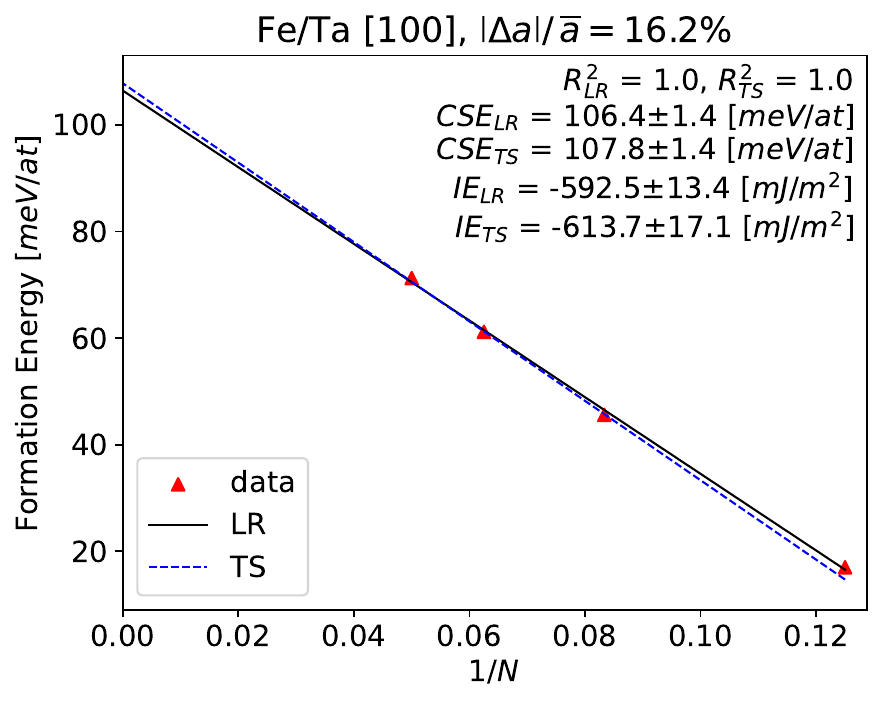}
    }
    \hfill
    \subfigure[]{%      
        \includegraphics[width=0.62\textwidth]{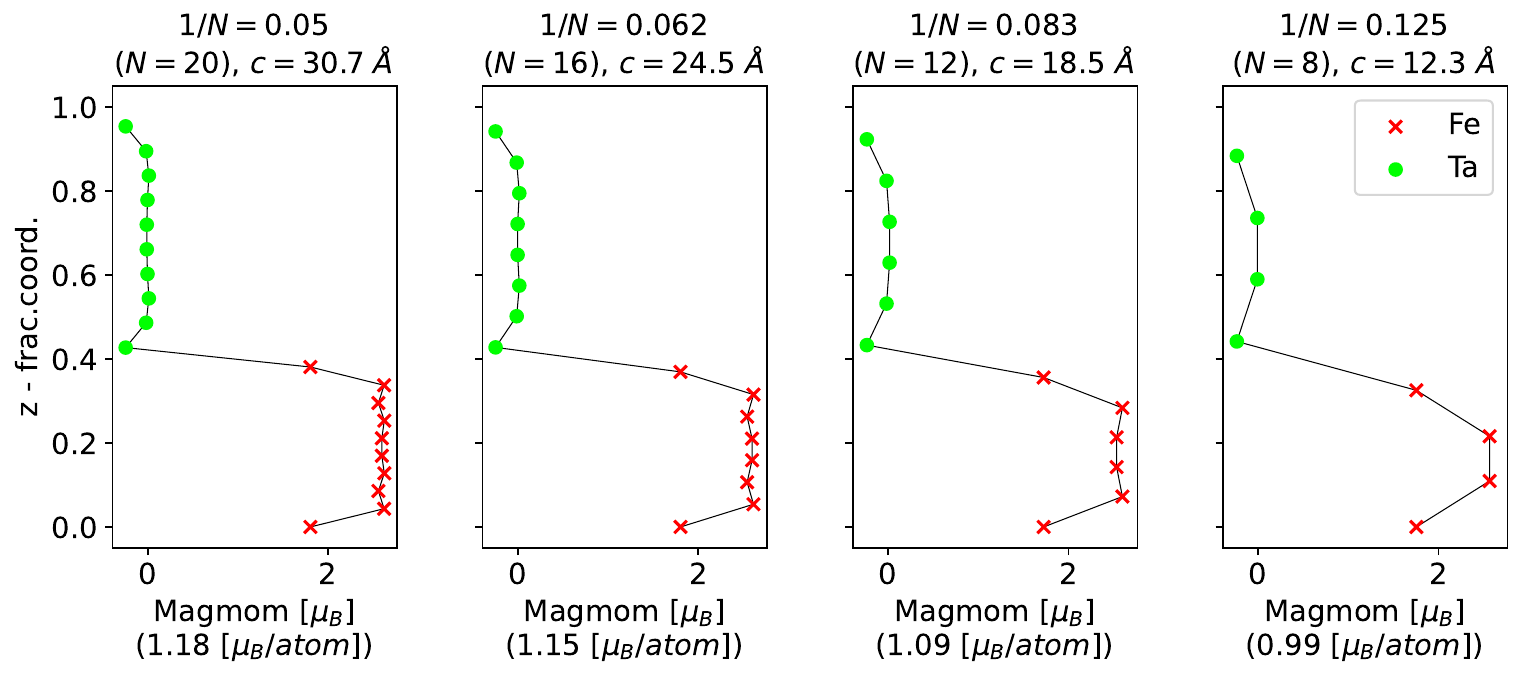}
    }
    \vspace{0.5em}
    \subfigure[]{%      
        \includegraphics[width=0.35\textwidth]{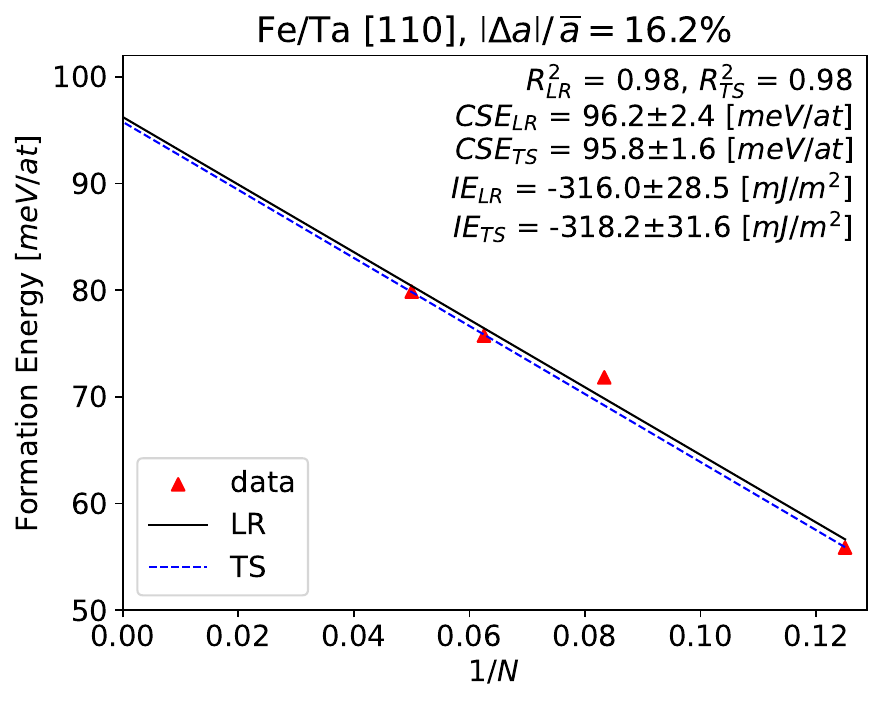}
    }
    \hfill
    \subfigure[]{%      
        \includegraphics[width=0.62\textwidth]{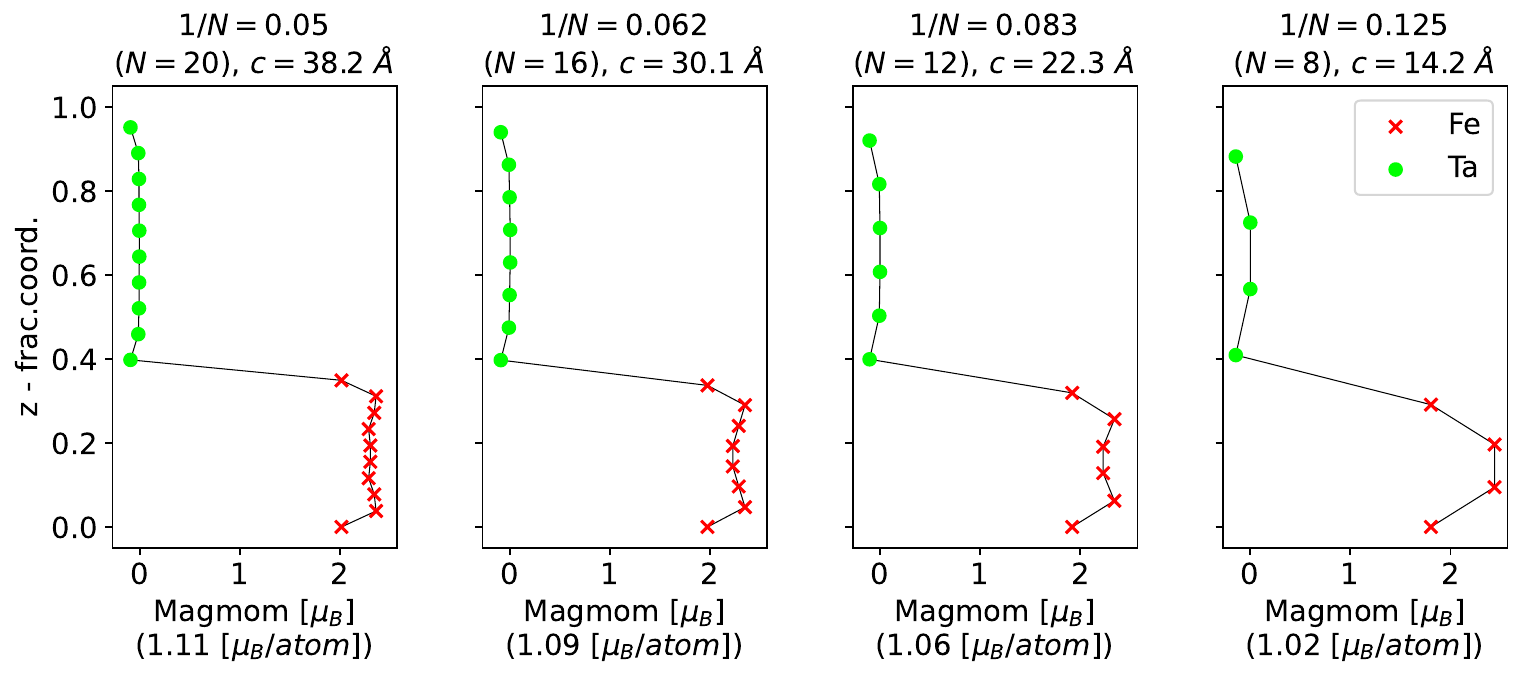}
    }
    \vspace{0.5em}
    \subfigure[]{%      
        \includegraphics[width=0.35\textwidth]{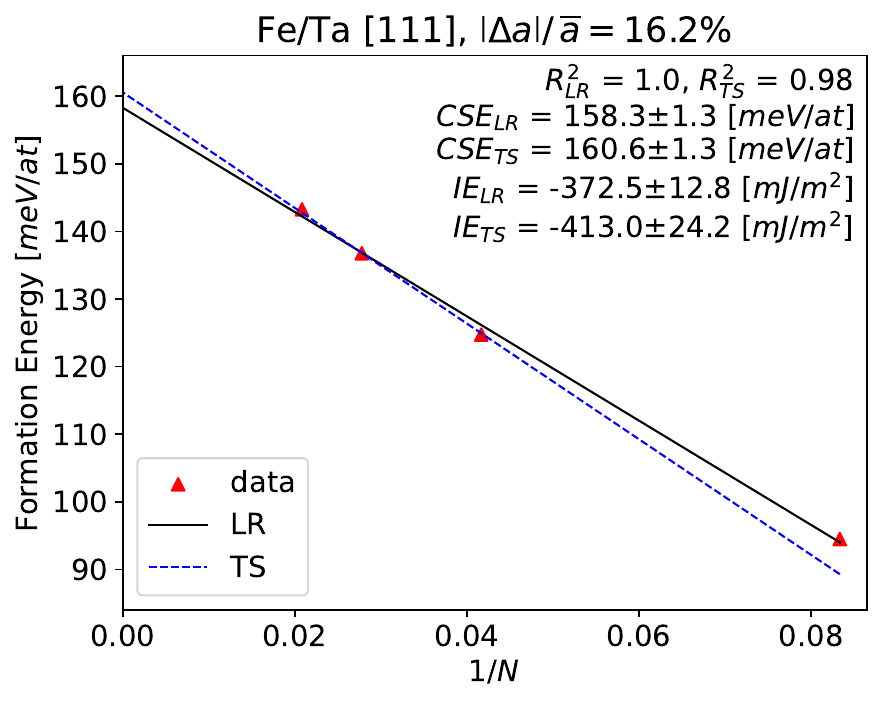}
    }
    \hfill
    \subfigure[]{%      
        \includegraphics[width=0.62\textwidth]{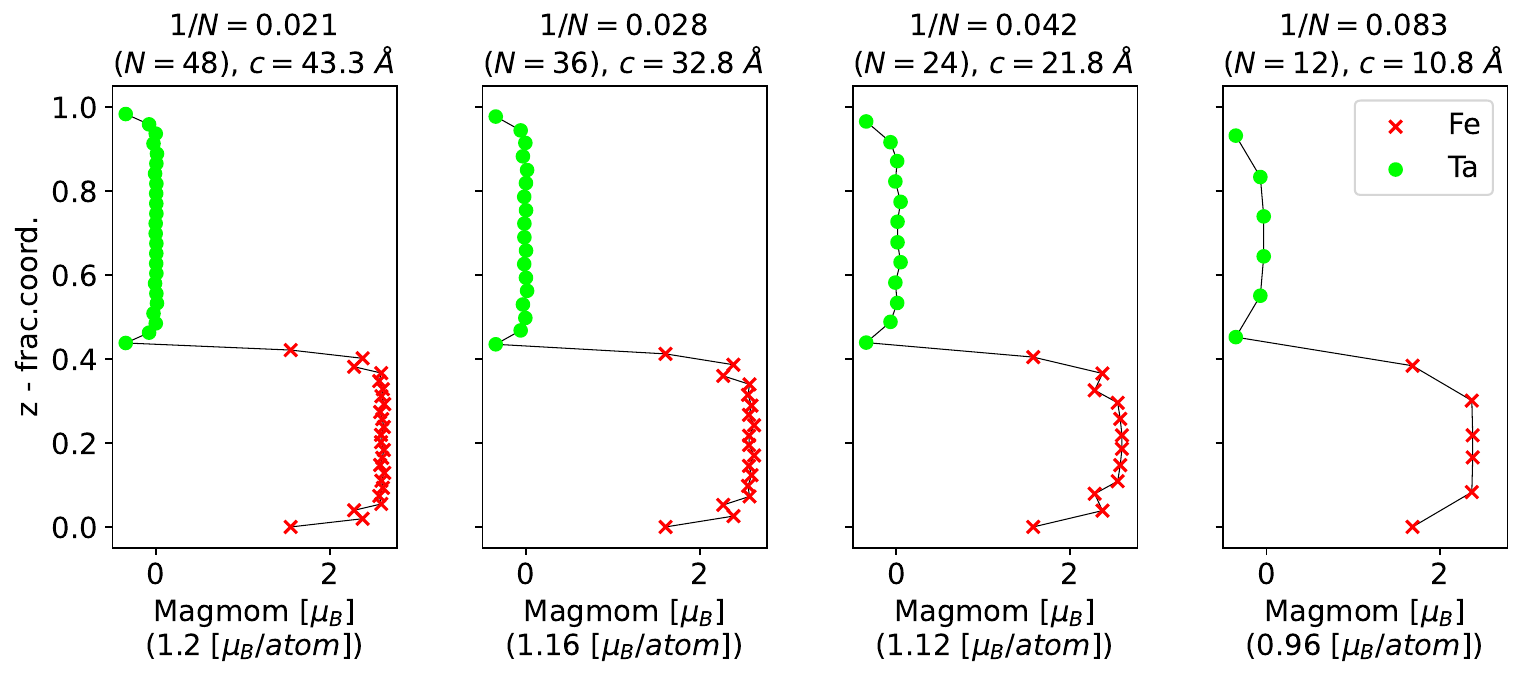}
    }

    \caption{ Example of converged results for the \textbf{ordering} Fe/Ta system. (a),(c),(e): formation-energy calculations used to estimate the interfacial energy (IE) and coherency strain energy (CSE) via Theil-Sen (TS) and linear (LR) regressions along the 3 calculated orientations: [100], [110], and [111]. (b),(d),(f): Magnetic-moment-per-atom profiles along the z-direction---perpendicular to the interface plane---for each superlattice in the formation energy plots. 
    }
    \label{fig:Example_BCC_OS}
\end{figure}

\begin{figure}[t]
\centering
\includegraphics[width=0.8\textwidth]{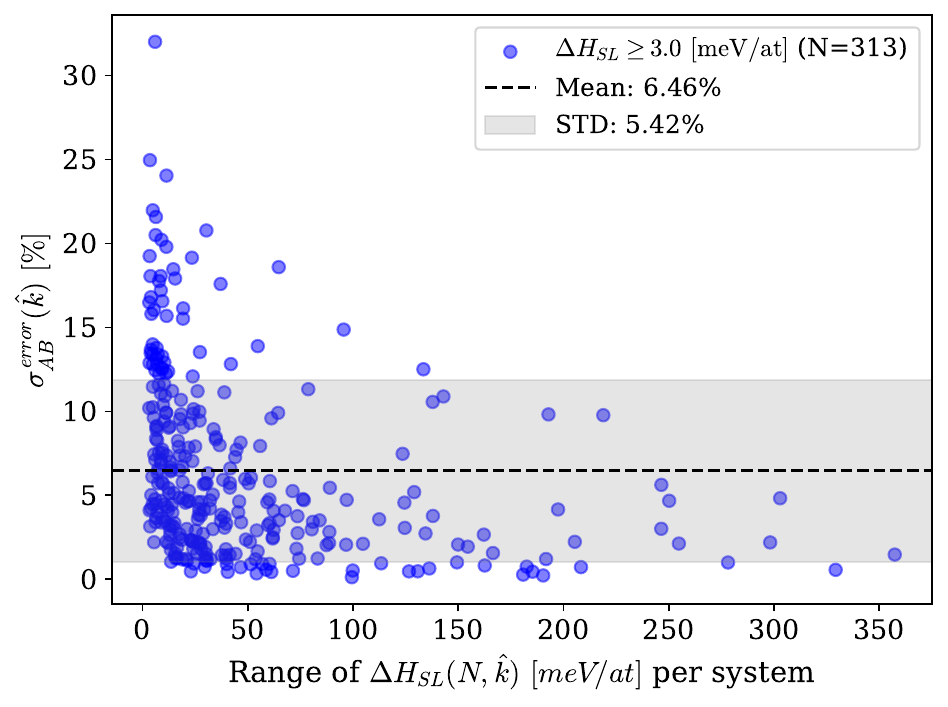}
\caption{Regression error or uncertainty in the database based on {Eq.(10)} of the main text. Includes 313 of 365 systems whose formation energy range exceeds \(3~meV/at\). } 
\label{fig:IE_error_DB}
\end{figure}

\clearpage
\begin{figure}[htbp]
    \centering

    % Left column: one centered figure
    \begin{minipage}[c]{0.44\textwidth}
        \centering
        \subfigure[]{%
            \includegraphics[width=\linewidth]{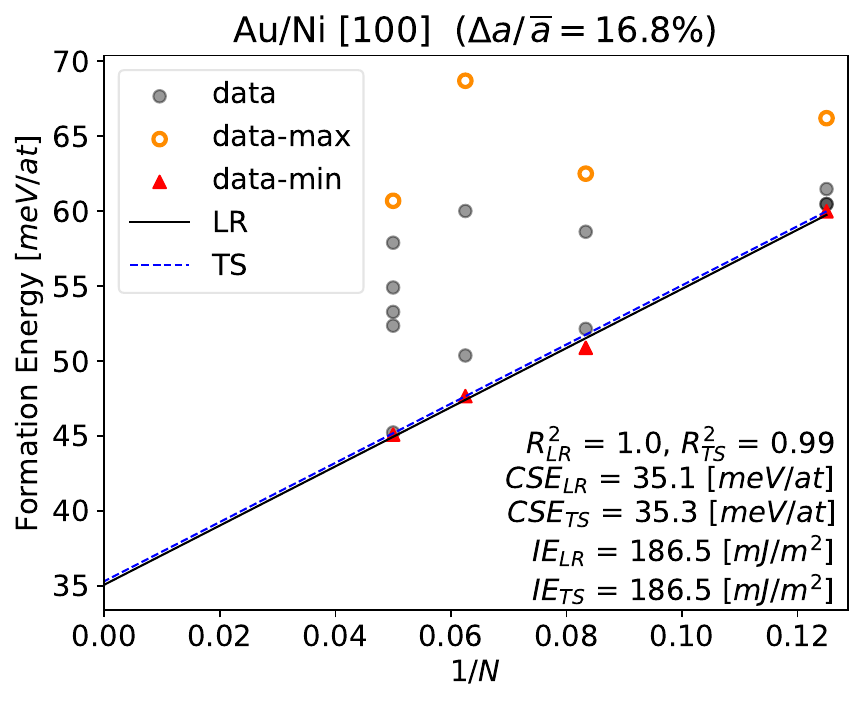}
        }
    \end{minipage}
    \hfill
    % Right column: two stacked figures
    \begin{minipage}[c]{0.55\textwidth}
        \centering
        \subfigure[]{%
            \includegraphics[width=\linewidth]{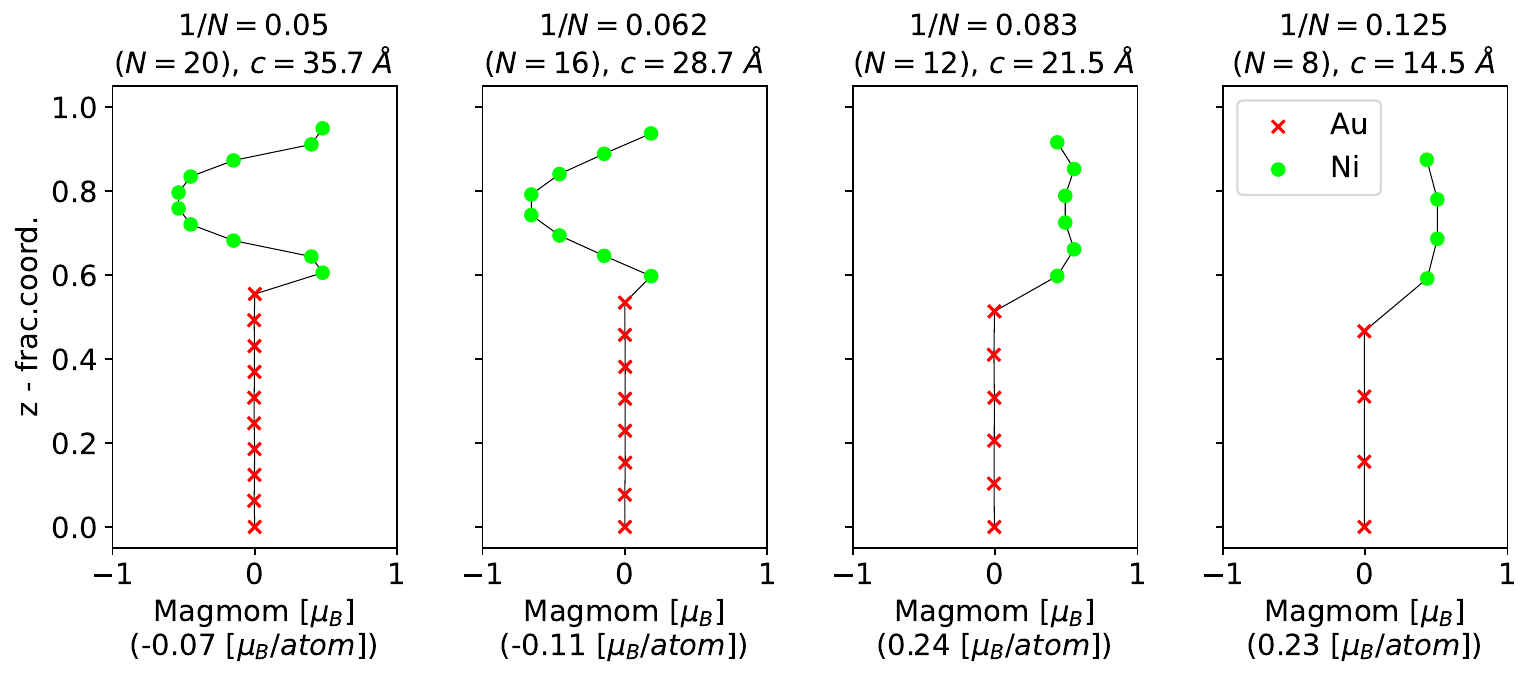}
        }
        \vspace{0.5em}
        \subfigure[]{%
            \includegraphics[width=\linewidth]{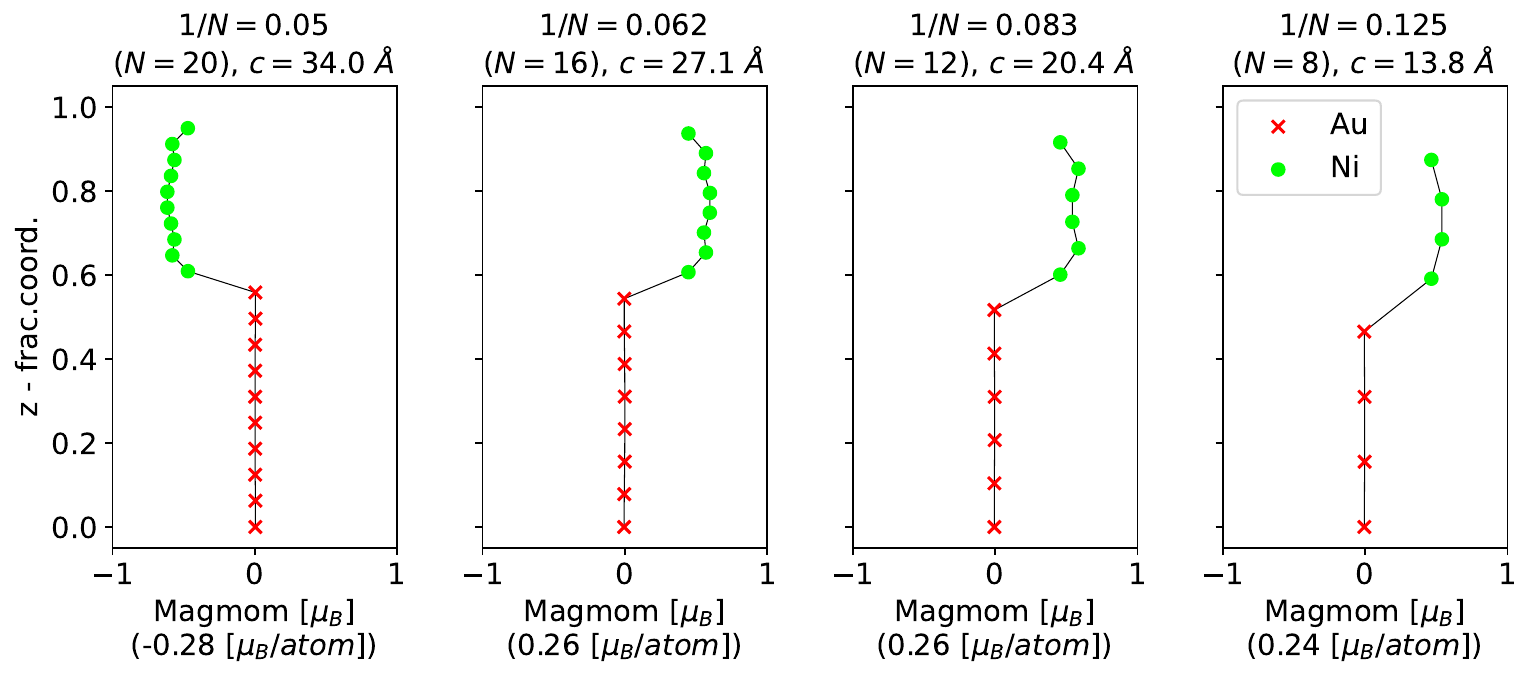}
        }
    \end{minipage}

    \caption{
    Example of the phase-separating Au/Ni system along \(\langle100\rangle\), with all formation energies from DFT-converged calculations plotted in (a). 
    The magnetic-moment-per-atom profiles along the superlattice z-direction for the discarded data points with maximum formation energy---data-max in (a)---and for the data points used for the linear regression---data-min in (a)---are plotted in (b) and (c), respectively. Superlattices with an oscillatory magnetic moment for Ni have higher formation energies, yielding slopes and y-intercepts that differ from the data-min points and affecting the interfacial energies and coherency strain results. This variability motivates revising each system to reduce calculation uncertainty, as discussed in Section 2.2 of the main text. To this end, particular emphasis was placed on spin-polarized calculations, as in the Au/Ni system shown in this figure.
    }
    \label{fig:three_panel_layout}
\end{figure}

\clearpage
\begin{figure}[t]
\centering
\includegraphics[width=0.68\textwidth]{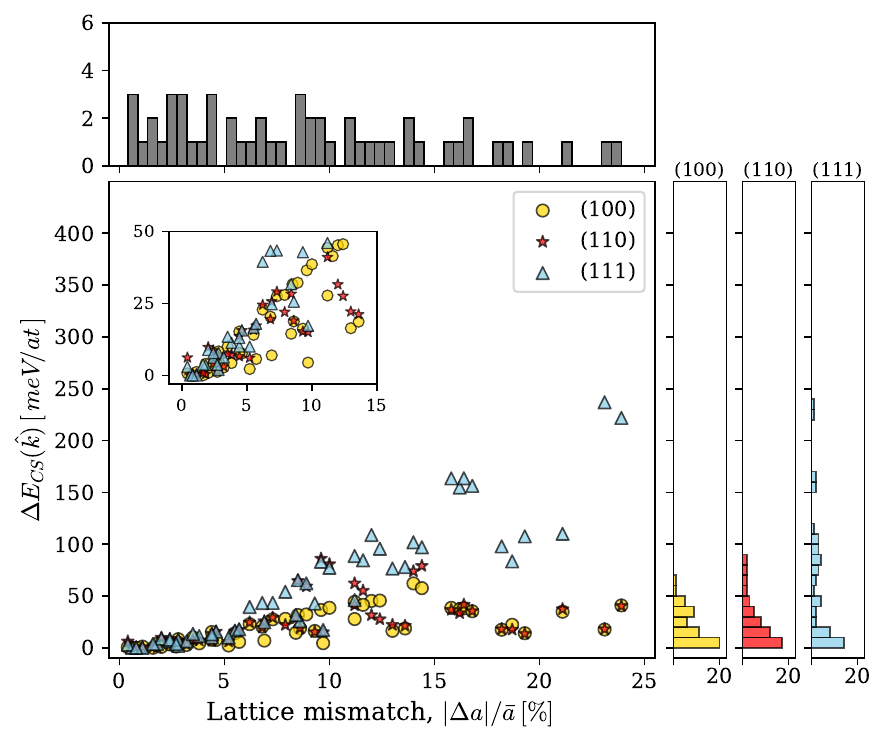}
\caption{Coherency strain energy vs. lattice mismatch for all FCC coherent interface structures in this database. Inset shows the initial range up to 15\% lattice mismatch.} 
\label{fig:cse_vs_lat_FCC}
\end{figure}

\begin{figure}[t]
\centering
\includegraphics[width=0.68\textwidth]{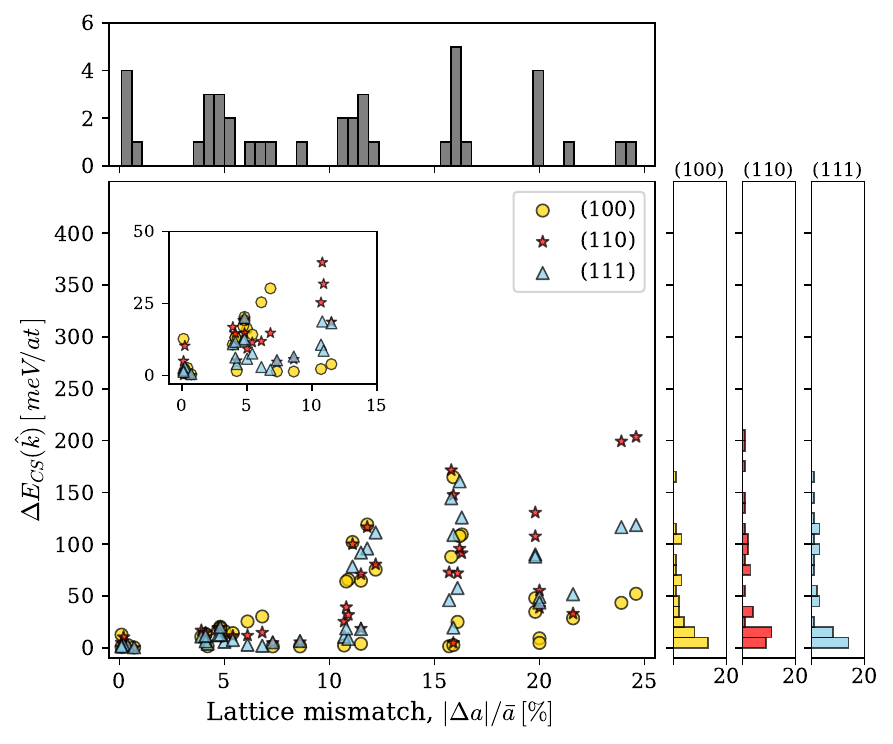}
\caption{Coherency strain energy vs. lattice mismatch for all BCC coherent interface structures in this database. Inset shows the initial range up to 15\% lattice mismatch.} 
\label{fig:cse_vs_lat_BCC}
\end{figure}

\begin{figure}[t]
\centering
\includegraphics[width=0.7\textwidth]{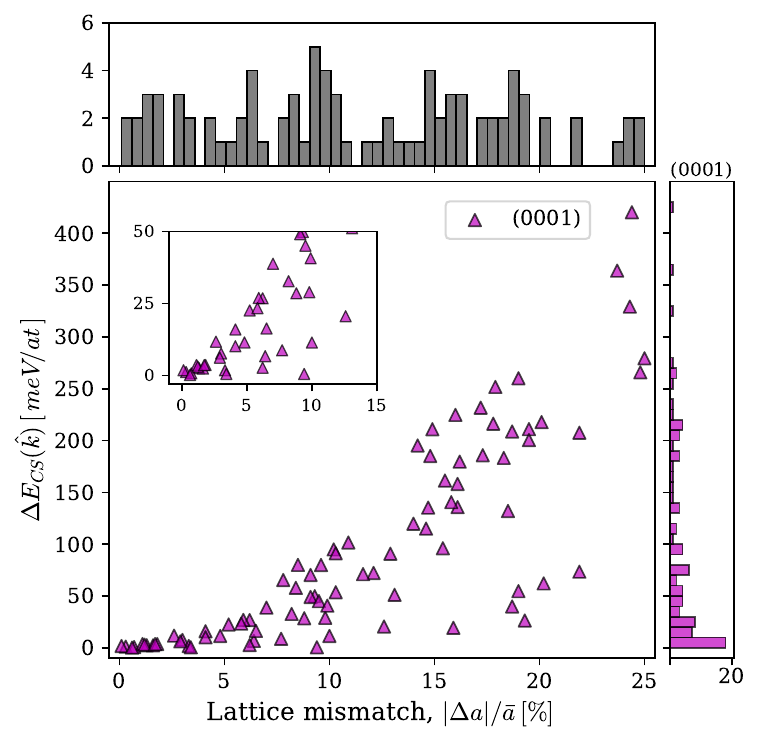}
\caption{Coherency strain energy vs. lattice mismatch for all HCP coherent interface structures in this database. Inset shows the initial range up to 15\% lattice mismatch.} 
\label{fig:cse_vs_lat_HCP}
\end{figure}

\begin{figure}[t]
\centering
\includegraphics[width=1.0\textwidth]{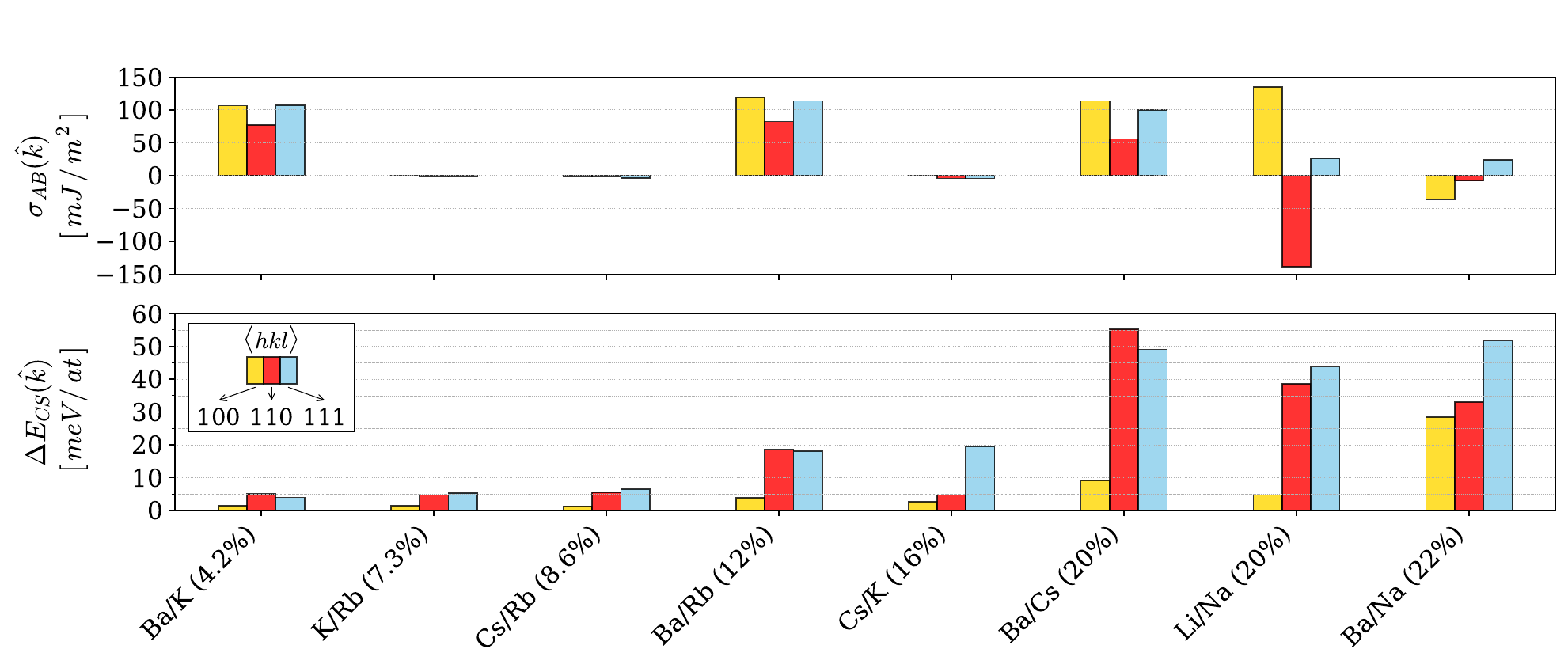}
\caption{Interfacial energies (top) and coherency strain energy (bottom) for the interface structures containing only alkali and alkaline earth metals with a BCC crystal structure in this database. The interfaces are sorted by lattice mismatch (value in parentheses), and the colored bar represents the results for each principal orientation $(\langle100\rangle, \langle110\rangle, \langle111\rangle)$, from left to right. The elements in the \(A/B\) labels are alphabetically sorted.} 
\label{fig:BBC_alkalis}
\end{figure}

\clearpage
\begin{figure}[htbp]
    \centering

    \subfigure{%
        \begin{overpic}[width=0.8\textwidth]{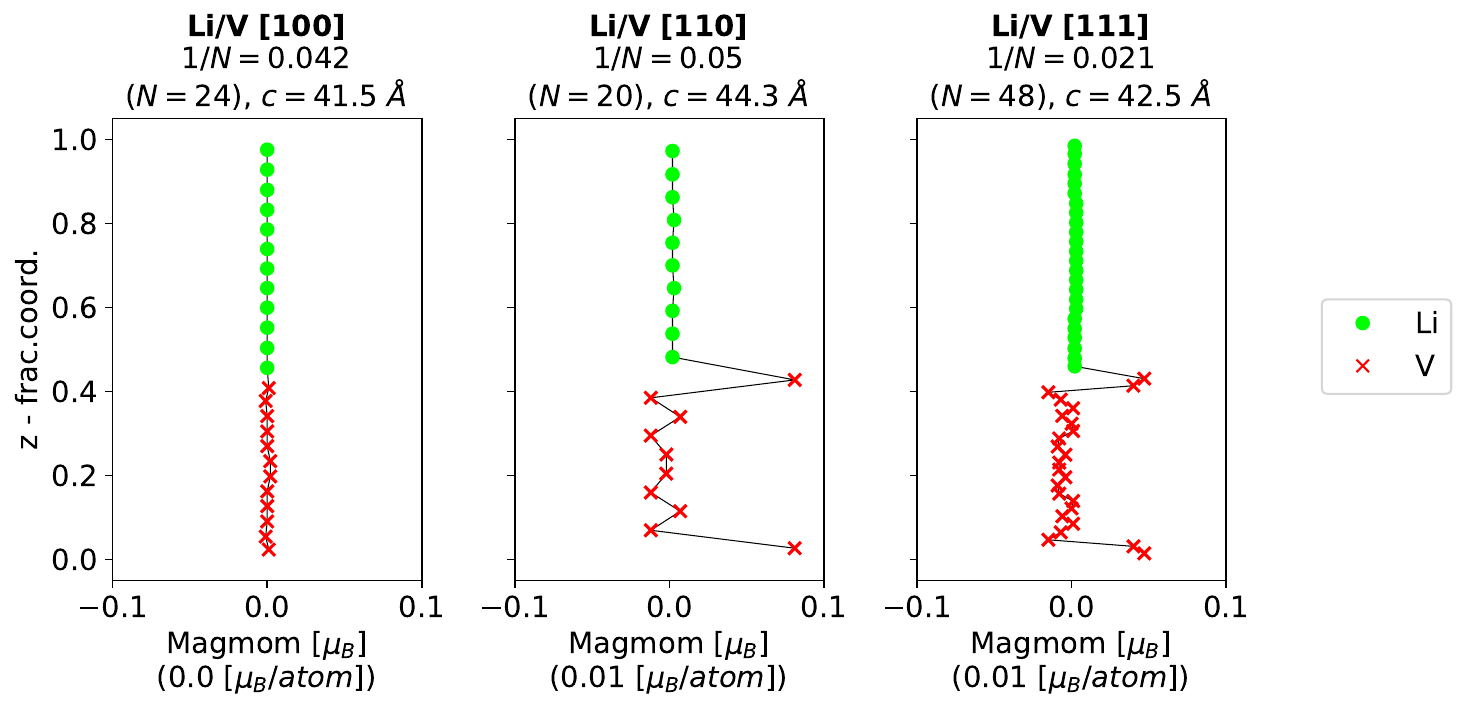}
            \put(-10,40){(a)}%{\textbf{(a)}}
        \end{overpic}
    }
    \vspace{0.5em}
    \subfigure{%
        \begin{overpic}[width=0.8\textwidth]{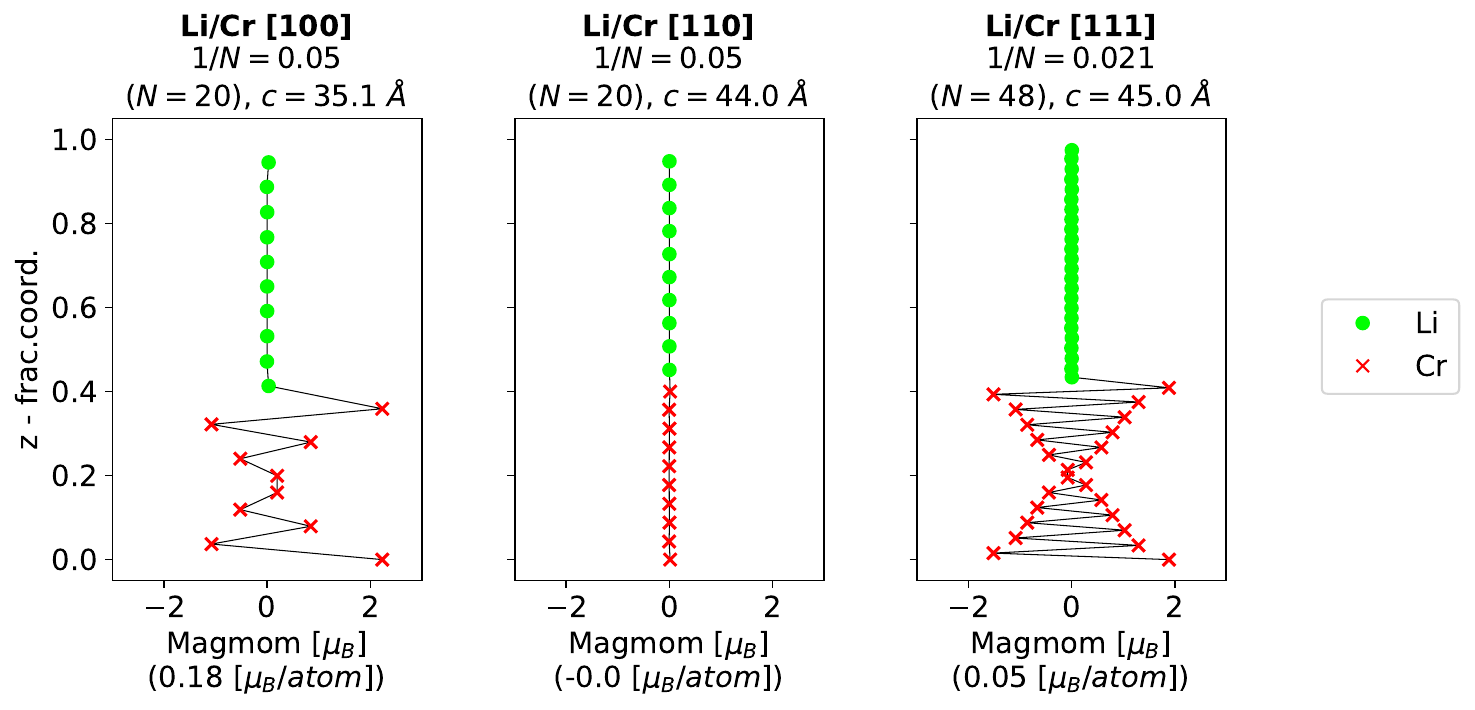}
            \put(-10,40){(b)}
        \end{overpic}
    }
    \vspace{0.5em}
    \subfigure{%
        \begin{overpic}[width=0.8\textwidth]{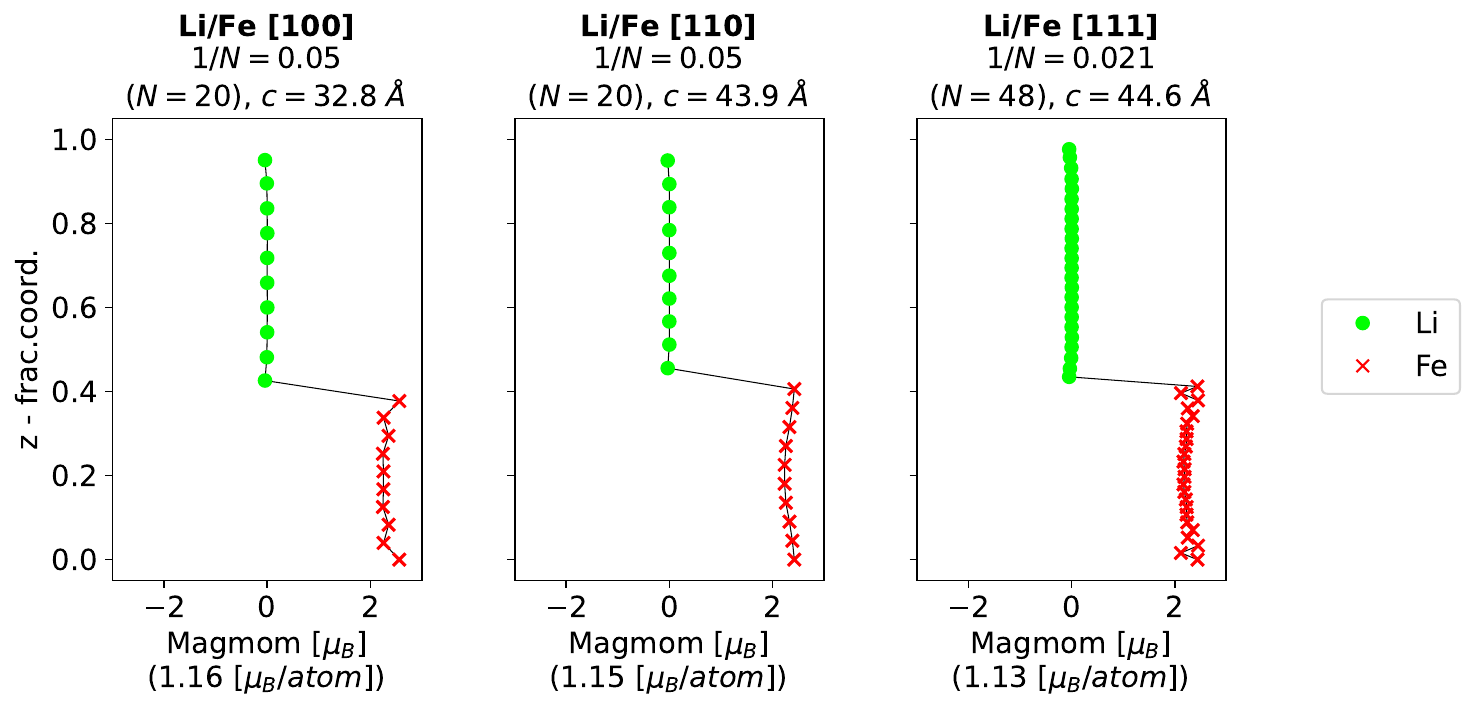}
            \put(-10,40){(c)}
        \end{overpic}
    }

    \caption{Magnetic moment profiles for the largest superlattices by orientation in the Li/V, Li/Cr, and Li/Fe systems. Clearly, the interfacial interaction affects the magnetic moments of the TMs. This effect is orientation-dependent. Note: All elements with partially filled d and f orbitals are initialized with a ferromagnetic alignment. Cr antiferromagnetism naturally emerges for certain interfaces.}   
    \label{fig:Example_magmoms_alk-TM}
\end{figure}

\begin{figure}[t]
\centering
\includegraphics[width=0.98\textwidth]{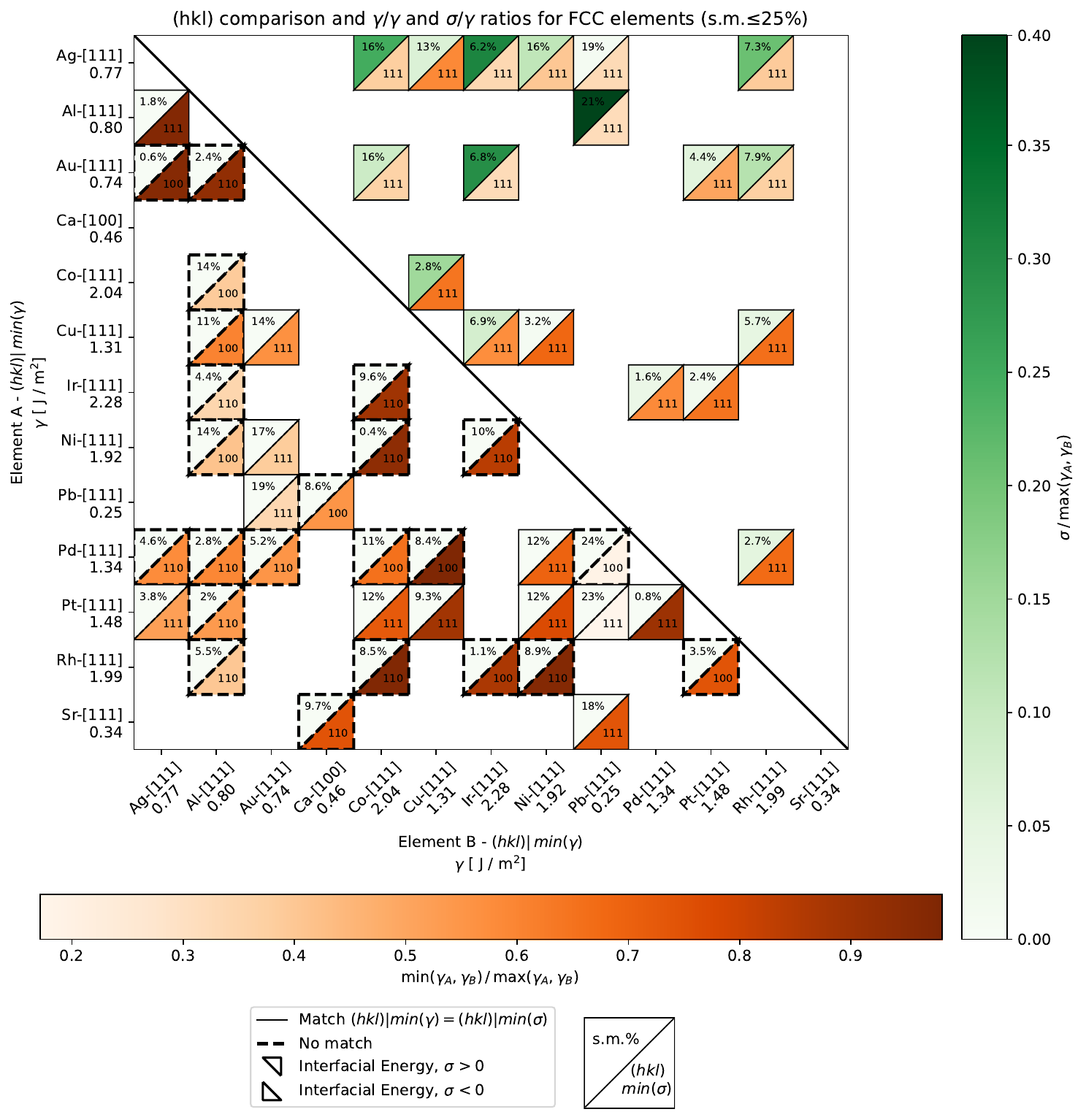}
\caption{Crossover between the \textbf{lowest-surface-energy-per-element}, \(\gamma\), from Crystalium\cite{Tran:2016} and the \textbf{lowest interfacial energy}, \(\sigma\), for the \textbf{FCC systems} from our database. The top-right triangle of the plot contains systems with positive interfacial energy. The opposite is true in the bottom-left triangle. Both axes have the FCC elements, the orientation at which the surface energy is the lowest, and the corresponding surface energy value in \([J/m^2]\). The vertical colorbar shows the ratio of interfacial to surface energy, while the horizontal one shows the ratio of the surface energies of the elements comprising the interface. Triangular contours per system represent the match/unmatch between the (hkl) planes of the surface and the interfacial energy. Inside each box, there is the size mismatch (s.m.\%) and the (hkl) plane along which the interfacial energy, \(\sigma\), is minimum.  } 
\label{fig:IE_app_FCC_vs_SurfaceE}
\end{figure}

\begin{figure}[t]
\centering
\includegraphics[width=0.99\textwidth]{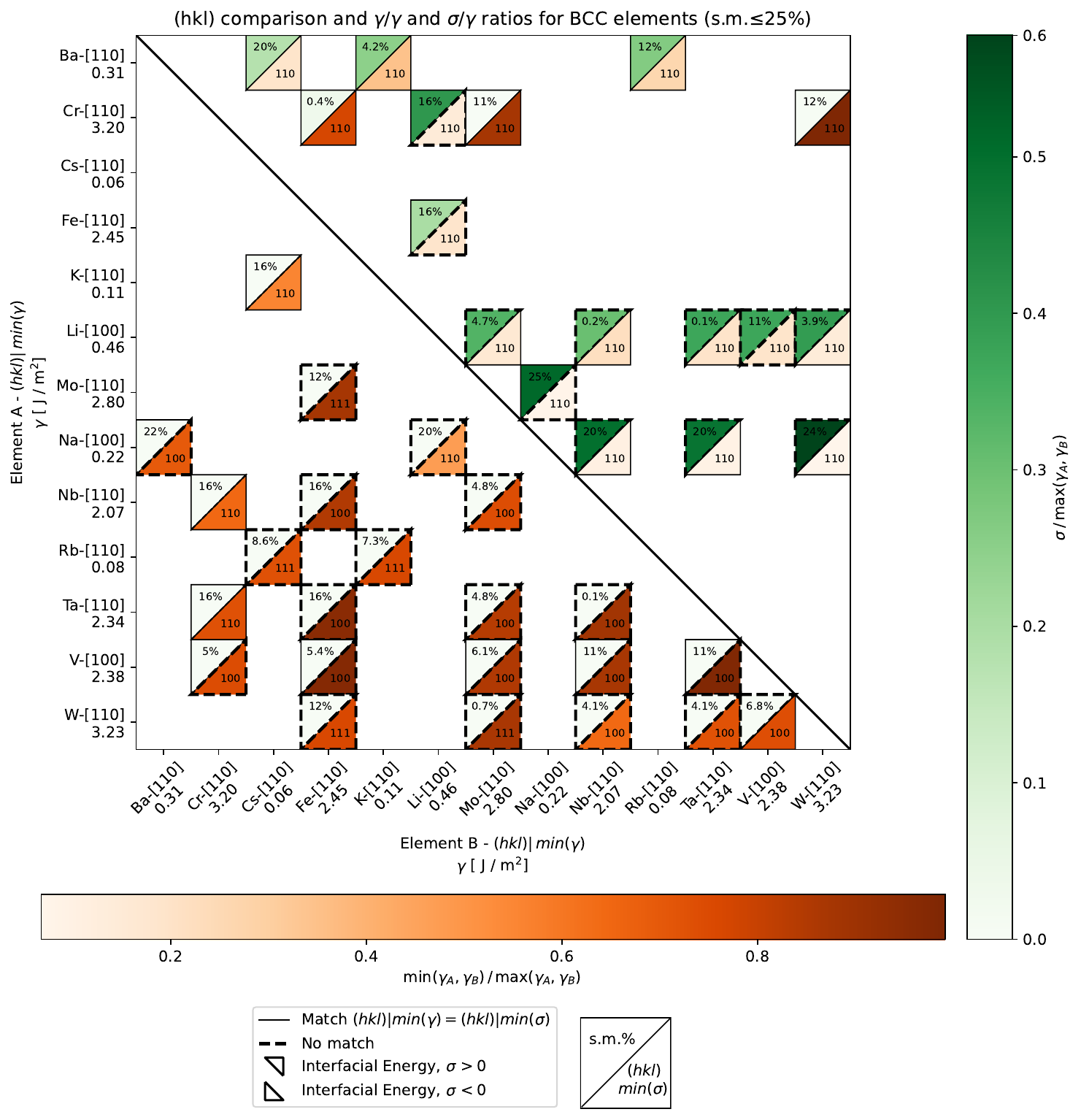}
\caption{Crossover between the Crystalium \textbf{lowest-surface-energy-per-atom} database~\cite{Tran:2016} and the \textbf{lowest interfacial energy} of the \textbf{BCC systems} in our database. Details of this figure are analogous to Fig.~\ref{fig:IE_app_FCC_vs_SurfaceE}.} 
\label{fig:IE_app_BCC_vs_SurfaceE}
\end{figure}

\begin{figure}[t]
\centering
\includegraphics[width=1.0\textwidth]{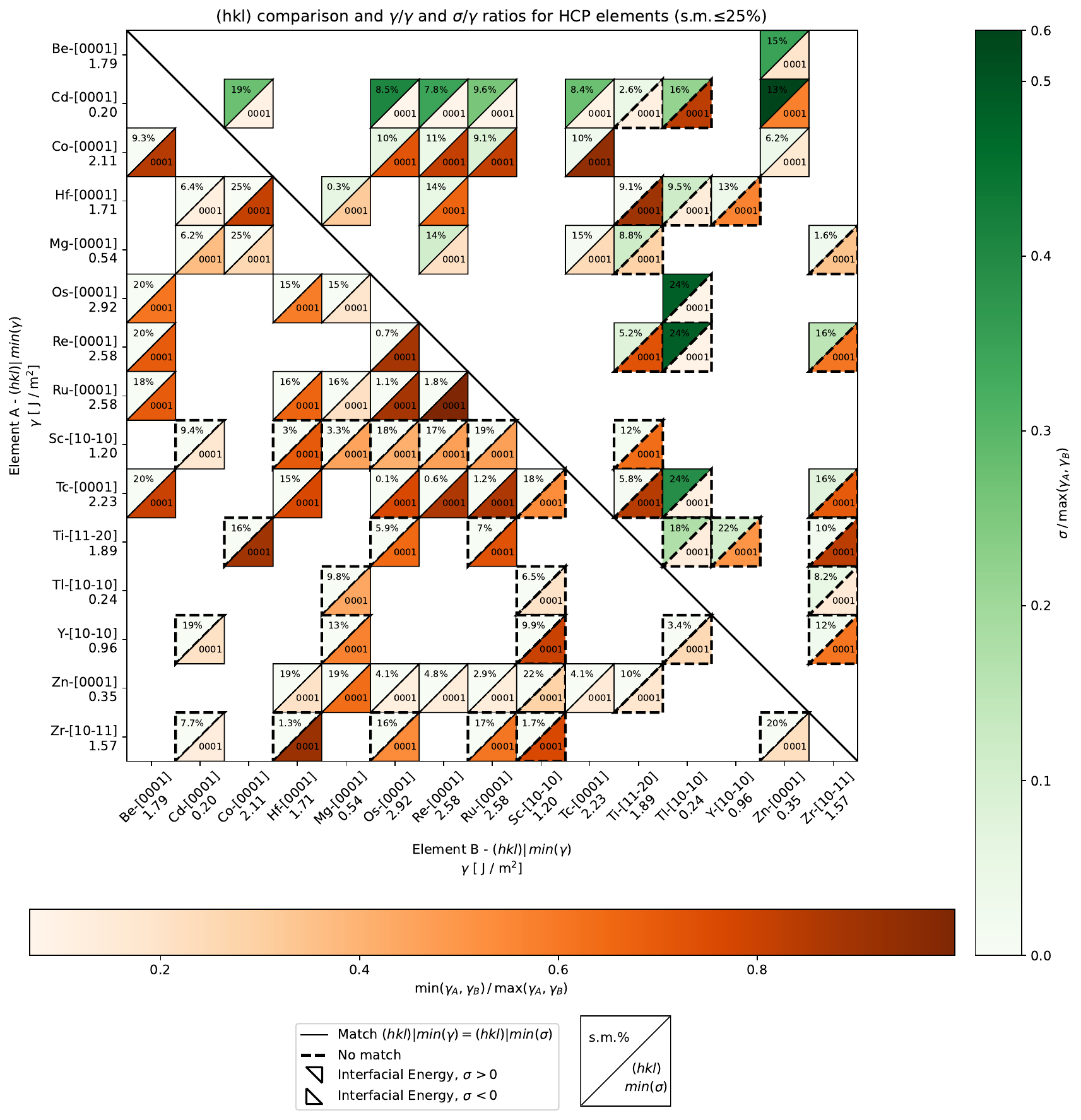}
\caption{Crossover between the Crystalium \textbf{lowest-surface-energy-per-atom} database~\cite{Tran:2016} and the \textbf{lowest interfacial energy} of the \textbf{HCP systems} in our database. Details of this figure are analogous to Fig.~\ref{fig:IE_app_FCC_vs_SurfaceE}.} 
\label{fig:IE_app_HCP_vs_SurfaceE}
\end{figure}

\begin{figure}[htbp]
    \centering

    % -------- First row: one large figure --------
    \refstepcounter{subfigure}
    \label{fig:panel-a}

    \begin{minipage}{0.55\textwidth}
        \centering
        \makebox[0pt][r]{%
            \raisebox{13em}{\footnotesize(a)}%
            \hspace{1em}%
        }%
        \includegraphics[
            width=0.90\linewidth
        ]{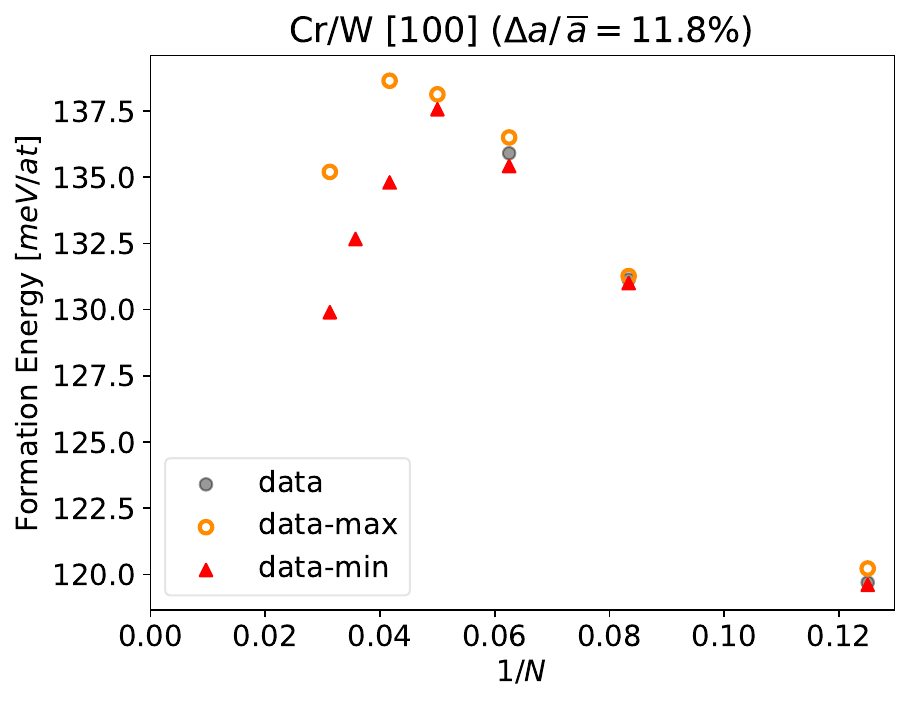}
    \end{minipage}

    \vspace{0.3cm}

    % -------- Second row: two figures --------
    \subfigure[]{%
        \label{fig:panel-b}
        \includegraphics[
            width=0.35\textwidth
        ]{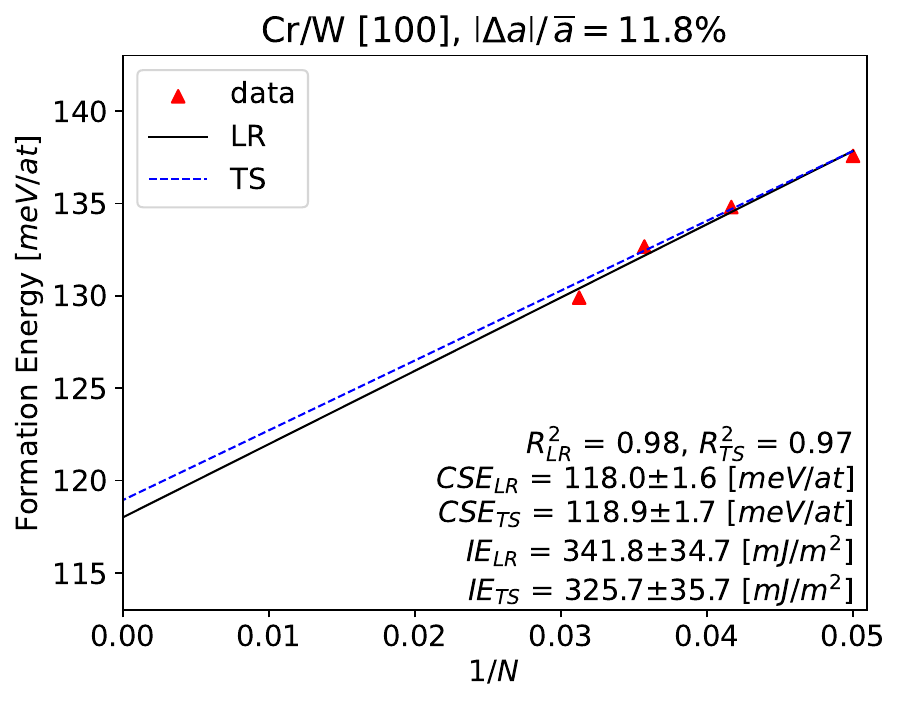}
    }
    \hfill
    \subfigure[]{%
        \label{fig:panel-c}
        \includegraphics[
            width=0.6\textwidth
        ]{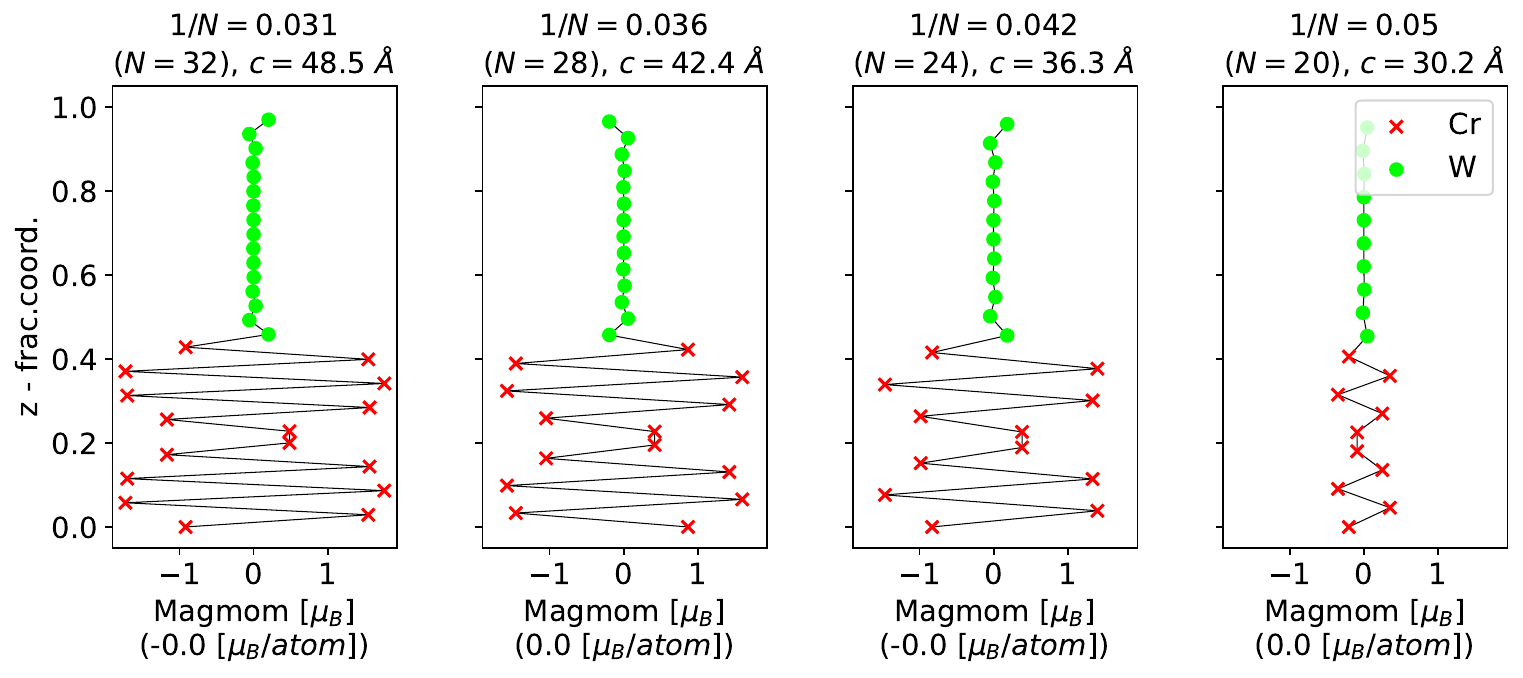}
    }

    \vspace{0.3cm}

    % -------- Third row: two figures --------
    \subfigure[]{%
        \label{fig:panel-d}
        \includegraphics[
            width=0.35\textwidth
        ]{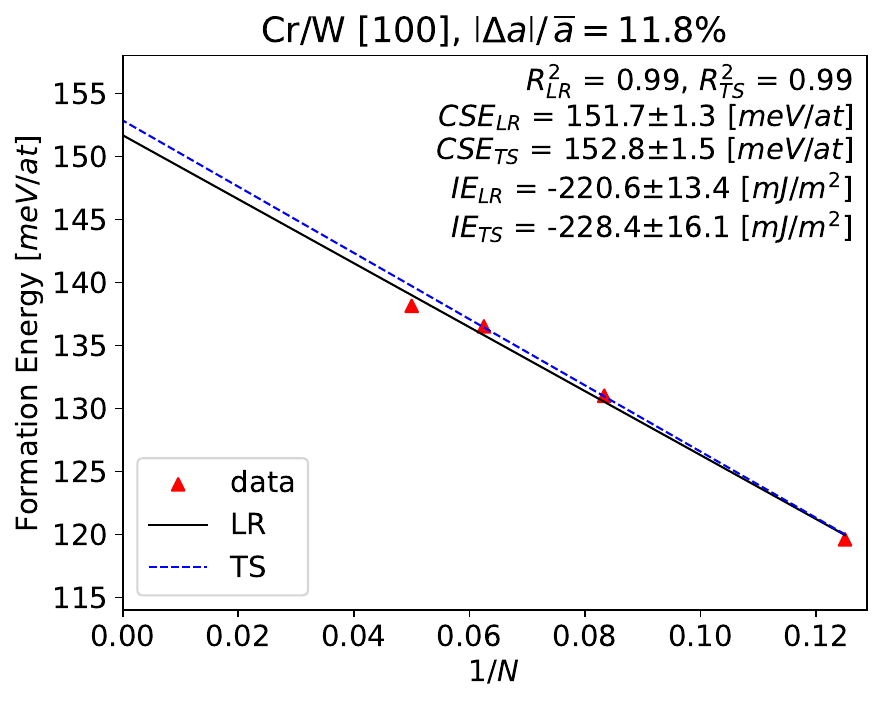}
    }
    \hfill
    \subfigure[]{%
        \label{fig:panel-e}
        \includegraphics[
            width=0.6\textwidth
        ]{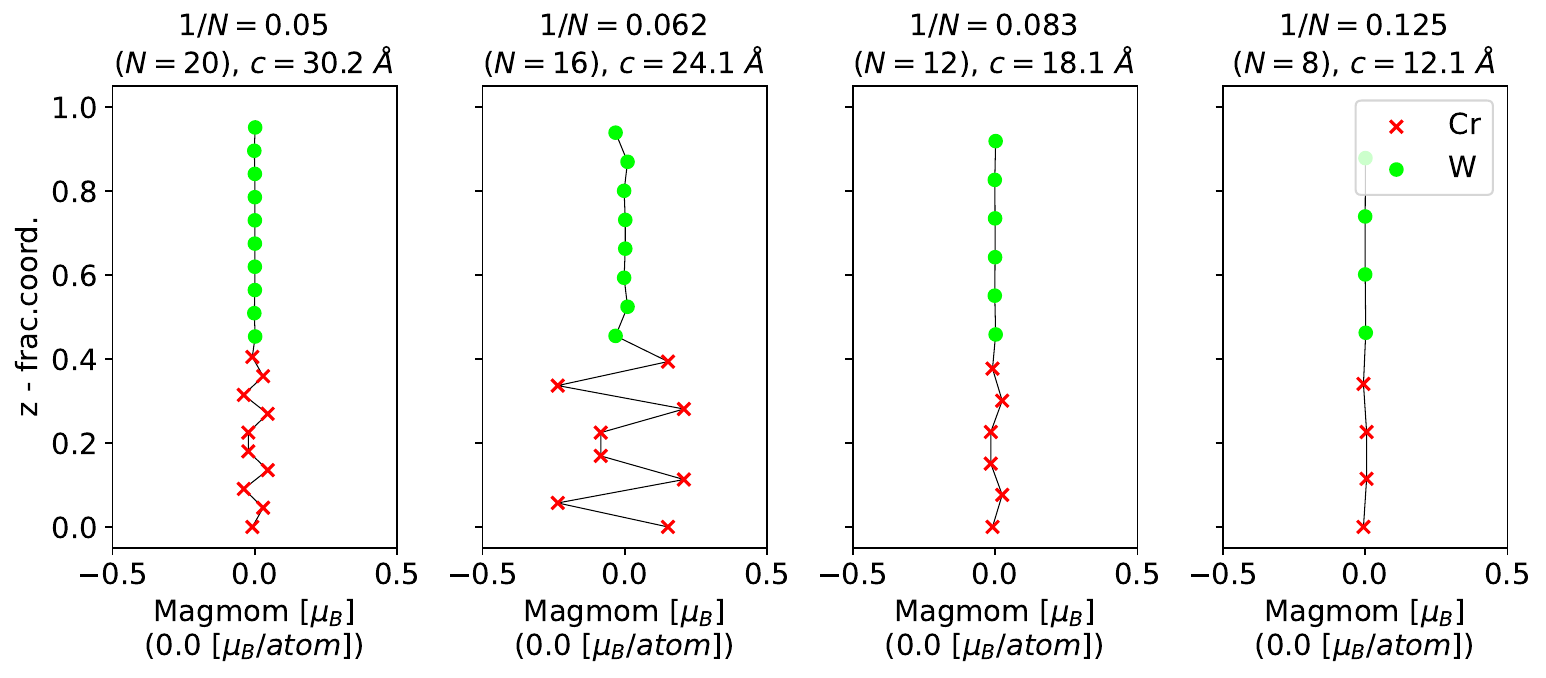}
    }

    \caption{Cr/W [100]: an example of the strain-magnetic relationship and its impact on the interfacial energies. (a): Formation energies for seven superlattice sizes ranging from 8 to 32 atoms. (b)\&(d): Regression for antiferromagnetic and less magnetic formation energies of (a), respectively. (c)\&(e): Magnetic moment profiles of superlattices in (b) and (d), respectively.}
    \label{fig:IE_app_Cr_W_100_magmom_example}
\end{figure}

\begin{figure}
\centering
\includegraphics[width=1.0\textwidth]{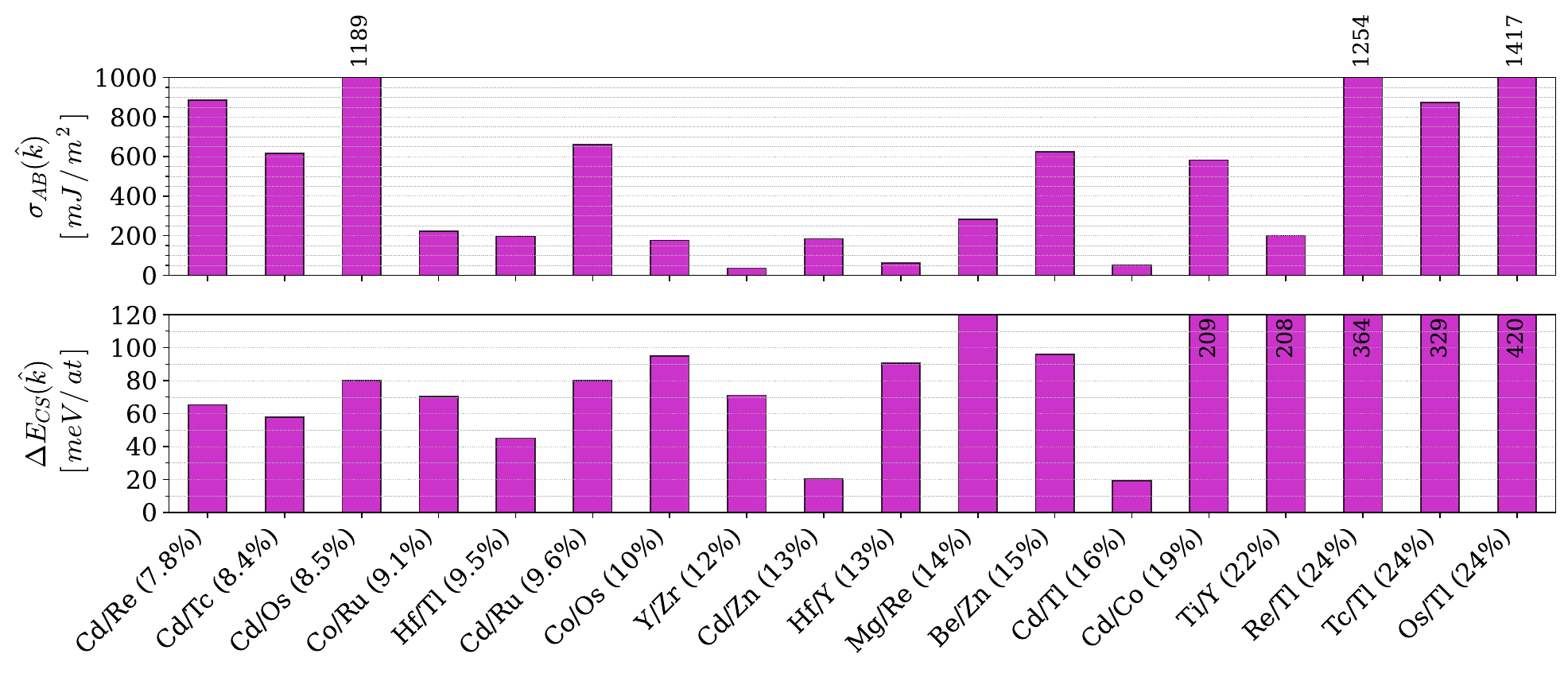}
\caption{Interfacial energies (top) and coherency strain energy (bottom) for the \textbf{phase-separating HCP} systems with positive \(\sigma_{AB}(0001)\) (Q-I in Fig.~2). This basically excluded only the Sc/Y system. The interfaces are sorted by lattice mismatch (value in parentheses). The elements in the \(A/B\) labels are alphabetically sorted.}
\label{fig:IE_app_PS_HCP_IE_CSE}
\end{figure}

\begin{figure}
    \centering
    \includegraphics[width=1.0\linewidth]{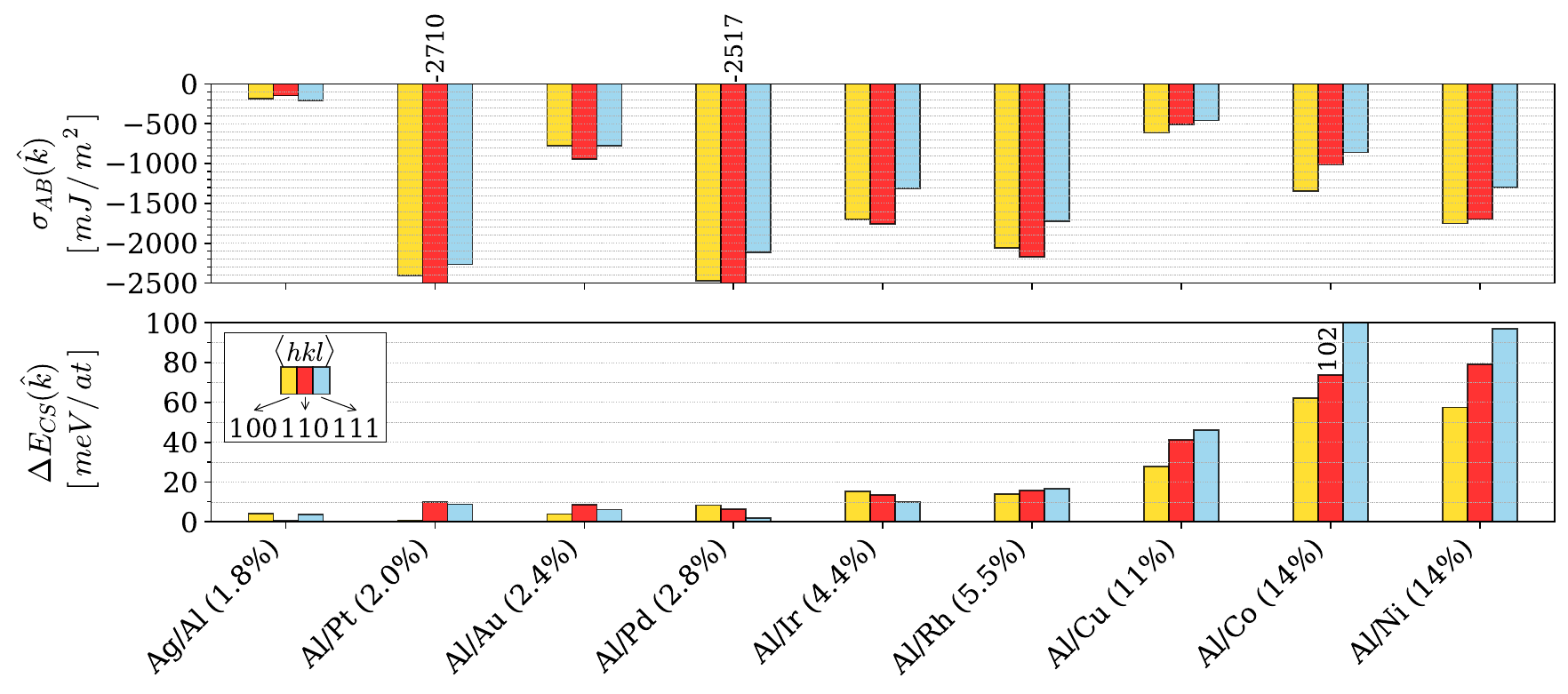}
    \caption{ Interfacial energies (top) and coherency strain energy (bottom) for \textbf{ordering FCC Al-based systems} for the three calculated orientations \(\hat{k}\) in Q-III of Fig.~2. 
    This figure is complementary to Fig.~6 in the main text.
    The interfaces are sorted by lattice mismatch (value in parentheses), and the colored bar represents the results for each principal orientation, $(\langle100\rangle,\langle110\rangle,\langle111\rangle)$, from left to right. The elements in the \(A/B\) labels are sorted alphabetically.}
    \label{fig:O_Al_IE_CSE}
\end{figure}

\begin{figure}[t]
\centering
\includegraphics[width=1.0\textwidth]{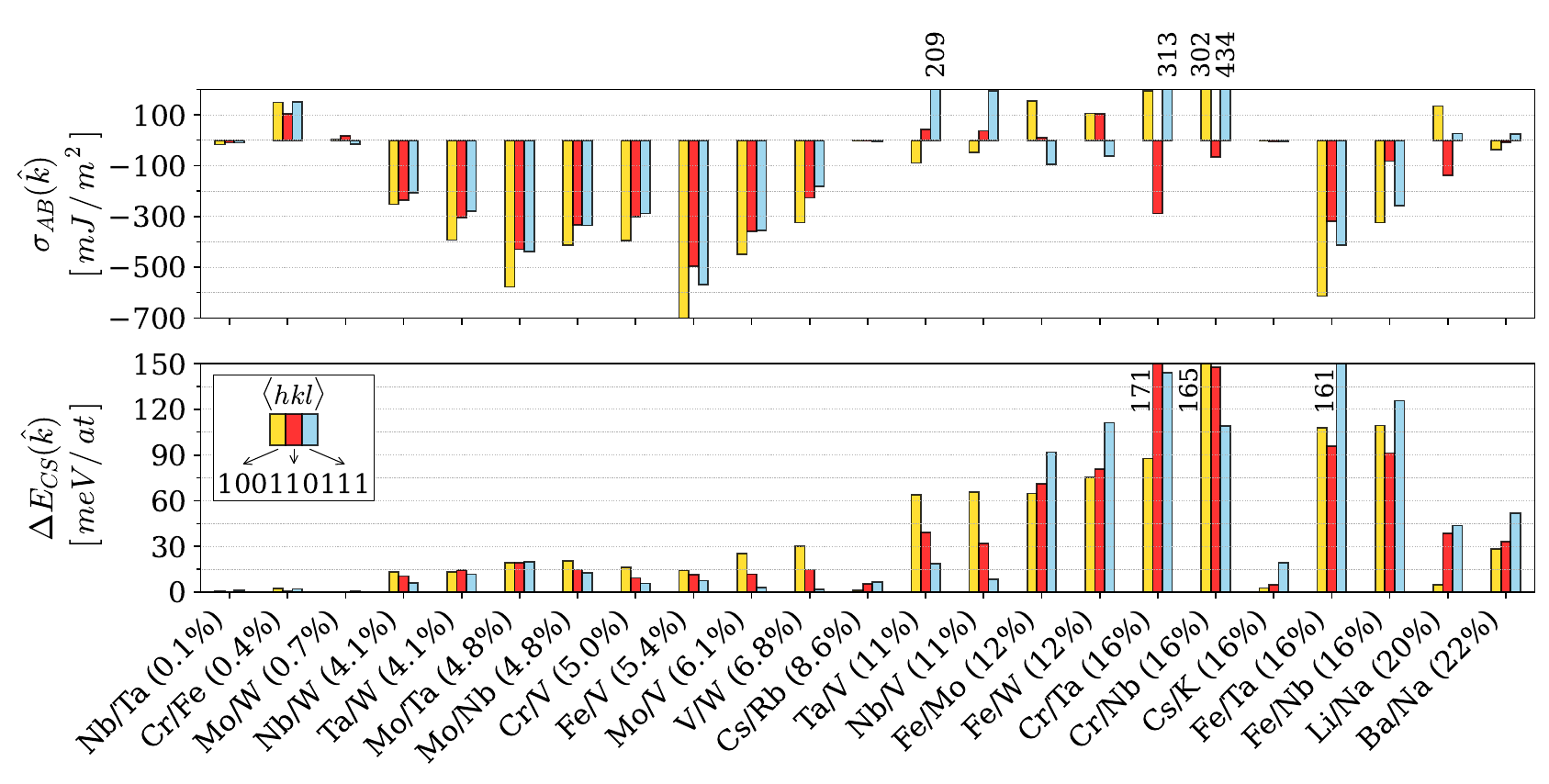}
\caption{Interfacial energies (top) and coherency strain energy (bottom) for the \textbf{ordering BCC} systems. They represent the 69 \(\sigma_{AB}(\hat{k})\) values for the BCC systems in quadrants Q-II and Q-III of Fig.~2. The interfaces are sorted by lattice mismatch (values in parentheses), and the colored bar represents the results for each principal orientation $(\langle100\rangle, \langle110\rangle, \langle111\rangle)$, from left to right. The elements in the \(A/B\) labels are alphabetically sorted.} 
\label{fig:BBC_ordering}
\end{figure}

\begin{figure}
\centering
\includegraphics[width=0.9\textwidth]{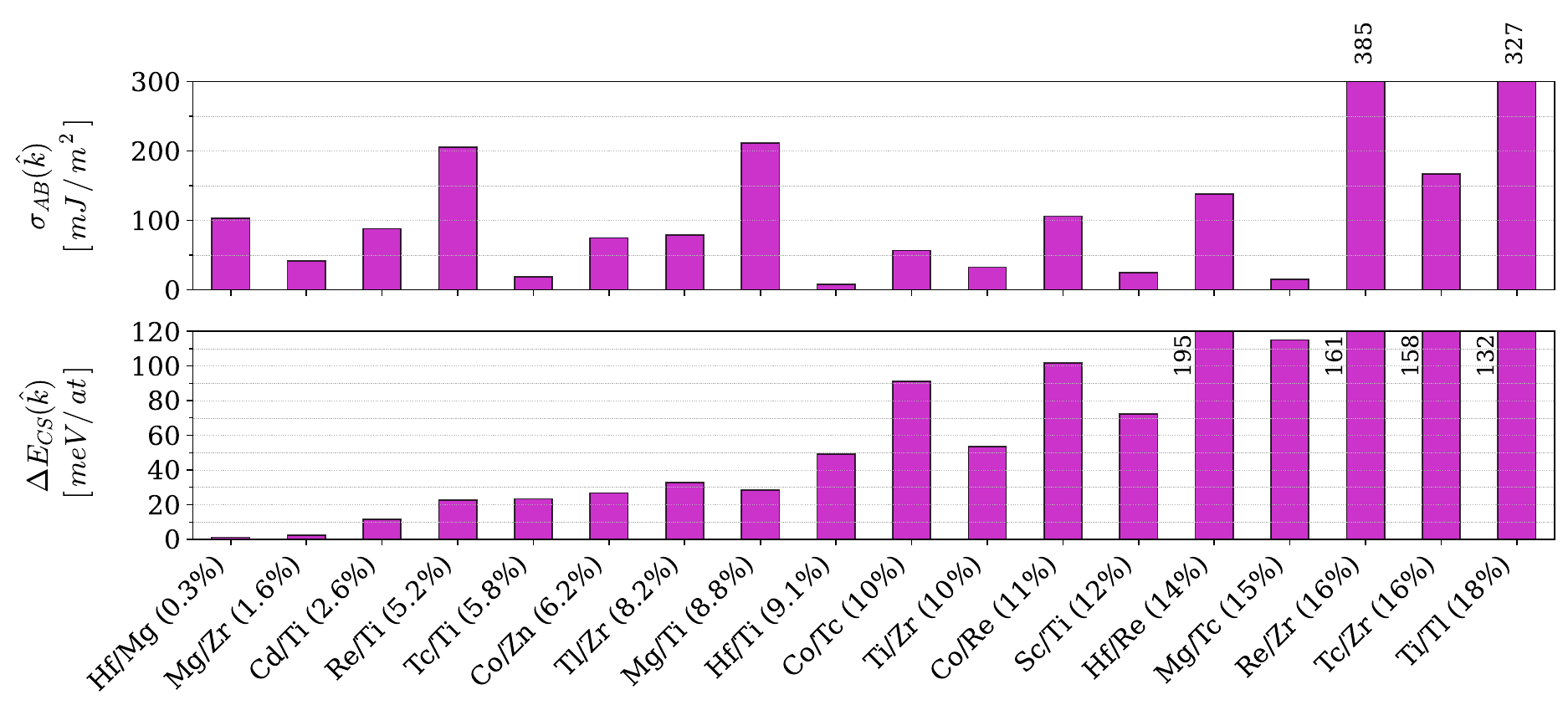}
\caption{Interfacial energies (top) and coherency strain energy (bottom) for the 18 \textbf{ordering HCP} systems in Q-II of Fig.~2. The interfaces are sorted by lattice mismatch (values in parentheses), and the elements in the \(A/B\) labels are alphabetically sorted.} 
\label{fig:HCP_ordering_QII}
\end{figure}

\begin{figure}
\centering
\includegraphics[width=0.95\textwidth]{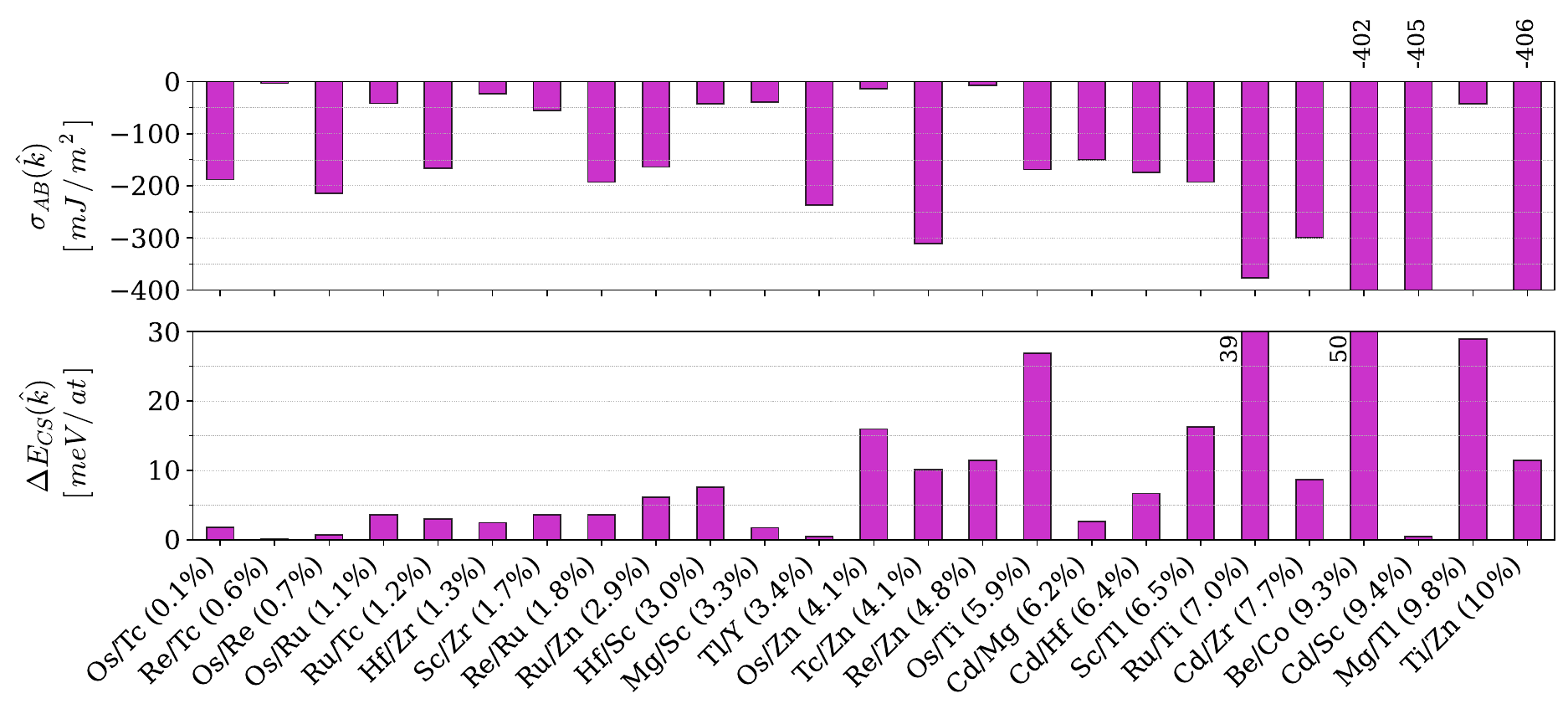}
\caption{Interfacial energies (top) and coherency strain energy (bottom) for the \textbf{ordering HCP} systems with size mismatch \(\leq10\%\). They represent 25 HCP systems in Q-III of Fig.~2. The interfaces are sorted by lattice mismatch (values in parentheses), and the elements in the \(A/B\) labels are alphabetically sorted.} 
\label{fig:HCP_ordering_QIII_1}
\end{figure}

\begin{figure}
\centering
\includegraphics[width=0.94\textwidth]{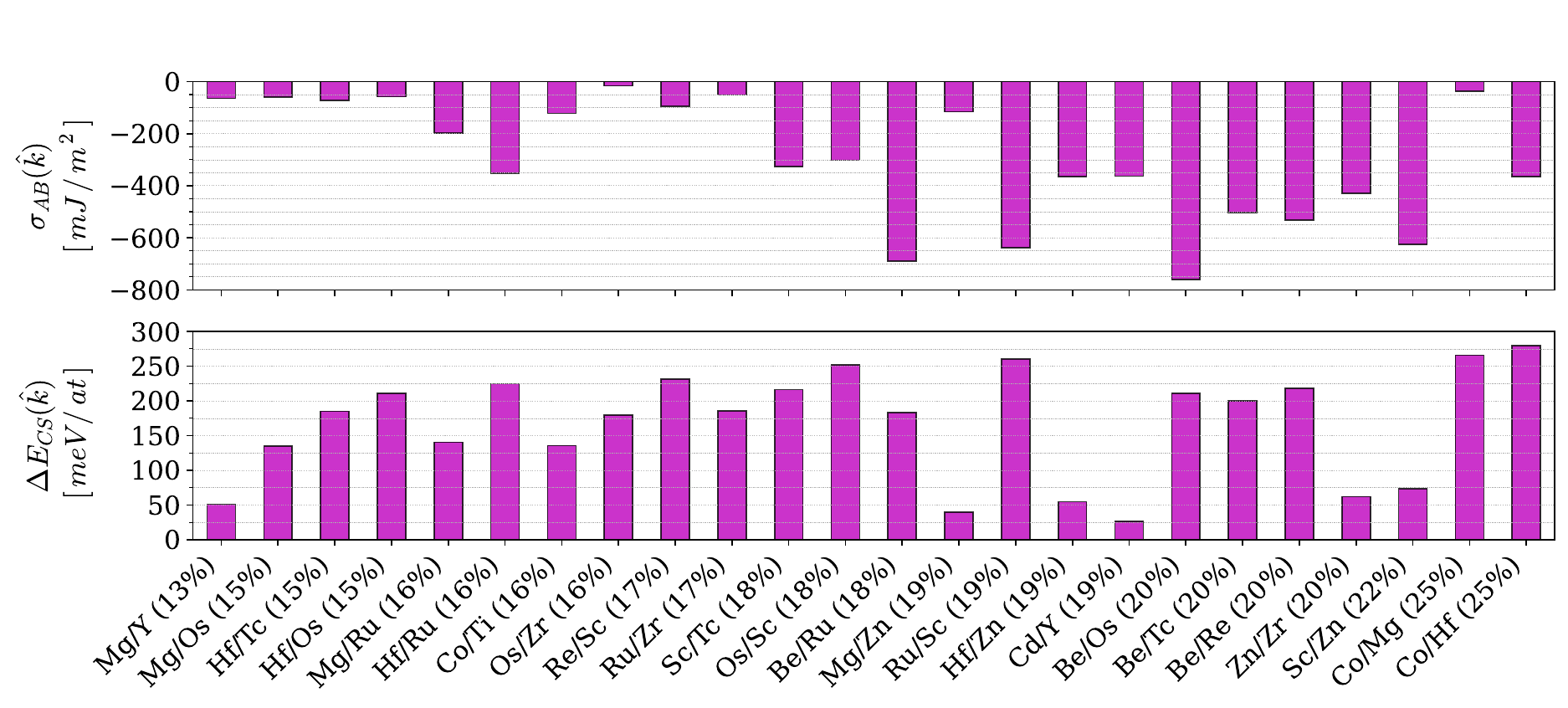}
\caption{Interfacial energies (top) and coherency strain energy (bottom) for the \textbf{ordering HCP} systems with size mismatch \(>10\%\). They represent 24 HCP systems in Q-III of Fig.~2. The interfaces are sorted by lattice mismatch (values in parentheses), and the elements in the \(A/B\) labels are alphabetically sorted.} 
\label{fig:HCP_ordering_QIII_2}
\end{figure}

\begin{figure}[t]
\centering
\includegraphics[width=0.99\textwidth]{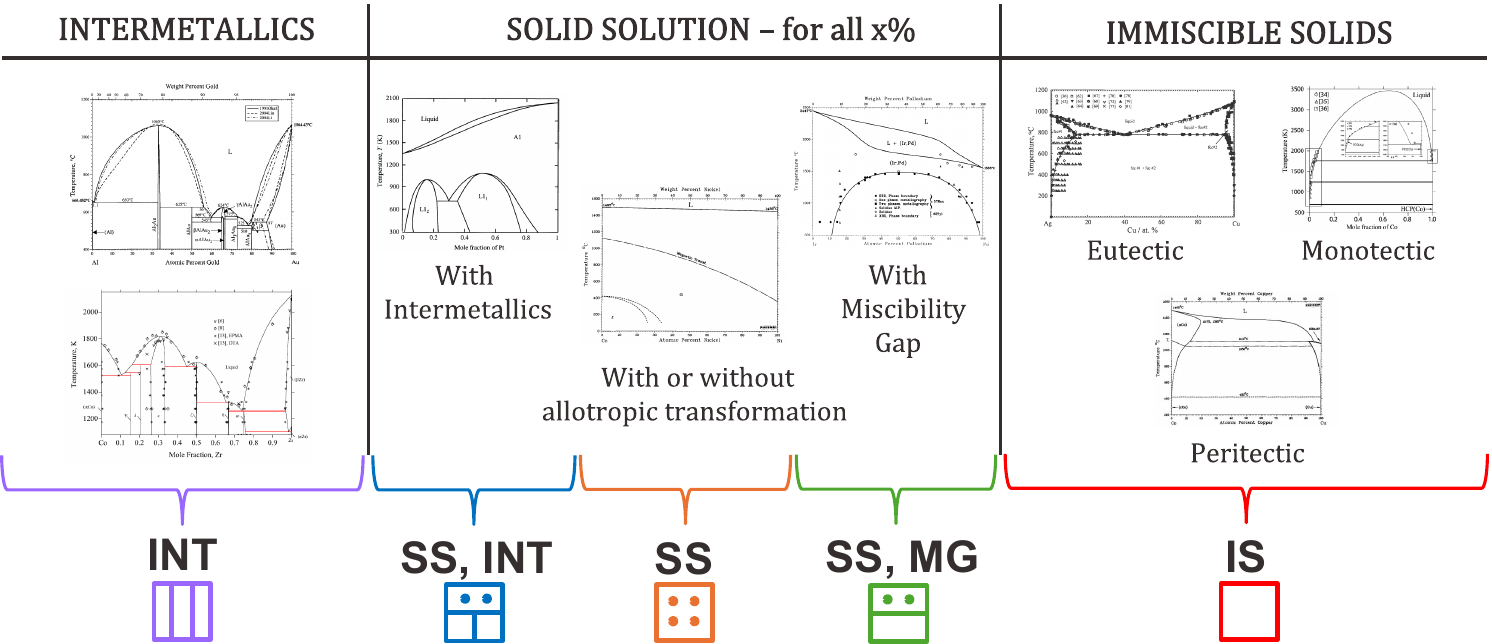}
\caption{Proposed simple phase-diagram classification based on the presence or absence of a solid solution across the entire composition range, and on the presence of intermetallics or immiscibility at lower temperatures.} 
\label{fig:phase_diag_class}
\end{figure}

\begin{figure}[t]
\centering
\includegraphics[width=0.95\textwidth]{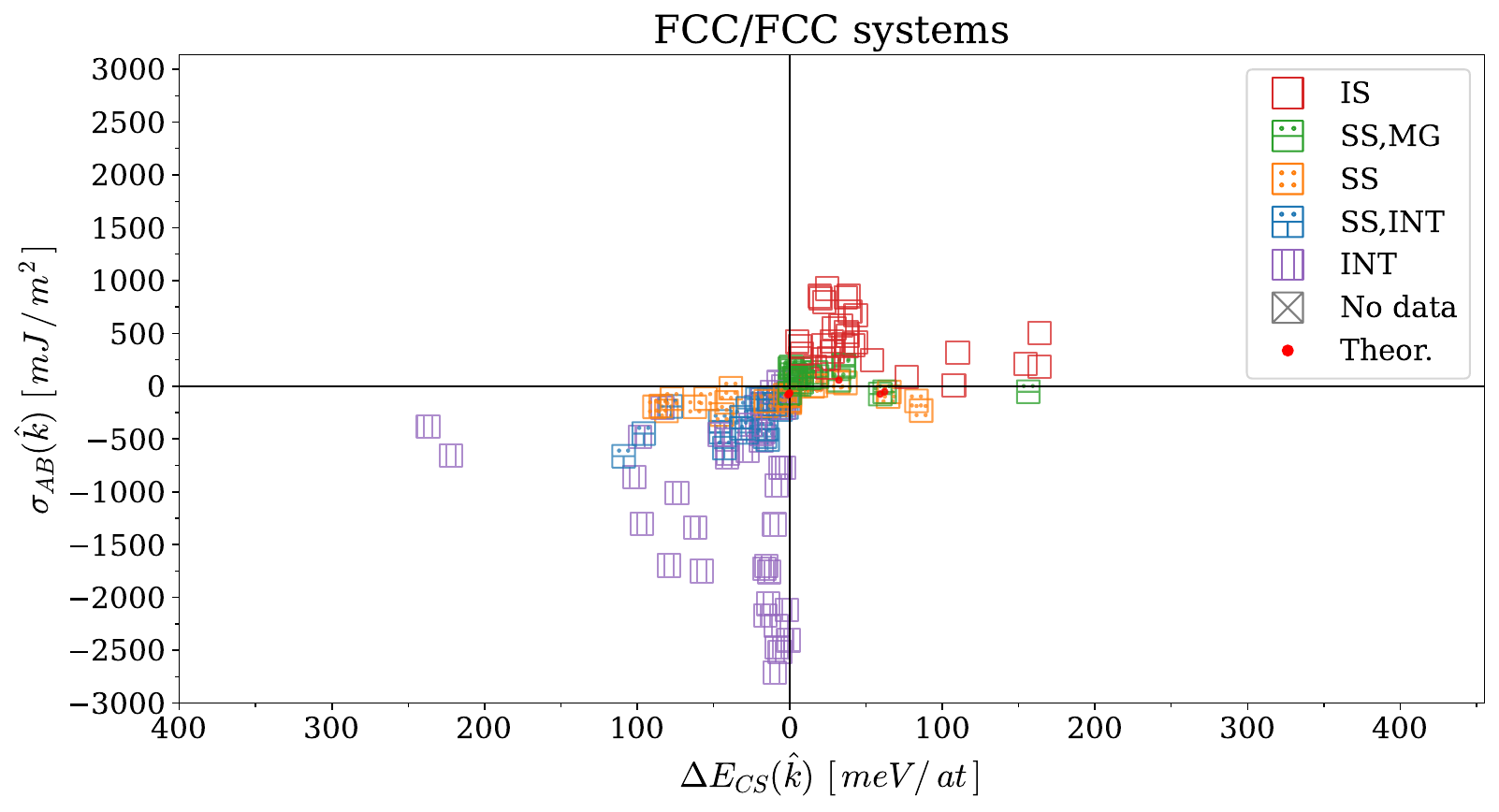}
\caption{Quadrants for all FCC/FCC systems, similar to Fig. 2 in the main text, together with the phase diagram labeling of Fig.~\ref{fig:phase_diag_class} generated from the data accumulated in Table~\ref{tab:phase_diagrams}. There is a global trend in the systems. All systems forming intermetallic compounds (INT) fall in quadrant Q-III with negative interfacial energy, while all immiscible solids (IS) are in Q-I, with positive interfacial energies. Systems with all labeling types with solid solution (SS) are present in all quadrants, but with a narrower interfacial energy range.} 
\label{fig:phase_diag_vs_IE_FCC}
\end{figure}

%%%%%%%%%%%%%%%%%%%%
%%%%%% TABLES %%%%%%
%%%%%%%%%%%%%%%%%%%%
\clearpage
\subsection{Supplementary Tables}

\begin{table}[H]
\centering
\caption{Supercell sizes used for each interface type and orientation.}
\label{tab:supercell-sizes}
\begin{tabular}{lccc}
\hline
(hkl)/(hkil) & \textbf{FCC/FCC} & \textbf{BCC/BCC} & \textbf{HCP/HCP} \\
\hline
(001) / ---  & 8,12,16,20  & 8,12,16,20  &  --- \\
(110) / --- & 8,16,24,32  & 8,12,16,20  &  --- \\
(111) / --- & 6,12,18,24  & 12,24,36,48 &  --- \\
--- / (0001) & ---         & ---   & 8,12,16,20 \\
\hline
\end{tabular}

\vspace{8pt}
\begin{minipage}{0.8\linewidth}
\centering
\footnotesize
\textit{Note:} Numbers represent atoms per supercell. In case of strong interface/interface interaction, the next supercell size was calculated and included in the final regression.
\end{minipage}

\end{table}

% \clearpage

\begin{longtable}{lccccccccr}
\caption{Elemental reference structures used in this database.}
\label{tab:elem_ref_oqmd}\\

\toprule
SG & Element & Energy (eV/at) & $\alpha$ & $\beta$ & $\gamma$ &
$a$ (\AA) & $b$ (\AA) & $c$ (\AA) & OQMD-ID \\
\midrule
\endfirsthead

\multicolumn{10}{c}{\tablename~\thetable{} -- continued from previous page}\\
\toprule
SG & Element & Energy (eV/at) & $\alpha$ & $\beta$ & $\gamma$ &
$a$ (\AA) & $b$ (\AA) & $c$ (\AA) & OQMD-ID \\
\midrule
\endhead

\midrule
\multicolumn{10}{r}{\textit{Continued on next page}}\\
\endfoot

\bottomrule
\endlastfoot

Fm-3m & Ag & -2.821 & 90 & 90 & 90 & 4.11 & 4.11 & 4.11 & 1214502 \\
Fm-3m & Al & -3.745 & 90 & 90 & 90 & 4.04 & 4.04 & 4.04 & 8100 \\
Fm-3m & Au & -3.267 & 90 & 90 & 90 & 4.14 & 4.14 & 4.14 & 636382 \\
Fm-3m & Ca & -1.978 & 90 & 90 & 90 & 5.44 & 5.44 & 5.44 & 594524 \\
Fm-3m & Co & -7.090 & 90 & 90 & 90 & 3.51 & 3.51 & 3.51 & 594128 \\
Fm-3m & Cu & -3.717 & 90 & 90 & 90 & 3.61 & 3.61 & 3.61 & 635950 \\
Fm-3m & Ir & -8.855 & 90 & 90 & 90 & 3.87 & 3.87 & 3.87 & 7521 \\
Fm-3m & Ni & -5.567 & 90 & 90 & 90 & 3.50 & 3.50 & 3.50 & 1214550 \\
Fm-3m & Pb & -3.704 & 90 & 90 & 90 & 5.00 & 5.00 & 5.00 & 1214556 \\
Fm-3m & Pd & -5.177 & 90 & 90 & 90 & 3.93 & 3.93 & 3.93 & 1214557 \\
Fm-3m & Pt & -6.056 & 90 & 90 & 90 & 3.96 & 3.96 & 3.96 & 7522 \\
Fm-3m & Rh & -7.269 & 90 & 90 & 90 & 3.82 & 3.82 & 3.82 & 7514 \\
Fm-3m & Sr & -1.683 & 90 & 90 & 90 & 6.00 & 6.00 & 6.00 & 1214573 \\
Im-3m & Ba & -1.924 & 90 & 90 & 90 & 5.03 & 5.03 & 5.03 & 603869 \\
Im-3m & Cr & -9.508 & 90 & 90 & 90 & 2.83 & 2.83 & 2.83 & 1215141 \\
Im-3m & Cs & -0.853 & 90 & 90 & 90 & 6.14 & 6.14 & 6.14 & 1215142 \\
Im-3m & Fe & -8.308 & 90 & 90 & 90 & 2.82 & 2.82 & 2.82 & 8568 \\
Im-3m & K  & -1.096 & 90 & 90 & 90 & 5.24 & 5.24 & 5.24 & 8500 \\
Im-3m & Li & -1.893 & 90 & 90 & 90 & 3.31 & 3.31 & 3.31 & 17585 \\
Im-3m & Mo & -10.848 & 90 & 90 & 90 & 3.16 & 3.16 & 3.16 & 1215167 \\
Im-3m & Na & -1.300 & 90 & 90 & 90 & 4.05 & 4.05 & 4.05 & 8534 \\
Im-3m & Nb & -10.094 & 90 & 90 & 90 & 3.32 & 3.32 & 3.32 & 1215170 \\
Im-3m & Rb & -0.962 & 90 & 90 & 90 & 5.64 & 5.64 & 5.64 & 8533 \\
Im-3m & Ta & -11.853 & 90 & 90 & 90 & 3.32 & 3.32 & 3.32 & 1215197 \\
Im-3m & V  & -8.941 & 90 & 90 & 90 & 2.98 & 2.98 & 2.98 & 8082 \\
Im-3m & W  & -12.960 & 90 & 90 & 90 & 3.18 & 3.18 & 3.18 & 1215207 \\
P6$_3$/mmc & Be & -3.755 & 90 & 90 & 120 & 2.26 & 2.26 & 3.57 & 622128 \\
P6$_3$/mmc & Cd & -0.900 & 90 & 90 & 120 & 2.99 & 2.99 & 5.63 & 593871 \\
P6$_3$/mmc & Co & -7.106 & 90 & 90 & 120 & 2.48 & 2.48 & 4.01 & 637416 \\
P6$_3$/mmc & Hf & -9.955 & 90 & 90 & 120 & 3.19 & 3.19 & 5.06 & 1215332 \\
P6$_3$/mmc & Mg & -1.542 & 90 & 90 & 120 & 3.19 & 3.19 & 5.20 & 9226 \\
P6$_3$/mmc & Os & -11.226 & 90 & 90 & 120 & 2.75 & 2.75 & 4.35 & 1215354 \\
P6$_3$/mmc & Re & -12.423 & 90 & 90 & 120 & 2.77 & 2.77 & 4.47 & 7248 \\
P6$_3$/mmc & Ru & -9.202 & 90 & 90 & 120 & 2.72 & 2.72 & 4.29 & 7247 \\
P6$_3$/mmc & Sc & -6.328 & 90 & 90 & 120 & 3.29 & 3.29 & 5.11 & 8146 \\
P6$_3$/mmc & Tc & -10.361 & 90 & 90 & 120 & 2.75 & 2.75 & 4.41 & 1215377 \\
P6$_3$/mmc & Ti & -7.762 & 90 & 90 & 120 & 2.92 & 2.92 & 4.63 & 8079 \\
P6$_3$/mmc & Tl & -2.359 & 90 & 90 & 120 & 3.51 & 3.51 & 5.68 & 1215381 \\
P6$_3$/mmc & Y  & -6.464 & 90 & 90 & 120 & 3.63 & 3.63 & 5.68 & 1215387 \\
P6$_3$/mmc & Zn & -1.266 & 90 & 90 & 120 & 2.64 & 2.64 & 5.03 & 122929 \\
P6$_3$/mmc & Zr & -8.547 & 90 & 90 & 120 & 3.24 & 3.24 & 5.17 & 67408 \\

\end{longtable}

\vspace{0.1pt}
\noindent
\begin{minipage}{\phasediagtablewidth}
%{0.8\linewidth}
\centering
\footnotesize
\textit{Note:} Energy values in Table~\ref{tab:elem_ref_oqmd} are rounded to 3 decimal places for visualization purposes.
The real reference table is attached as \texttt{info\_elements\_data}.
\end{minipage}

\begin{table}[htbp]
\centering
\caption{Cases with slightly negative coherency strain energy from the supercell approach. Since the negative values fall within the uncertainty, they were set to 0.0 for plotting purposes.}
\label{tab:IE_app_ecs_uncertain}
\vspace{0.5em}
\begin{tabular}{lccccc}
\toprule
System & $~\langle hkl \rangle$ & \twolinecell{Lattice}{mismatch\([\%]\)} & \twolinecell{\(\Delta E_{CS}(\hat{k})\)}{\([meV/at]\)} & \twolinecell{\(~\Delta E^{error}_{CS}(\hat{k})\)}{\([meV/at]\)} & \twolinecell{\(\Delta H_{SL}\text{-range}\)}{\([meV/at]\)} \\
\midrule
Ag/Au & $~\langle 100 \rangle$ & 0.6 & \(-0.246\) & 0.506 & 11.2\\
Ag/Au & $~\langle 110 \rangle$ & 0.6 & \(-0.588\) & 0.888 & 20.3\\
Ag/Au & $~\langle 111 \rangle$ & 0.6 & \(-0.596\) & 0.902 & 15.5\\
Mo/W & $~\langle 110 \rangle$ & 0.7 & \(-0.489\) & 2.117 & 2.1\\
Pd/Pt & $~\langle 100 \rangle$ & 0.8 & \(-0.242\) & 0.635 & 6.5\\
Pd/Pt & $~\langle 111 \rangle$ & 0.8 & \(-0.350\) & 0.554 & 9.8\\
Ir/Pd& $~\langle 100 \rangle$ &  1.6 & \(-0.242\) & 0.458 & 6.5\\
\bottomrule
\end{tabular}
\end{table}

\newcounter{pdrow}
\setcounter{pdrow}{0}
\newcommand{\pdnum}{\stepcounter{pdrow}\arabic{pdrow}}

\clearpage
\small
\setlength{\LTcapwidth}{\phasediagtablewidth}
\renewcommand{\arraystretch}{1.15}

\begin{longtable}{rlcccccccc}
\caption{Phase diagrams references for FCC/FCC systems.}
\label{tab:phase_diagrams}\\

\toprule
No. & System & \headcell{1.2cm}{OQMD\\LRO} & Group$^{*}$ & \headcell{1.42cm}{General\\Invariant} & Source & \headcell{1.3cm}{T-range\\ \degree C} & Year & Ref. & Comment\\
\midrule
\endfirsthead

\multicolumn{10}{c}{\tablename~\thetable{} -- continued from previous page}\\
\toprule
No. & System & \headcell{1.2cm}{OQMD\\LRO} & Group$^{*}$ & \headcell{1.42cm}{General\\Invariant} & Source & \headcell{1.3cm}{T-range\\ \degree C} & Year & Ref. & Comment\\
\midrule
\endhead

\midrule
\multicolumn{10}{r}{\textit{Continued on next page}}\\
\endfoot

\bottomrule
\endlastfoot

\pdnum) & Ag--Al & OS & INT & \twolinecell{Eutectic}{Peritectic} & Calc. & 200--1000 & 1987 & 
\cite{McAlister:1987} & \notecell{Assessed P. Diag.}\\
\pdnum) & Ag--Pt & OS & INT & Peritectic & \twolinecell{Exp.}{Calc.} & 600--1800 & \twolinecell{1996}{2017} & \twolinecell{\cite{Durussel:1996}}{\cite{Hart:2017}} & \notecell{Later reference includes correction.}\\
\pdnum) & Al--Au & OS & INT & Eutectic & Calc. & 0--1200 & 2005 & \cite{Okamoto:2005} & \notecell{Assessed P. Diag.}\\
\pdnum) & Al--Co & OS & INT & Peritectic & Calc. & 0--1800 & 2019 & \cite{Ostrowska:2019} & \notecell{Reassessed P. Diag. See reference Fig.~1}\\
\pdnum) & Al--Cu & OS & INT & \twolinecell{Eutectic}{Peritectic} & Calc. & 300--1200 & 2015 & \cite{Liang:2015} & \notecell{Reassessed P. Diag. See reference Fig.~2}\\
\pdnum) & Al--Ir & OS & INT & \twolinecell{Eutectic}{Peritectic} & Exp. & 200--2500 & 2009 & \cite{Okamoto:2009} & \notecell{Reassessed P. Diag. from exp. data.}\\
\pdnum) & Al--Ni & OS & INT & \twolinecell{Eutectic}{Peritectic} & \twolinecell{Exp.}{Calc.} & 400--1800 & \twolinecell{2004}{2023} & \twolinecell{\cite{Okamoto:2004}}{\cite{Shao:2023}} & \notecell{Later reference includes corrections.}\\
\pdnum) & Al--Pd & OS & INT & \twolinecell{Eutectic}{Peritectic} & Calc. & 200--1800 & 2017 & \cite{Duriska:2017} & --\\
\pdnum) & Al--Pt & OS & INT & \twolinecell{Eutectic}{Peritectic} & Calc. & 127--1927 & 2016 & \cite{Liu:2016} & \notecell{Reassessed P. Diag. See reference Fig.~7}\\
\pdnum) & Al--Rh & OS & INT & \twolinecell{Eutectic}{Peritectic} & Exp. & 600--2100 & 2015 & \cite{Grushko:2015} & \notecell{Reassessed P. Diag. See reference Fig.~1}\\
\pdnum) & Au--Pb & OS & INT & \twolinecell{Eutectic}{Peritectic} & Calc. & 27--1227 & 2004 & \cite{Wang:2004} & \notecell{Assessed P. Diag. See reference Fig.~2}\\
\pdnum) & Pb--Pd & OS & INT & \twolinecell{Eutectic}{Peritectic} & Calc. & 27--1627 & 1999 & \cite{Ghosh:1999} & --\\
\pdnum) & Pb--Pt & OS & INT & \twolinecell{Eutectic}{Peritectic} & Calc. & -273--1827 & 2009 & \cite{Long:2009} & --\\
\pdnum) & Pb--Sr & OS & INT & \twolinecell{Eutectic}{Peritectic} & Calc. & 27--1227 & 2008 & \cite{Idbenali:2008} & \notecell{Reassessed P. Diag. See reference Fig.~3}\\
\pdnum) & Ca--Sr & OS & INT & \twolinecell{Eutectic}{Peritectic} & Exp. & 0--1200 & 1981 & \cite{Bruzzone:1981} & --\\
\pdnum) & Au--Cu & OS & SS, INT & -- & Calc. & 127--427 & 2007 & \cite{Cao:2007} & \notecell{Assessed P. Diag. See reference's Fig.~10.}\\
\pdnum) & Au--Pd & OS & SS, INT & -- & Calc. & 0--1600 & 1985 & \cite{Okamoto:1985b} & \notecell{SS and ordering has been confirmed.}\\
\pdnum) & Co--Pt & OS & SS, INT & -- & Calc. & 27--1827 & 2011 & \cite{Kim:2011} & \notecell{SS, ordering, and allotropic transf. observed.}\\
\pdnum) & Cu--Pd & OS & SS, INT & -- & Exp. & 200--700 & 2019 & \cite{Popov:2019} & \notecell{Reassessed P. Diag.}\\
\pdnum) & Cu--Pt & OS & SS, INT & -- & Calc. & 27--927 & 2006 & \cite{Abe:2006} & \notecell{Assessed P. Diag. See reference's Fig.~2.}\\
\pdnum) & Ni--Pt & OS & SS, INT & -- & Calc. & -73--1927 & 2009 & \cite{Lu:2009} & \notecell{Assessed P. Diag. See reference's Fig.~1\&2.}\\
\pdnum) & Pt--Rh & OS & SS, INT & -- & \twolinecell{Calc.}{Calc.} & \twolinecell{--}{-250--2000} & \twolinecell{2012}{2013} & \twolinecell{\cite{Maisel:2012}}{\cite{Okamoto:2013}}& \notecell{Order-disorder transf. observed, and discarded older prediction of MG.}\\
\pdnum) & Ag--Au & OS & SS & -- & Calc. & 950--1075 & 1983 & \cite{Okamoto:1983} & \notecell{Assessed P. Diag. See reference's Fig.~1. No LRO confirmed yet.}\\
\pdnum) & Ag--Pd & OS & SS & -- & Calc. & 900--1600 & 2010 & \cite{Sopousek:2010} & \notecell{Reassessed P. Diag. with new experimental data. See reference's Fig.~3.}\\
\pdnum) & Ca--Sr & \textbf{PS} & SS & -- & Calc. & 300--900 & 2008 & \cite{Aljarrah:2008} & \notecell{Reassessed P. Diag. See reference's section 2.3 and Fig.~9.}\\
\pdnum) & Co--Ni & OS & SS & -- & Calc. & 0--1600 & 1983 & \cite{Nishizawa:1983} & \notecell{Assessed P. Diag. See reference's Fig.~1\&2. Allotropic transformation observed, but not LRO.}\\
\pdnum) & Co--Ir & \textbf{PS} & SS & -- & \twolinecell{Exp.}{Exp.} & \twolinecell{400--2500}{--} & \twolinecell{1952}{2015} & \twolinecell{\cite{Koster:1952}}{\cite{Patel:2015}} & \notecell{Assessed P. Diag. See first reference's Fig.~5\&17. Nanoparticle synthesis showed similar results. No ordered compounds were observed. Bernard and Kuprina's ordered compounds~\cite{Bernard:1986} are not broadly accepted.}\\
\pdnum) & Co--Pd & OS & SS & -- & Exp. & -273--1600 & 1991 & \cite{Ishida:1991} & \notecell{Assessed P. Diag. See reference's Fig.~1. Allotropic transformation observed, but not LRO.}\\
\pdnum) & Co--Rh & \textbf{PS} & SS & -- & \twolinecell{Exp.}{Exp.} & \twolinecell{400--2500}{--} & \twolinecell{1952}{2004} & \twolinecell{\cite{Koster:1952}}{\cite{Fromen:2004}} & \notecell{Assessed P. Diag. See first reference's Fig.~4\&16. Nanoparticle synthesis showed similar results.}\\
\pdnum) & Ir--Ni & OS & SS & -- & Exp. & 500--900 & 2022 & \cite{Shubin:2022} & \notecell{Reevaluation of P. Diag. Ruled out the presence of previously proposed ordered compounds.}\\

\pdnum) & Ni--Pd & OS & SS & -- & \twolinecell{Exp.}{Calc.} & \twolinecell{-220--1600}{1177--1577} & \twolinecell{1984}{1999} & \twolinecell{\cite{Nash:1984}}{\cite{Ghosh:1999b}} & \notecell{Assessed P. Diag. See references Fig.~1\&14. MG was ruled out by X-ray, but it remains an open question.}\\
\pdnum) & Pd--Pt & OS & \twolinecell{SS, MG}{\twolinecell{SS}{SS}} & -- & \twolinecell{Calc.}{\twolinecell{Calc.}{Exp.}} & \twolinecell{400--1800}{\twolinecell{--}{400--600}} & \twolinecell{1991}{\twolinecell{1996}{2022}} & \twolinecell{\cite{Bharadwaj:1991}}{\twolinecell{\cite{Klein:1996}}{\cite{Lee:2022}}} & \notecell{Predicted SS+MG, but MG was ruled out later with theory and experiments.}\\
\pdnum) & Au--Ni & PS & SS, MG & -- & \twolinecell{Exp.}{\twolinecell{Calc.}{Calc.}} & \twolinecell{127--1527}{\twolinecell{327--1027}{127--1527}} & \twolinecell{1995}{\twolinecell{2000}{2005}} & \twolinecell{\cite{Bienzle:1995}}{\twolinecell{\cite{Colinet:2000}}{\cite{Wang:2005}}} & \notecell{All of them show evidence of MG. Last assessed Ph. Diag. obtained a critical point at \textbf{816.4}\degree C and \textbf{73.0} at.\% Ni. }\\
\pdnum) & Au--Pt & PS & SS, MG & -- & \twolinecell{Exp.}{Calc.} & \twolinecell{400--1800}{200--1800} & \twolinecell{1985}{2008} & \twolinecell{\cite{Okamoto:1985c}}{\cite{Grolier:2008}} & \notecell{The critical point of the miscibility gap is at about \textbf{1260}\degree C and \textbf{61} at.\% Pt. }\\
\pdnum) & Cu--Ni & PS & SS, MG & -- & Exp. & 397--565 & 2018 & \cite{Iguchi:2018} & \notecell{New critical proposed with nano-layered films experiment at \textbf{507}\degree C and \textbf{50.0} at.\% Ir. }\\
\pdnum) & Cu--Rh & PS & SS, MG & -- & \twolinecell{Calc.}{Calc.} & \twolinecell{500--2100}{427--1227} & \twolinecell{1982}{2000} & \twolinecell{\cite{Chakrabarti:1982}}{\cite{Priya:2000}} & \notecell{ Both assessed Ph. Diag. with MG. Most recent critical point at \textbf{1135}\degree C and \textbf{59} at.\% Rh. }\\
\pdnum) & Ir--Pd & PS & SS, MG & -- & \twolinecell{Exp.}{Exp.} & \twolinecell{500--1600}{500--2500} & \twolinecell{1959}{1991} & \twolinecell{\cite{Raub:1959}}{\cite{Tripathi:1991}} & \notecell{See Fig. 8 and Table III of first ref. Assessed Ph. Diag. Observation of MG. Critical point at \textbf{1480}\degree C and \textbf{49.6} at.\% Ir. }\\
\pdnum) & Ir--Pt & PS & SS, MG & -- & \twolinecell{Exp.}{Exp.+Calc.} & \twolinecell{500--1600}{500--2500} & \twolinecell{1959}{2014} & \twolinecell{\cite{Raub:1959}}{\cite{Okamoto:2014}} & \notecell{See Fig. 8 and Table III of first ref. Later research indicates that the previous results were not in equilibrium. Critical point recalculated with ab initio\cite{Yamabe:2009}: \textbf{1027}\degree C and \textbf{50} at.\% Ir. }\\
\pdnum) & Ir--Rh & PS & SS, MG$^{*}$ & -- & \twolinecell{Calc.}{Calc.} & \twolinecell{500--1600}{500--2500} & \twolinecell{1959}{2014} & \twolinecell{\cite{Raub:1959}}{\cite{Tripathi:1991b}} & \notecell{First ref. has a MG prediction based on difference of melting points. Second ref. calculate a critical point: \textbf{1335}\degree C and \textbf{62} at.\% Ir. Also suggest an intermetallic phase at 78 at.\% Ir}\\
\pdnum) & Ni--Rh & PS & SS, MG$^{*}$ & -- & \twolinecell{Calc.}{Calc.} & \twolinecell{-200--2200}{-273--727} & \twolinecell{1984}{2015} & \twolinecell{\cite{Nash:1984b}}{\cite{Teeriniemi:2015}} & \notecell{First ref. suggested MG. Second ref. obtain a critical point via ab initio calcs.: \textbf{492}\degree C and \textbf{73.0} at.\% Rh. }\\
\pdnum) & Pd--Rh & PS & SS, MG & -- & \twolinecell{Exp.}{\twolinecell{Exp.}{Calc.}} & \twolinecell{500--1600}{\twolinecell{500--1200}{527--2127}} & \twolinecell{1959}{\twolinecell{1994}{2022}} & \twolinecell{\cite{Raub:1959}}{\twolinecell{\cite{Tripathi:1994}}{\cite{Laurin:2022}}} & \notecell{See Fig. 8 and Table III of first ref. Assessed Ph. Diag. obtain a critical point at \textbf{937}\degree C and \textbf{45.0} at.\% Ir. }\\
\pdnum) & Ag--Co & PS & IS & Monotectic & Calc. & 227--3227 & 2009 & \cite{Zhu:2009} & --\\
\pdnum) & Ag--Cu & PS & IS & Eutectic & Calc. & 0--1200 & 2018 & \cite{Wang:2018} & \notecell{Assessed P. Diag. See reference's Fig.~4.}\\
\pdnum) & Ag--Ir & PS & IS & Eutectic & Exp. & 0--1200 & 1986 & \cite{Karakaya:1986} & -- \\
\pdnum) & Ag--Ni & PS & IS & \begin{tabular}[t]{@{}c@{}} Monotectic\\ Eutectic \end{tabular} & Calc. & 0--3500 & 2018 & \cite{Wang:2018} & \notecell{Assessed P. Diag. See reference's Fig.~8.}\\
\pdnum) & Ag--Pb & PS & IS & Eutectic & Calc. & -73--1127 & 2010 & \cite{Gierlotka:2010} & --\\
\pdnum) & Ag--Rh & PS & IS & Monotectic & Calc. & 200--2000 & 1986 & \cite{Karakaya:1986b} & -- \\
\pdnum) & Al--Pb & PS & IS & Monotectic & Calc. & 127-1527 & 2001 & \cite{Shim:2001} & -- \\
\pdnum) & Au--Co & PS & IS & Eutectic & Exp. & 0---1600 & 1985 & \cite{Okamoto:1985} & --\\
\pdnum) & Au--Ir & PS & IS & -- & Exp. & -- & 1984 & \cite{Okamoto:1984} & \notecell{No P. Diag. found. Reference highlights system immiscibility.}\\
\pdnum) & Au--Rh & PS & IS & Monotectic & Calc. & 400---3200 & 1984 & \cite{Okamoto:1984b} & --\\
\pdnum) & Cu--Ir & PS & IS & Peritectic & Calc. & 700--2500 & 1984 & \cite{Chakrabarti:1987} & --\\
\pdnum) & Co--Cu & PS & IS & Peritectic & Calc. & 200--1600 & \twolinecell{1984}{2007} & \twolinecell{\cite{Nishizawa:1984}}{\cite{Turchanin:2007}} & \twolinecell{--}{--}\\

\pdnum) & Cr--Fe & \textbf{OS} & SS,MG & -- & \twolinecell{Exp.}{Exp.} & \twolinecell{227--727}{27--927}& \twolinecell{1987}{2023} & \twolinecell{\cite{Andersson:1987}}{\cite{Dubiel:2023}} & \notecell{Experimentally confirmed MG. The OQMD-LRO originates from a very diluted compound, CrFe\(_{15}\), which is on the hull by \(6~meV/at\).} \\

\pdnum) & Cr--Mo & PS & SS,MG* & -- & Calc. & 0--3000 & 1987 & \cite{Venkatraman:1987} & \notecell{ Suggested and predicted, but not experimentally confirmed. ``[Q]uantitative information concerning the position of the miscibility gap is based solely on calculations."} \\

\pdnum) & Cr--W & PS & SS,MG & -- & Exp. & 500--3500 & 1984 & \cite{Naidu:1984} & \notecell{ Experimentally confirmed MG, using electron probe results. } \\

\end{longtable}

\vspace{3pt}
\noindent
\begin{minipage}{\phasediagtablewidth}
\footnotesize
$^{*}$Groups are classified and sorted based on the acronyms of Fig.~\ref{fig:phase_diag_class}: 
\begin{itemize}
  \item INT: Intermetallics
  \item SS, INT: Intermetallics + Solid Solution
  \item SS: Solid Solution
  \item SS, MG: Miscibility Gap + Solid Solution
  \item IS: Immiscible Solids
  \item $^{*}$: Theoretically predicted, or uncertain
\end{itemize}
\end{minipage}

\bibliographystyle{apsrev4-2}
\bibliography{supporting/references}